\documentclass[preprint,nofootinbib,aps,superscriptaddress,eqsecnum]{revtex4-1}
 \pdfoutput=1
\usepackage{soul}
 \usepackage{graphicx}
 \usepackage{amsmath}
\usepackage{amsfonts}
\usepackage{amssymb}
\usepackage{hyperref}
\usepackage[titletoc]{appendix}
\usepackage{caption}
\usepackage{subcaption}
\def\bea{\begin{eqnarray}}
\def\eea{\end{eqnarray}}
 \def\be{\begin{equation}}
\def\ee{\end{equation}}

\begin{document}
  \begin{flushright}
\preprint{\textbf{KIAS-P19026}}  
\preprint{\textbf{OU-HEP-1004}}
\end{flushright}  
  
\title{Constraining a general U(1)$^\prime$ inverse seesaw model from vacuum stability, dark matter and collider}

 \author{Arindam Das}
\email[Email Address: ]{arindam.das@het.phys.sci.osaka-u.ac.jp}
\affiliation{Department of Physics, Osaka University, Toyonaka, Osaka 560-0043, Japan}

\author{Srubabati Goswami}
\email[Email Address: ]{sruba@prl.res.in}
\affiliation{Theoretical Physics Division, 
Physical Research Laboratory, Ahmedabad - 380009, India}

 \author{Vishnudath K. N.}
\email[Email Address: ]{vishnudath@prl.res.in}
\affiliation{Theoretical Physics Division, 
Physical Research Laboratory, Ahmedabad - 380009, India}
\affiliation{Discipline of Physics, Indian Institute of Technology, Gandhinagar - 382355, India}

\author{Takaaki Nomura}
\email[Email Address: ]{nomura@kias.re.kr}
\affiliation{School of Physics, KIAS, Seoul 02455, Korea}

\begin{abstract}

We consider a class of gauged $U(1)$ extensions of the Standard Model (SM), where the light neutrino masses are generated by an inverse seesaw mechanism. In addition to the three right handed neutrinos, we add three singlet fermions and demand an extra $Z_2$ symmetry under which, the third generations of both of the neutral fermions are odd, which in turn gives us a stable dark matter candidate. We express the $U(1)$ charges of all the fermions in terms of the U(1) charges of the standard model Higgs and the new complex scalar. We study the bounds on the parameters of the model from vacuum stability, perturbative unitarity, dark matter relic density and direct detection constraints. We also obtain the collider constraints on the $Z'$ mass and the $U(1)'$ gauge coupling. Finally we compare all the bounds on the $Z'$ mass versus the $U(1)'$ gauge coupling plane.

 \end{abstract}
 
\maketitle

 \section{Introduction}

The discovery of the Higgs boson with a mass of $125$ GeV at the Large Hadron Collider (LHC) \cite{Chatrchyan:2012xdj,Aad:2012tfa} has placed the SM on a firm footing. However, the SM still does not have answers to some of the very fundamental questions like the origin of the neutrino masses and the existence of dark matter (DM). A straight forward way to include the generation of the sub-eV scale neutrino masses and the presence of the DM into the SM is by adding extra particles, which may or may not involve the extension of the SM gauge group.

Among the various beyond standard model (BSM) scenarios that have been proposed in the literature, the models in which the SM is extended by a $U(1)$ gauge group has received some attention. The models with an extra $U(1)$ gauge group naturally contain three right handed neutrinos as a result of the conditions for the gauge anomaly cancellation. Thus, the active light neutrino masses can be generated via the canonical type-I seesaw mechanism \cite{Minkowski:1977sc,seesaw1,seesaw2,Mohapatra:1979ia}. 
However, in canonical type-I seesaw model, which is considered in most of the $U(1)$ extended models, one either has to go for extremely large Majorana masses ($\sim 10^{14}$ GeV) or very small Yukawa couplings ($\sim 10^{-6}$), making it difficult to probe the heavy neutrinos at the colliders. Motivated by testability in colliders, various TeV scale extensions of the type-I seesaw model have been considered in the literature (for recent reviews, see 
\cite{Boucenna:2014zba,Deppisch:2015qwa,Cai:2017mow,Das:2018hph}). One of the most popular TeV scale seesaw models is the inverse seesaw model \cite{Mohapatra:1986bd} where the smallness of the neutrino mass can then be attributed to a small lepton 
number violating term. A tiny value  of this lepton number violating term is deemed natural, since when this parameter is zero, the global U(1) lepton number symmetry is reinstated  and neutrinos are massless.
Especially, an inverse seesaw mechanism in the context of a $U(1)_{B-L}$ extension of the SM has been studied in reference \cite{Datta:2013mta}. In these models, the presence of extra singlet fermions (in addition to the right handed neutrinos) helps us to bring down the seesaw scale (which is the $U(1)$ breaking scale) to $\sim O(\textrm{TeV})$, simultaneously allowing for large Yukawa couplings, $Y_\nu \sim O(\textrm{0.1})$.

An important aspect of the $U(1)$ extended models which has been scrutinized recently is the implications for the stability of the electroweak (EW) vacuum \cite{Iso:2009nw,Iso:2009ss,Iso:2012jn,Chakrabortty:2013zja,Coriano:2014mpa,Coriano:2015sea,Das:2015nwk,Accomando:2016sge,Oda:2015gna,Das:2016zue}.  The measured values of the SM parameters, especially the top mass $M_t$ and strong coupling constant $\alpha_s$ implies that there exists an extra deeper minima near the Planck scale ($M_{Planck}$), which threatens the stability of the present EW vacuum \cite{Alekhin:2012py,Buttazzo:2013uya}, since this may tunnel into that true vacuum. The calculation of the decay probability  suggests that the present EW vacuum is metastable at $~3\sigma$ which means that the decay time is greater than the age of the universe. It is well known that the scalar couplings pull the vacuum towards stability whereas the Yukawa couplings push it towards instability. The EW vacuum stability in the context of a class of minimal $U(1)$ extensions containing extra scalars and fermions have been studied by the authors of \cite{Chakrabortty:2013zja,Coriano:2014mpa,Coriano:2015sea,
Accomando:2016sge} and they have shown that the behaviour of the EW vacuum depends also on the $U(1)$ quantum numbers chosen, since the renormalization group equations (RGEs) depend on these quantum numbers. The conformal symmetric versions of such models have been considered in references \cite{Oda:2015gna,Das:2016zue}.

As already mentioned, the existence of the DM is another major motivation for going beyond the standard model. Measurements  by  Planck  and  WMAP  demonstrate  that  nearly  
85 percent of  the  Universe's matter density is dark \cite{Ade:2015xua}. Hence, it is very important to study models that can simultaneously explain neutrino mass as well as DM and their theoretical as well as phenomenological implications. The models with an extra $U(1)$ gauge group can accommodate a DM candidate even in the minimal version (with type-I seesaw), by adding an additional $Z_2$ symmetry \cite{Basak:2013cga,Oda:2017kwl}, where the third generation of the right handed neutrinos act as the DM candidate. Other versions of the $U(1)_{B-L}$ extension with scalar DM have been studied in \cite{Rodejohann:2015lca,Cao:2017ydw,Singirala:2017see,Camargo:2019ukv}.Also, there are various realizations of the grand unified theories (GUTs) that predict the existence of extra $Z'$ boson \cite{Langacker:1980js,Hewett:1988xc}.
The presence of the extra $Z'$ boson that couples to the quarks and the leptons also gives rise to a rich collider phenomenology in the $U(1)$ models\cite{Basso:2008iv,Lindner:2016bgg,Ekstedt:2016wyi,Das:2016zue,Accomando:2017qcs,Accomando:2016sge}. Searches for such $Z'$ boson through it's decay dileptons have been conducted by the ATLAS and the CMS collaborations and lower limits on the $Z'$ mass has been obtained \cite{Aaboud:2016cth,Khachatryan:2016zqb,Aad:2019fac}. 

In this paper, we consider a class of gauged $U(1)$ extensions of the SM, where active light neutrino masses are generated by an inverse seesaw mechanism. In addition to the three right handed neutrinos, we add three singlet fermions and demand an extra $Z_2$ symmetry under which, the third generations of both the neutral fermions are odd, which in turn gives us a stable DM candidate. This allows us to consider large neutrino Yukawa couplings and at the same time, keeping the $U(1)'$ symmetry breaking scale to be of the order of $\sim \, O(1)$ TeV. The main difference of this inverse seesaw model from that considered in \cite{Datta:2013mta} is that the extra neutral fermions that we are adding are singlets under the gauge group and hence we do not have to worry about anomaly cancellation. Also, instead of considering one particular model, we express the $U(1)$ charges of all the fermions in terms of the U(1) charges of the SM Higgs and the new complex scalar. We perform a comprehensive study of the bounds on the model parameters from low energy neutrino data, vacuum stability, perturbative unitarity and DM as well as collider constraints. The rest of the paper is organized as follows. In sections II and III, we introduce the class of the $U(1)$ models under consideration and discuss the fermionic and the scalar sectors. We discuss the fitting of the neutral fermion mass matrix in section IV, by taking all the experimental constraints into account. In section V, we discuss the RGE evolution of the couplings and present the parameter space allowed by vacuum stability and perturbative unitarity in various planes. This is followed by a discussion on the DM scenario in these models, where we present the parameter space giving the correct relic density and satisfying the direct detection bounds at the same time. In section VII, we discuss the combined bounds from vacuum stability, unitarity, DM relic density and the collider constraints and finally, we summarize in section VIII.

\section{Model and Neutrino Mass at the tree level}\label{sect2}

 The model considered is based on the gauge group $SU(3)_c \times  SU(2)_L \times U(1)_Y \times U(1)'  $. In addition to the SM particles, we have three right handed neutrinos ${\nu_R}_i$, a complex scalar $\Phi$ required to break the $U(1)'$ symmetry and three gauge singlet Majorana fermions $S_i$. An extra $Z_2$ symmetry is imposed to have a stable fermionic dark matter. The matter and Higgs sector field content  along with their transformation properties under $SU(3)_c \times  SU(2)_L \times U(1)_Y \times U(1)'  $ are given below.
\be Q_L \, = \, \begin{bmatrix}
    u_L \\
    d_L
\end{bmatrix} \sim \, (3,~2,~\frac{1}{6},~x_q) \,\,\,; \,\,\, d_R \sim \, (3,~1,~-\frac{1}{3},~x_d)\,\,\,; \,\,\, u_R \sim \, (3,~1,~\frac{2}{3},~x_u) \label{qua},\ee
\be l_L \, = \, \begin{bmatrix}
    \nu_L \\
    e_L
\end{bmatrix} \sim \, (1,~2,~-\frac{1}{2},~x_l) \,\,\,; \,\,\,e_R \sim \, (1,~1,~-1,~x_e)\,\,\,; \,\,\, \nu_R \sim \, (1,~1,~0,~x_\nu) ,\label{lep} \ee
\be H \, = \,\frac{1}{\sqrt{2}}\begin{pmatrix}
  G^{+}\\
 v+h+iG^0
\end{pmatrix}   
\sim \, (1,~2,~\frac{1}{2},~\frac{x_H}{2}) \,\,\,; \,\,\,\Phi = \frac{1}{\sqrt{2}}(\phi + u + i\chi) \sim \, (1,~1,~0,~-x_\Phi) \label{hig}, \ee
\be  S \sim (1,~1,~0,~0). \ee

Note that the generation indices have been suppressed here. Under $Z_2$, the third generation of $\nu_R$ and $S$, i.e., ${\nu_R}_3$ and $S_3$ are odd whereas all the other particles are even and we assume that this $Z_2$ is not broken. 

The $U(1)'$ charges of the fermions are defined to satisfy the gauge and gravitational anomaly-free conditions:
\be U(1)'\times [SU(3)_c]^2 \,\,\,\,\, :\,\,\,\,\, 2x_q -x_u -x_d = 0 \nonumber , \ee
\be U(1)'\times [SU(2)_L]^2   \,\,\,\,\, :\,\,\,\,\,3x_q + x_l = 0 \nonumber ,\ee
\be \nonumber  U(1)'\times [U(1)_Y]^2   \,\,\,\,\, :\,\,\,\,\, x_q - 8x_u - 2x_d +3x_l -6x_e = 0, \ee
\be \nonumber  [U(1)']^2\times U(1)_Y  \,\,\,\,\, :\,\,\,\,\, x_q^2 - 2x_u^2 +x_d^2 - x_l^2 + x_e^2 = 0, \ee
\be \nonumber  [U(1)']^3  \,\,\,\,\, :\,\,\,\,\, 6x_q^3 - 3x_u^3 -3x_d^3+2x_l^3 - x_\nu^3 - x_e^3 = 0 ,\ee
\be U(1)'\times [\textrm{grav}]^2  \,\,\,\,\, :\,\,\,\,\, 6x_q - 3x_u -3x_d +2x_l - x_\nu - x_e = 0. \ee

The most general Yukawa Lagrangian (along with the Majorana mass for $S$) invariant under $SU(3)_c \times  SU(2)_L \times U(1)_Y \times U(1)' $ that could be written using the fields given above is,
\be -L_{\textrm{Yukawa}} =  Y_e \overline{l}_L H e_R + Y_\nu \overline{l}_L \tilde{H} \nu_R + Y_u \overline{Q}_L\tilde{H} u_R + Y_d \overline{Q}_L H d_R  \,\,\,+ \,\,\, y_{NS} \overline{\nu}_R \Phi S + \frac{1}{2}  \overline{S^c}M_\mu S   \,\,\, +\,\,\, \textrm{h.c.}, \label{yuk1}\ee
where $\tilde{H} = i\sigma_2 H^*$. The invariance of this Yukawa Lagrangian under the $U(1)'$ symmetry gives us the following conditions : 
\be \frac{x_H}{2} = -x_q + x_u = x_q-x_d =-x_l+x_\nu =x_l-x_e \,\,\,\,;\,\,\,\, -x_\Phi = x_\nu  .\ee
Using these conditions and the anomaly-free conditions, the $U(1)'$ charges of all the fermions could be determined in terms of $x_H$ and $x_\Phi$ as,
\be  x_\nu = -x_\Phi \,\,\,\,;\,\,\,\, x_l = -x_\Phi-\frac{x_H}{2} \,\,\,\,; \,\,\,\, x_e = -x_\Phi - x_H  ,\nonumber \ee
\be  x_q = \frac{1}{6}(2 x_\Phi + x_H) \,\,\,\,;\,\,\,\, x_u = \frac{1}{3}(2 x_H + x_\Phi) \,\,\,\, ; \,\,\,\, x_d= \frac{1}{3}(x_\Phi - x_H)  ,      \ee

Note that the choice $x_\Phi = 1$ and $x_H = 0$ correspond to the well known $U(1)_{B-L}$ model. From Eq.(\ref{yuk1}), after symmetry breaking, the terms relevant for neutrino mass are,
  \be
 -L_{mass} = \overline{\nu}_L M_D \nu_R + \overline{\nu_R} \,M_R S \,\,+\,\, \frac{1}{2}\,\overline{S^c} M_\mu S \,\,\, +\,\,\, \textrm{h.c.} ,\ee
 where, $M_D = Y_\nu \langle H \rangle \,  $ and $M_R = y_{NS} \langle \Phi \rangle \,$. The neutral fermion mass matrix $M_\nu$ can be defined as,
 \be -L_{mass} = \,\,\ \frac{1}{2}( \, \overline{\nu_L^c} \,\, \, \overline{\nu}_R\,\,\, \overline{S^c}\,) \begin{pmatrix}
  0 & M_D^* & 0 \\
 M_D^\dagger & 0 & M_R \\
 0 & M_R^T & M_\mu
\end{pmatrix}  \begin{pmatrix}
 \nu_L \\
 \nu_R^c \\
 S
\end{pmatrix}  \,\,\, + \,\,\,\textrm{h.c.}. \label{gen}\ee
 
The mass scales of the three 
sub-matrices of $M_\nu$ may naturally have a hierarchy  $ \, M_R >>  M_D >> M_\mu \,$. Then, the effective light neutrino mass matrix in the seesaw approximation is  given by, 
\be M_{light} \,\, = \,\, M_D^* (M_R^T)^{-1}M_\mu M_R^{-1} M_D^\dagger \label{eqISMmlight} . \ee

Because of the extra $Z_2$ symmetry, the Yukawa coupling matrices $Y_\nu$ and $y_{NS}$ and hence the mass matrices $M_D$ and $M_R$ will have the following textures,
\be M_R = y_{NS}\langle\Phi\rangle \sim \begin{pmatrix}
  \times & \times & 0 \\
 \times & \times & 0 \\
 0 & 0 & \times
\end{pmatrix} \,\,\,\,\, \textrm{and} \,\,\,\,\, M_D = Y_\nu\langle H\rangle \sim \begin{pmatrix}
  \times & \times & 0 \\
 \times & \times & 0 \\
 \times & \times & 0
\end{pmatrix} . \ee 
In addition, we will choose $M_\mu$ to be diagonal without loss of generality. Since ${\nu_R}_3$ and $S_3$ do not mix with other neutral fermions, they will not contribute to the seesaw mechanism and we will have a minimal inverse seesaw mechanism (3 $\nu_L$ + 2 $\nu_R$ + 2 S case) in which the lightest active neutrino will be massless. The two fermions ${\nu_R}_3$ and $S_3$ mix among themselves and the lightest mass eigenstate could be a stable DM candidate. In the heavy sector, we will have two pairs of degenerate pseudo-Dirac neutrinos of masses of the order  $\sim$ $M_R \, \pm \, M_\mu$ that mix with the active light neutrinos. 
Thus, we have an inverse seesaw mechanism in which the smallness of $M_{light}$ is naturally attributed to the smallness of both $M_\mu$ and $ \,\, \frac{M_D}{M_R}$. For instance, $M_{light} \sim \mathcal{O}\,(0.1)$ eV can easily be achieved by taking  $ \,\, \frac{M_D}{M_R} \sim 10^{-2}\,\,$ 
and $\,M_\mu \sim \mathcal{O}\,$(1) keV. Thus, the seesaw scale can be lowered down considerably for typical values of the parameters -- $Y_\nu \,\sim \, \mathcal{O}(0.1)$, $M_D \,\sim \, 10 $ GeV and $M_R \, \sim \, 1$ TeV.

\section{Scalar Potential of the Model and Symmetry Breaking}

The scalar potential of the model is given by,
\be  V(\Phi , H)\,\, = \,\, {m_1^2}H^\dagger H \,+\, \
{\lambda_1}(H^\dagger H)^2 \,+\, \lambda_3 H^\dagger H \,\Phi^\dag \Phi \,+ \, {m_2^2}\Phi^\dag \Phi \,
+\,{\lambda_{2}} (\Phi^\dag \Phi)^2 \, \label{eqpot1}. \ee

The trivial conditions that give a stable potential are,
\be  \lambda_1>0 \,\,\,\,\,;\,\,\,\,\, \lambda_2>0\,\,\,\,\, \textrm{and} \,\,\,\,\, \lambda_3>0,\ee
and if $\lambda_3<0$, the stability of the potential can still be achieved by satisfying the following conditions :
\be \lambda_1 > 0, \,\,\,\,\, \lambda_2 > 0 , \,\,\,\,\, 4 \lambda_1 \lambda_2 - \lambda_3^2 > 0.  \ee

The above conditions are obtained by demanding the Hessian matrix corresponding to the potential to be positive definite at large field values \cite{Chakrabortty:2013zja,Kannike:2012pe,Chakrabortty:2013mha}.

The two scalar fields acquire vacuum expectation values($vev$s) given by,
\be \langle H \rangle = \frac{1}{\sqrt{2}}\begin{pmatrix}
   0 \\
 v 
\end{pmatrix} \,\,\,\,\, ; \,\,\,\,\,\langle \Phi \rangle = \frac{u}{\sqrt{2}} . \ee
The values of $v$ and $u$ are determined by the minimization conditions and are given by,

\be v^2 = \frac{m_2^2\lambda_3/2 - m_1^2\lambda_2}{\lambda_1\lambda_2 - \lambda_3^2/4}\,\,\,\,\,;\,\,\,\,\, u^2 = \frac{m_1^2\lambda_3/2 - m_2^2\lambda_1}{\lambda_1\lambda_2 - \lambda_3^2/4} .  \ee

After symmetry breaking, the mixing between the fields $h$ and $\phi$ could be rotated away by an orthogonal transformation to get the physical mass eigenstates as,
The values of $v$ and $u$ are determined by the minimization conditions and are given by,

\be v^2 = \frac{m_2^2\lambda_3/2 - m_1^2\lambda_2}{\lambda_1\lambda_2 - \lambda_3^2/4}\,\,\,\,\,;\,\,\,\,\, u^2 = \frac{m_1^2\lambda_3/2 - m_2^2\lambda_1}{\lambda_1\lambda_2 - \lambda_3^2/4} .  \ee

After symmetry breaking, the mixing between the fields $h$ and $\phi$ could be rotated away by an orthogonal transformation to get the physical mass eigenstates as,
The values of $v$ and $u$ are determined by the minimization conditions and are given by,

\be v^2 = \frac{m_2^2\lambda_3/2 - m_1^2\lambda_2}{\lambda_1\lambda_2 - \lambda_3^2/4}\,\,\,\,\,;\,\,\,\,\, u^2 = \frac{m_1^2\lambda_3/2 - m_2^2\lambda_1}{\lambda_1\lambda_2 - \lambda_3^2/4} .  \ee

After symmetry breaking, the mixing between the fields $h$ and $\phi$ could be rotated away by an orthogonal transformation to get the physical mass eigenstates as,
\be   \begin{pmatrix}
   h_1 \\
 h_2
\end{pmatrix} = \begin{pmatrix}
   \textrm{cos}\theta & -\textrm{sin}\theta \\
 \textrm{sin}\theta & \textrm{cos}\theta 
\end{pmatrix}   \begin{pmatrix}
   h \\
 \phi
\end{pmatrix} ,\label{Higgsmixing} \ee
  The masses of the scalar eigenstates are, 
  \be m_{h_{1,2}}^2 = \lambda_1 v^2 + \lambda_2 u^2 \mp \sqrt{(\lambda_1 v^2 - \lambda_2 u^2)^2 + (\lambda_3 u v)^2} . \ee
From these, one can get the relations,
  \be \lambda_1= \frac{m_{h_{1}}^2}{4 v^2} (1 + \textrm{cos}2\theta) +  \frac{m_{h_{2}}^2}{4 v^2} (1 - \textrm{cos}2\theta), \nonumber \ee
  \be \lambda_2= \frac{m_{h_{1}}^2}{4 u^2} (1 - \textrm{cos}2\theta) +  \frac{m_{h_{2}}^2}{4 u^2} (1 + \textrm{cos}2\theta) ,\nonumber \ee
  \be\lambda_3 = \textrm{sin}2\theta \Big( \frac{m_{h_{2}}^2-m_{h_{1}}^2}{2uv}  \Big) . \label{scalarcouplings}\ee

  We use these equations to set the initial conditions on the scalar couplings $\lambda_1, \lambda_2$ and $\lambda_3$ while running the renormalization group equations. Also, from the above equations, one can get,
  \be \textrm{tan}2\theta = \frac{\lambda_3 u v}{\lambda_1 v^2 - \lambda_2 u^2}.  \ee

  \subsection{Perturbative Unitarity}
  
  In addition to the vacuum stability conditions, the constraints from the perturbative unitarity conditions also put bounds on the model parameters. By considering the $hh \rightarrow hh$ and $\phi \phi \rightarrow \phi \phi$ processes, one can derive combined constraints on the three couplings appearing in the scalar potential\cite{Huffel:1980sk,Duerr:2016tmh} :
  \be |\lambda_3| \leq 8 \pi \,\,\,\,\, ; \,\,\,\,\, 3(\lambda_1 + \lambda_2) \pm \sqrt{\lambda_3^2 + 9(\lambda_1-\lambda_2)^2} \leq 8\pi \ee
  Demanding the other running couplings to remain in the perturbative regime gives us,
  \be g_i \leq \sqrt{4 \pi},  \ee
  where $g_i$ stands for SM gauge couplings. For the $U(1)$ gauge coupling $g'$, we require,
  
  \be (x_{q, d, u, l, e, \nu, \Phi}) g', ~  (x_{H}/2)g'   < \sqrt{4 \pi}. \ee

\section{Numerical Analysis and Parameter Scanning in the Neutrino Sector}

To study the parameter space allowed by vacuum stability as well as perturbativity bounds up to $M_{Planck}$ using the RGEs, we have to first fix the initial values for all the couplings. While setting the initial values for the neutrino Yukawa couplings $Y_\nu$ and $y_{NS}$, we have to make sure that they reproduce the correct oscillation parameters and satisfy all the experimental constraints. To do this, we find sample benchmark points for $Y_\nu$, $y_{NS}$ and $M_{\mu}$ and the vev of the extra scalar $\Phi $($u )$ by fitting them with all the constraints using the downhill simplex method \cite{Press:1996}. Note that here, $Y_\nu$ is a complex $3 \times 2$ matrix, $y_{NS}$ is a complex $2 \times 2$ matrix and $M_{\mu} $ is a $2 \times 2$ diagonal matrix with real entries. The various constraints we have taken are:

\begin{table}[ht]
 $$
 \begin{array}{|c|c|}
 \hline {\mbox {Parameter} }& NH  \\
 \hline
  \Delta m^2_{sol}/10^{-5} eV^{2}&            6.80 \rightarrow 8.02 \\
   \Delta m^2_{atm}/10^{-3} eV^{2}&             +2.399 \rightarrow +2.593 \\
          \sin^2\theta_{12}&            0.272\rightarrow 0.346  \\
          \sin^2\theta_{23}&            0.418 \rightarrow 0.613  \\
          \sin^2\theta_{13}&            0.01981 \rightarrow 0.02436\\
                
 \hline
 \end{array}
 $$
 \label{table f}\caption{\small{ The oscillation parameters in their $3\sigma$ range,  for NH as given by the global analysis
of neutrino oscillation data with three 
light active neutrinos \cite{Esteban:2016qun}. 
 }}\label{oscdata}\end{table}

\begin{itemize}
\item Cosmological constraint on the sum of light neutrino masses as given by the Planck 2018 results \cite{Aghanim:2018eyx}. This puts an upper limit on the sum of active light neutrino masses to be,
  \be\Sigma \, = \, m_1 + m_2 + m_3 < 0.14 \, \textrm{eV} .\ee
 Note that in our case, the lightest active neutrino is massless and also we are restricting our analysis only to the normal hierarchy (NH) of the active neutrino masses since the vacuum stability, dark matter and collider analyses are independent of the hierarchy of the light neutrino masses.  In addition, it has been found that the best fit of the data is for the NH and IH is disfavored with a $ \Delta \chi^2 = 4.7 (9.3)$ without (with)   Super-Kamiokande  atmospheric neutrino data \cite{Esteban:2018azc}. Thus we have,
  \be  m_1 = 0 \,\,,\,\, m_2 = \sqrt{\Delta {m_{sol}}^2} \,\, ; \,\, m_3 =  \sqrt{\Delta m_{atm}^2}  \ee
  
  \item The constraints on the oscillation parameters in their $3\sigma$ range, given by the global analysis \cite{Capozzi:2016rtj,Esteban:2016qun} 
of neutrino oscillation data with three 
light active neutrinos following NH are given in Table \ref{oscdata}. We use the standard parametrization of the PMNS matrix in which,
\be U_\nu = \begin{pmatrix}
  c_{12}c_{13} & s_{12}c_{13} & s_{13} e^{-i\delta} \\
 -c_{23}s_{12}-s_{23}s_{13}c_{12}e^{i\delta} & c_{23}c_{12}-s_{23}s_{13}s_{12}e^{i\delta} & s_{23}c_{13} \\
 s_{23}s_{12}-c_{23}s_{13}c_{12}e^{i\delta} & -s_{23}c_{12}-c_{23}s_{13}s_{12}e^{i\delta} & c_{23}c_{13}
\end{pmatrix} P \ee
where $ c_{ij} = cos\theta_{ij} \,\, , \,\, s_{ij} = sin\theta_{ij}$ and the phase matrix 
$P = \textrm{diag}\, (1,\,e^{i\alpha_2}, \, e^{i(\alpha_3 + \delta)})$  contains the Majorana phases.

\item The constraints on the non-unitarity of $U_{PMNS}$ = $U_L$  as given by the analysis of electroweak precision observables
  along with various other low energy precision observables \cite{Antusch:2016brq}. At 90$\%$ confidence level,  we have,
   \be |U_L U_L^\dagger|\, =\,  \begin{pmatrix}
  0.9979-0.9998 & <10^{-5} & <0.0021 \\
 <10^{-5} & 0.9996-1.0 & <0.0008 \\
 <0.0021 & <0.0008 & 0.9947-1.0
\end{pmatrix}. \ee
  
   This also takes care of the constraints coming from various charged lepton flavor violating decays like $ \, l_i \rightarrow l_j \, \gamma $. For example, the branching ratio for the decay $\mu \, \rightarrow e \, \gamma $ is constrained as \cite{TheMEG:2016wtm}, \be \textrm{Br}(\mu \, \rightarrow e \, \gamma)  \,\, < \, 4.2 \times 10^{-13} .\ee 
  
  In addition, it has been shown in reference \cite{Coy:2018bxr} that the $\mu \rightarrow e$ conversion in nuclei can give the strongest bound out of all the flavor violating observables in the case of type-I seesaw models. The bound on the branching ratio for the $\mu \rightarrow e$ conversion in Gold ($Au$) nucleus reads as \cite{Bertl:2006up},
  \be  \rm{Br}(\mu ~Au \rightarrow e~ Au) < 7 \times 10^{-3}.\ee
  This has been converted into a bound on the parameter $\hat{R}_{e \mu}$  in reference \cite{Coy:2018bxr} as, 
  \be  \hat{R}_{e \mu} < 9.7 \times 10^{-6},\ee
  
  where,
  \be  \hat{R}_{e \mu} = 2 \sum_{j}( {Y_\nu})_{e j }^* ({Y_\nu})_{\mu j} ~ \Big( \frac{m_W^2}{M_j^2} \Big) ~ \textrm{Log} \Big (\frac{M_j}{m_W} \Big)\ee

where $j = 1,2$, $M_1, M_2$ are the heavy neutrino masses such that $M_1 \neq M_2$ and the factor of 2 takes care of the degeneracy in mass spectrum. In our fitting, we have made sure that the parameter sets that we consider satisfy all these bounds. 

\end{itemize}

In table (\ref{benchmark}), we give two benchmark points consistent with all the experimental data discussed above. As a consistency check, we also give the value of  $Br(\mu \rightarrow e~\gamma)$ obtained at the two benchmark points.

\begin{table}[ht]
 $$
 \begin{array}{|c|c|c|c|}
 \hline {\mbox {Parameter} }& BM-I & BM-II\\
 \hline

    Tr[Y_{\nu}Y^{\dag}_{\nu}]&             0.0898 & 0.4000 \\

    [Y_\nu]_{3\times 2} \,\, &

     \begin{pmatrix}
     0.0694 - i\, 0.1182 & 0 - i\, 0.0499   \\
 0.0038 - i\, 0.0022  & 0.0778 +i\, 0.0442  \\
-0.0008 - i\, 0.2183  & -0.0071 -i\, 0.1128 
\end{pmatrix}  &

\begin{pmatrix}
  -0.0210 +i\, 0.2269 &  -0.0329 +i\, 0.0036   \\
 0.0495 -i\, 0.0352 & - 0.2321 - i\,
   0.3021  \\
 -0.1081 -i\, 0.3771  & 0.1450 
+i\, 0.1526
\end{pmatrix}    \\

    Tr[y_{NS}Y^{\dag}_{R}]&            0.0101 & 0.1472 \\

    [y_{NS}]_{2\times 2} \,\, &

     \begin{pmatrix}
     
  0.0031 - i\, 0.0082 & 0.0375 -i\, 0.0351   \\
 0.0821 +i\, 0.0093 &  -0.0002 -i\, 0.0241
 
\end{pmatrix}  &

 \begin{pmatrix}
 0.2861 +i\, 0.0073 & -0.0025 +i\, 0.1521  \\
0.0623 -i\, 0.0545 &  -0.1596 - i\,0.0990 
\end{pmatrix}      \\

    [M_\mu]_{2\times 2} \,\,GeV &

     \begin{pmatrix}
1.0921 \times 10^{-6} & 0  \\
 0 & -2.2092\times 10^{-8}
\end{pmatrix}  &

 \begin{pmatrix}
 1.2655\times 10^{-8} & 0 \\
 0 & -2.5248\times 10^{-8}
\end{pmatrix}   \\

              M_j~GeV &      1766.82,\, 1766.82,\, 3085.87,\, 3085.87  & 2227.88, \,2227.88,\, 3659.58,\, 3659.58  \\

Br(\mu \rightarrow e~\gamma)&         4.0946\times 10^{-13}  & 2.2954\times 10^{-13}\\

u~ (TeV) &         50  & 12\\

 \hline
 \end{array}
 $$
 \label{benchmark}\caption{\small{Two sample benchmark points for the neutrino sector. The above parameters give the correct mixing angles and satisfies the non-unitarity constraints on $U_{PMNS}$. The value of $Br(\mu \rightarrow e~\gamma)$ is given as a check.
 }}\label{benchmark}\end{table}


\section{RGE Evolution}

The couplings in any quantum field theory get corrections from higher-order loop diagrams and as a result, the couplings run with the renormalization scale. We have the renormalization group equation (RGE) for a coupling $C$ as, 
\be \mu\frac{dC}{d\mu} \, = \, \sum_{i} \frac{\beta_{C}^{(i)}}{(16\pi^2)^i} , \ee
 where i stands for the $i^{th}$ loop and $\beta_{C}$ is the corresponding $\beta$ function.

We have evaluated the SM coupling constants at the  the top quark mass scale and then run them using the RGEs from $M_t$ to $ M_{Planck}$. For this, we have taken into account the various threshold corrections at $M_t$ \cite{Sirlin:1985ux,Melnikov:2000qh,Holthausen:2011aa}.  Then the SM RGEs are used to run all the couplings up to the $vev$ of the new scalar, after which, the new couplings enter. 
The modified RGEs for the  $SU(3)_c \times  SU(2)_L \times U(1)_Y \times U(1)'  $ have been used. These have  been generated using SARAH \cite{Staub:2013tta}. 
We have used two-loop RGEs for all the SM parameters and $g'$ and the new scalar couplings $\lambda_2$ and $\lambda_3$, whereas for the neutrino Yukawa couplings, we have used the one-loop RGEs. The one- and the two-loop RGEs of the model are given in the appendix. Throughout this paper, we have fixed the standard model parameters as $m_h = 125.7$ GeV, $M_t = 173.4$ GeV and $\alpha_s = 0.1184$. Also, we have kept the $U(1)$ gauge mixing to be $0$ at the scale $u$ throughout this paper.

\begin{figure}
\begin{center}
\includegraphics[scale=0.22]{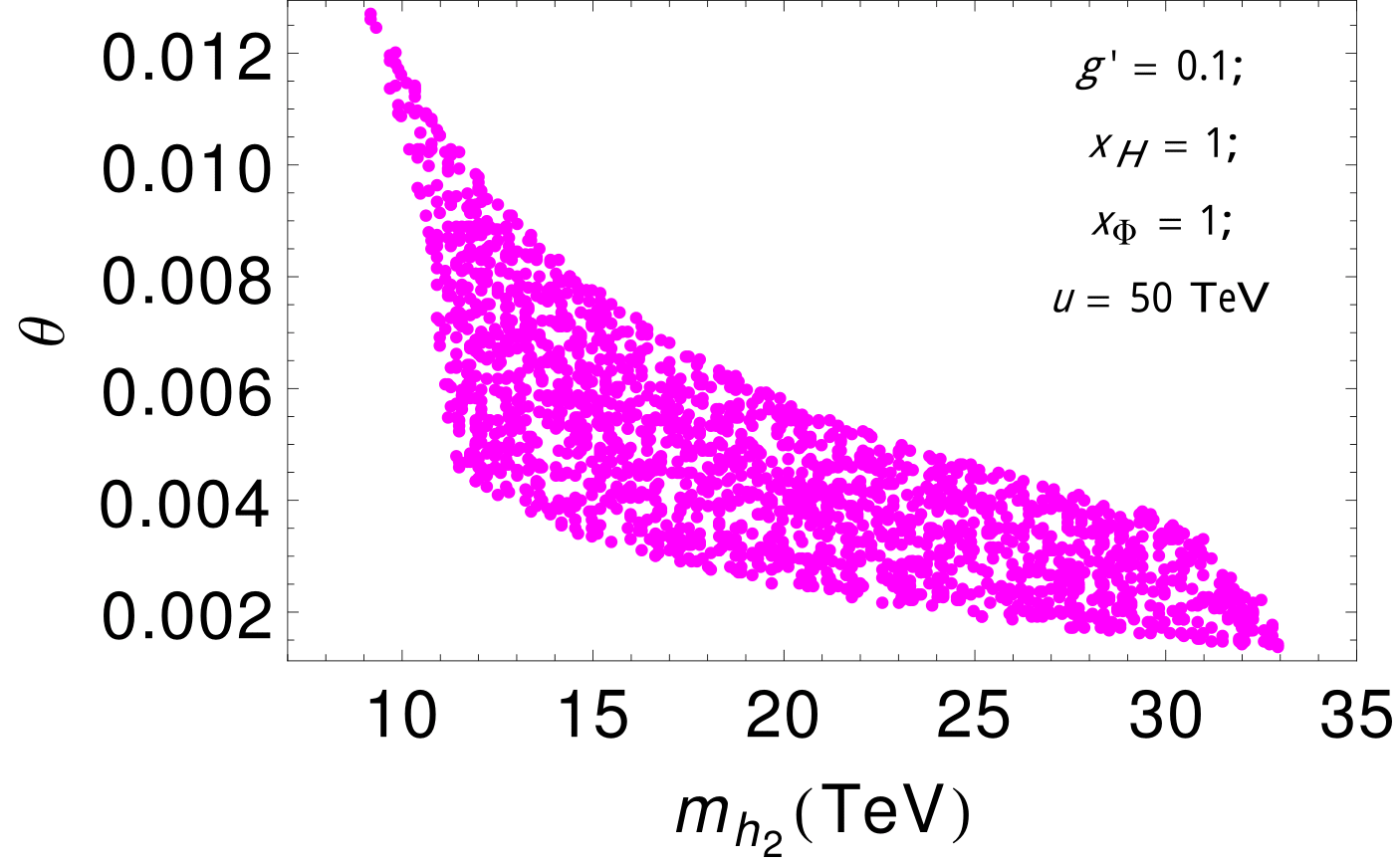}
\end{center}
\caption{Region in the $m_{h_2}-\theta$ plane allowed by both vacuum stability and perturbativity bounds up to $M_{Planck}$ for the model with $x_H = x_\Phi=1$. For the neutrino Yukawa couplings, we have used BM-I from the Table \ref{benchmark} and we have fixed $g'=0.1$ and $y^{33}_{NS} = 0.5$.}
\label{mH2thetaBL}
\end{figure}

Fig. \ref{mH2thetaBL} displays the allowed region in the  $m_{h_2}-\theta$ plane for the model with $x_H= x_\Phi=1$, keeping all the other parameters fixed. For the neutrino Yukawa couplings, we have used BM-I from the Table \ref{benchmark} and we have fixed $g'=0.1$ and $y^{33}_{NS} = 0.5$. From the figure, one can see that for higher values of $\theta$, only smaller values of $m_{h_2}$ are allowed whereas for smaller values of $\theta$, larger values of $m_{h_2}$ over a wider range are allowed. Also it can be seen that for this model with the considered set of parameters, the values of $m_{h_2}>33$ TeV and $\theta > 0.013$ are disallowed.

\begin{figure}[h!]
    \centering
    \begin{subfigure}[h]{0.49\textwidth}
        \includegraphics[width=\textwidth]{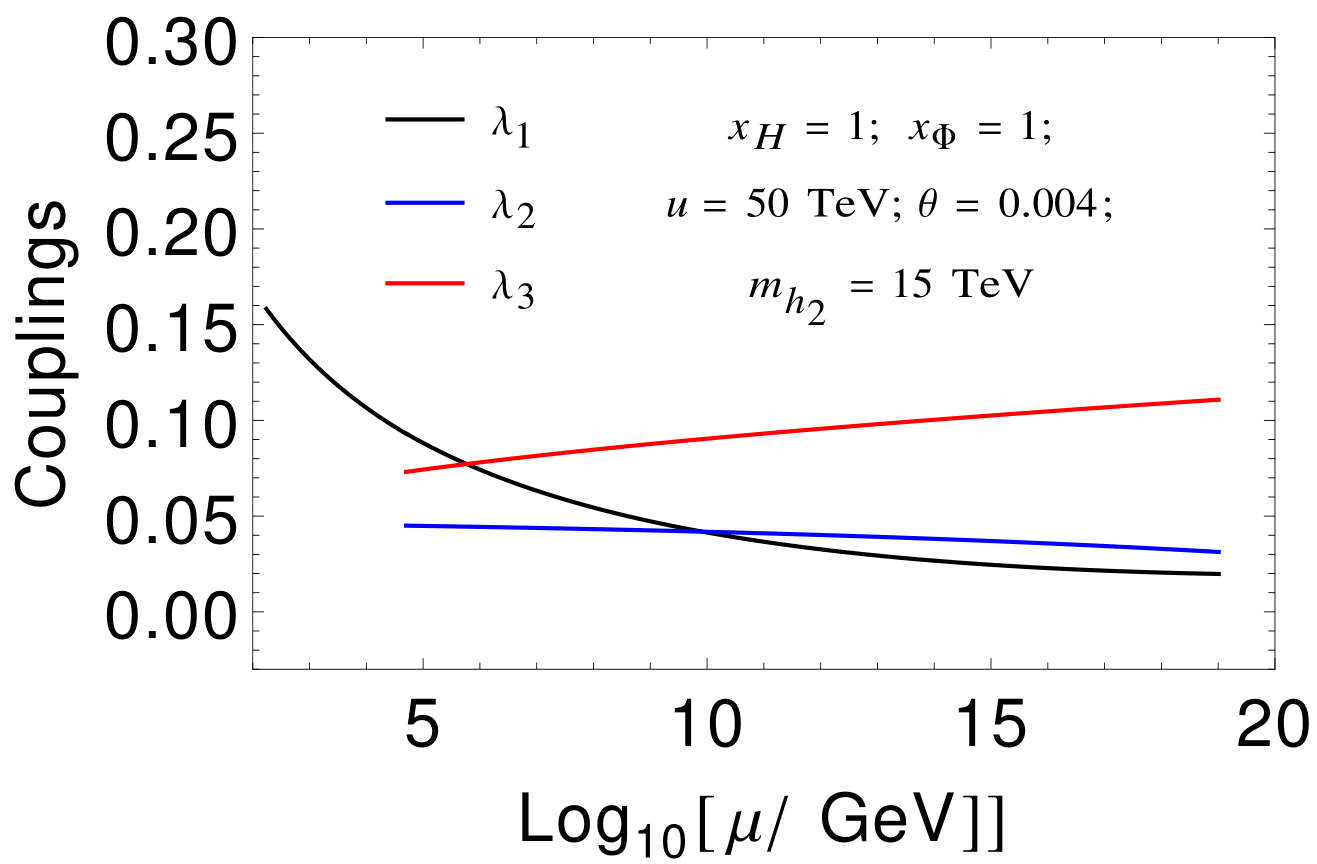}
        \caption{\centering $m_{h_2} = 15 $ TeV, $\theta = $ 0.004} \label{RGE1}
    \end{subfigure}
    ~ 
    \begin{subfigure}[h]{0.49\textwidth}
        \includegraphics[width=\textwidth]{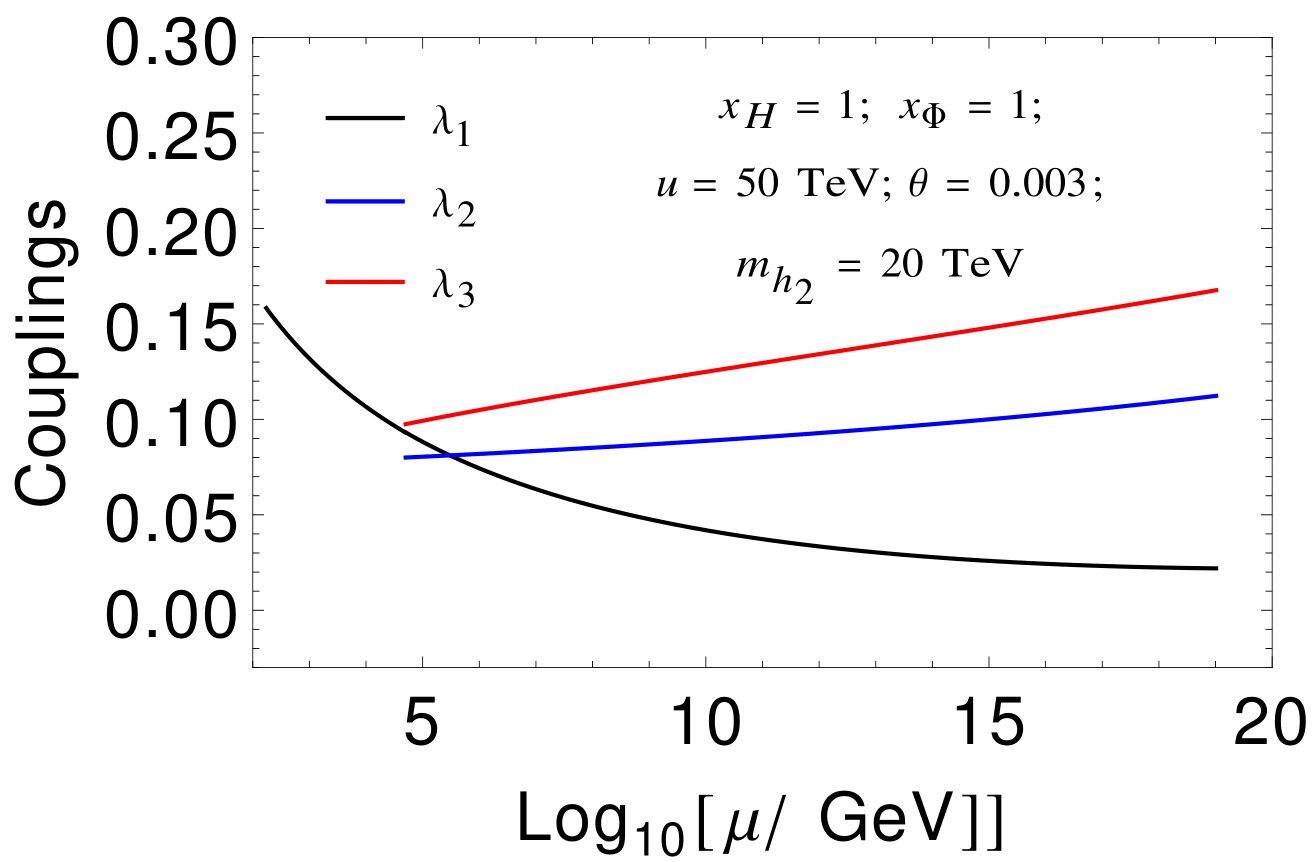}
        \caption{\centering $m_{h_2} = 20 $ TeV, $\theta = $ 0.003} \label{RGE2}
            \end{subfigure}

      \caption{Running of $\lambda_1$, $\lambda_2$, $\lambda_3$ and $4\lambda_1 \lambda_2 - \lambda_3^2$ for the model with $x_H=x_\Phi=1$ for two different values of $m_{h_2}$ and $\theta$.  For the neutrino Yukawa couplings, we have used BM-I from the Table \ref{benchmark} and we have fixed $g'=0.1$ and $y^{33}_{NS} = 0.5$.}\label{lambdarunning}
\end{figure}

  In Fig. \ref{lambdarunning}, we have plotted the running of $\lambda_1$, $\lambda_2$ and $\lambda_3$  for the model with $x_H = x_\Phi=1$ for two different values of $m_{h_2}$ and $\theta$. The figure in the left side is for $m_{h_2} = 15 	$ TeV and $\theta = $ 0.004 whereas the one in the right side is for $m_{h_2} = 20$ TeV and $\theta = $ 0.003.  For the neutrino Yukawa couplings, we have used BM-I from the Table \ref{benchmark} and we have fixed $g'=0.1$ and $y^{33}_{NS} = 0.5$. We can see that all the three quartic couplings remain positive up to $M_{Planck}$ for both the cases implying that the electroweak vacuum is absolutely stable. This can be seen from Fig. \ref{mH2thetaBL} as well where the above mentioned points fall in the stable region. Here, the presence of the extra scalar coupling helps in stabilizing the vacuum.

\begin{figure}[h!]
    \centering
    \begin{subfigure}[h]{0.49\textwidth}
        \includegraphics[width=\textwidth]{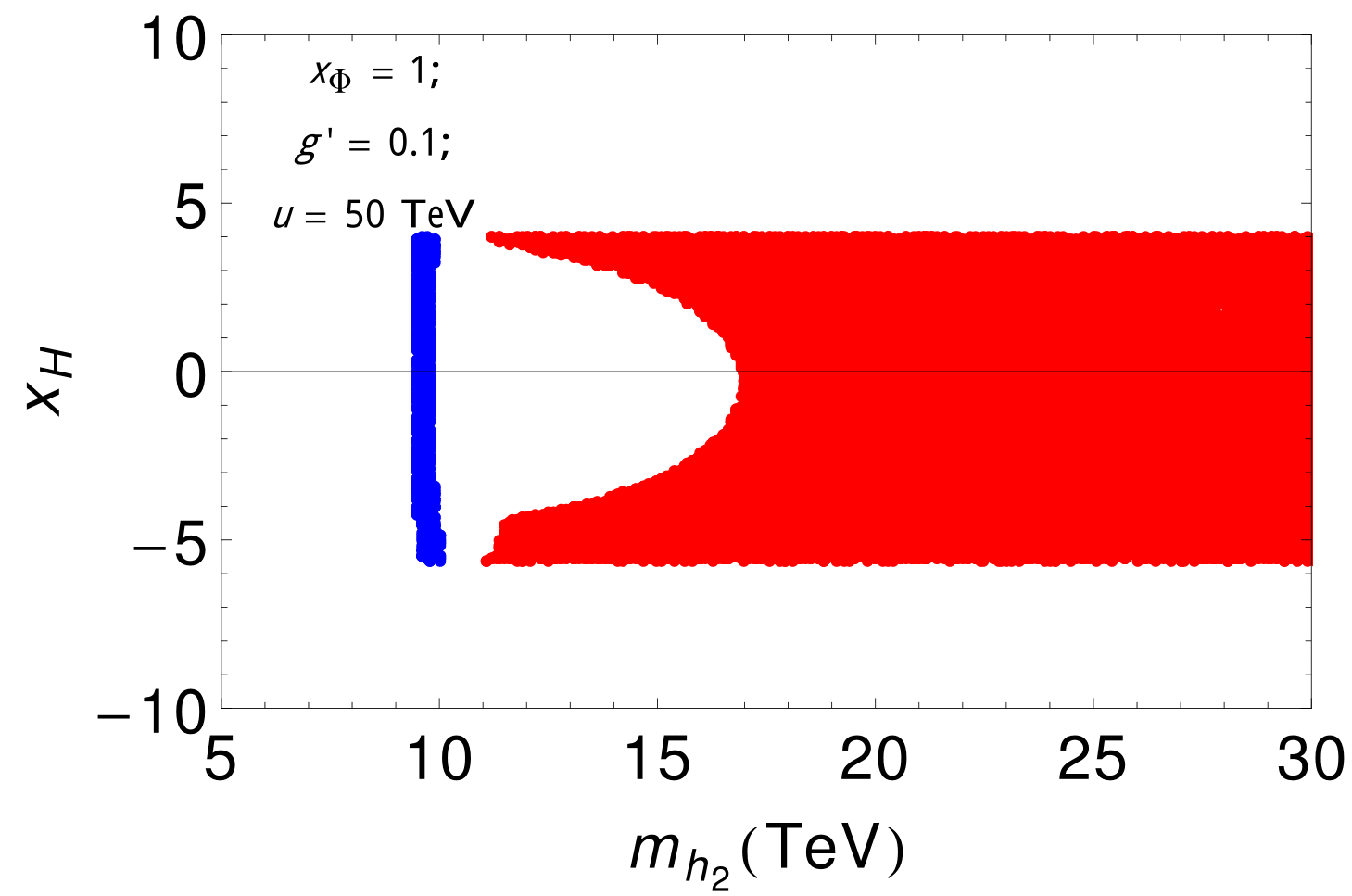}
         \caption{ \centering }
         \label{mh2xh}
    \end{subfigure}
    ~ 
    \begin{subfigure}[h]{0.49\textwidth}
        \includegraphics[width=\textwidth]{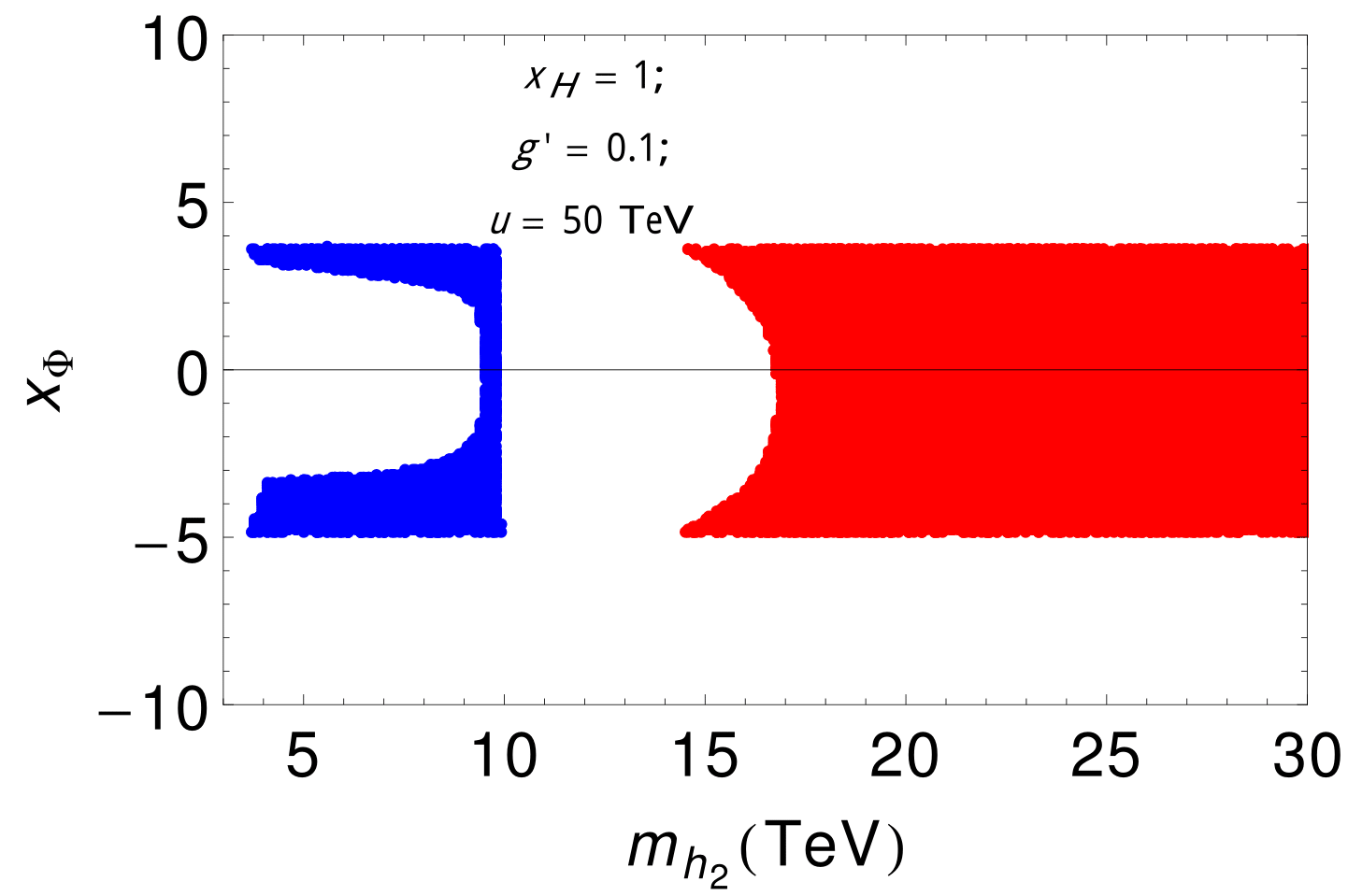}
         \caption{ \centering }
         \label{mh2xphi}
            \end{subfigure}

      \caption{Regions in the $m_{h_2}-x_H$ and $m_{h_2}-x_\Phi$ planes allowed by both vacuum stability and perturbativity bounds up to $M_{Planck}$ for two different values of $\theta$. For the left panel, we have fixed $x_\Phi = 1$ and for the right panel, we have fixed $x_H =1$. For the neutrino Yukawa couplings, we have used BM-I from the Table \ref{benchmark} and we have fixed $g'=0.1$ and $y^{33}_{NS} = 0.5$. The red region is for $\theta = 0.003$ and the blue region is for $\theta = 0.01$.}\label{mh2xhxphi}
\end{figure}

In Fig. \ref{mh2xhxphi}, we have plotted the regions allowed by both vacuum stability and perturbativity bounds up to $M_{Planck}$ in the $m_{h_2}-x_H$ and $m_{h_2}-x_\Phi$ planes, for two different values of $\theta$. The red regions are for $\theta = 0.003$ and the blue regions are for $\theta = 0.01$. 
Fig. \ref{mh2xh} shows the allowed regions in the  $m_{h_2}-x_H$ plane keeping all the other parameters fixed. For the neutrino Yukawa couplings, we have used BM-I from the Table \ref{benchmark} and we have fixed $x_\Phi=1$, $g'=0.1$ and and $y^{33}_{NS} = 0.5$.  It can be seen that for $\theta = 0.01$, a very  narrow region of $m_{h_2}$ in the range $\approx 9 - 10$ TeV is allowed by the stability and perturbativity constraints and the corresponding allowed range of $x_H$ is $\approx -5.7-4.1$. Here, the higher values of $m_{h_2}$ are disfavored by the perturbativity constraints whereas the lower values of $m_{h_2}$ are disfavored by the constraints from vacuum stability. At the same time, for $\theta= 0.003$, $m_{h_2} \approx 11 - 30 $ TeV is allowed depending on the value of $x_H$.

Similarly, in Fig. \ref{mh2xphi}, we have shown the allowed region in the  $m_{h_2}-x_\Phi$ plane keeping $x_H=1$ and all the other parameters fixed for two different values of $\theta$. Here also, for $\theta = 0.01$, the values of $m_{h_2}$ greater than 10 TeV are disfavored by unitarity constraints. The lower values of $m_{h_2}$ are disfavored by the stability constraints depending on the value of $x_\Phi$. For $-3\leq x_\Phi \leq 3 $, values of $m_{h_2}$ less than $\sim 9$ TeV are disallowed, whereas for  $-5.5\leq x_\Phi \leq -3 $ and $3\leq x_\Phi \leq 4 $, values of $m_{h_2}$ as low as $\sim 3$ TeV are allowed. For $\theta = 0.003$, values of $m_{h_2} < 14 - 15.5 $ TeV are disallowed depending on the values of $x_H$, but values as high as 30 TeV are allowed for $ -5 \leq x_H \leq 4 $. These results are consistent with the observations from Fig. \ref{mH2thetaBL} where we have seen that for $x_H = x_\Phi = 1$, larger(smaller) values of $m_{h_2}$ are disfavored for larger(smaller) values of $\theta$.

\begin{figure}[h!]
    \centering
    \begin{subfigure}[h]{0.49\textwidth}
        \includegraphics[width=\textwidth]{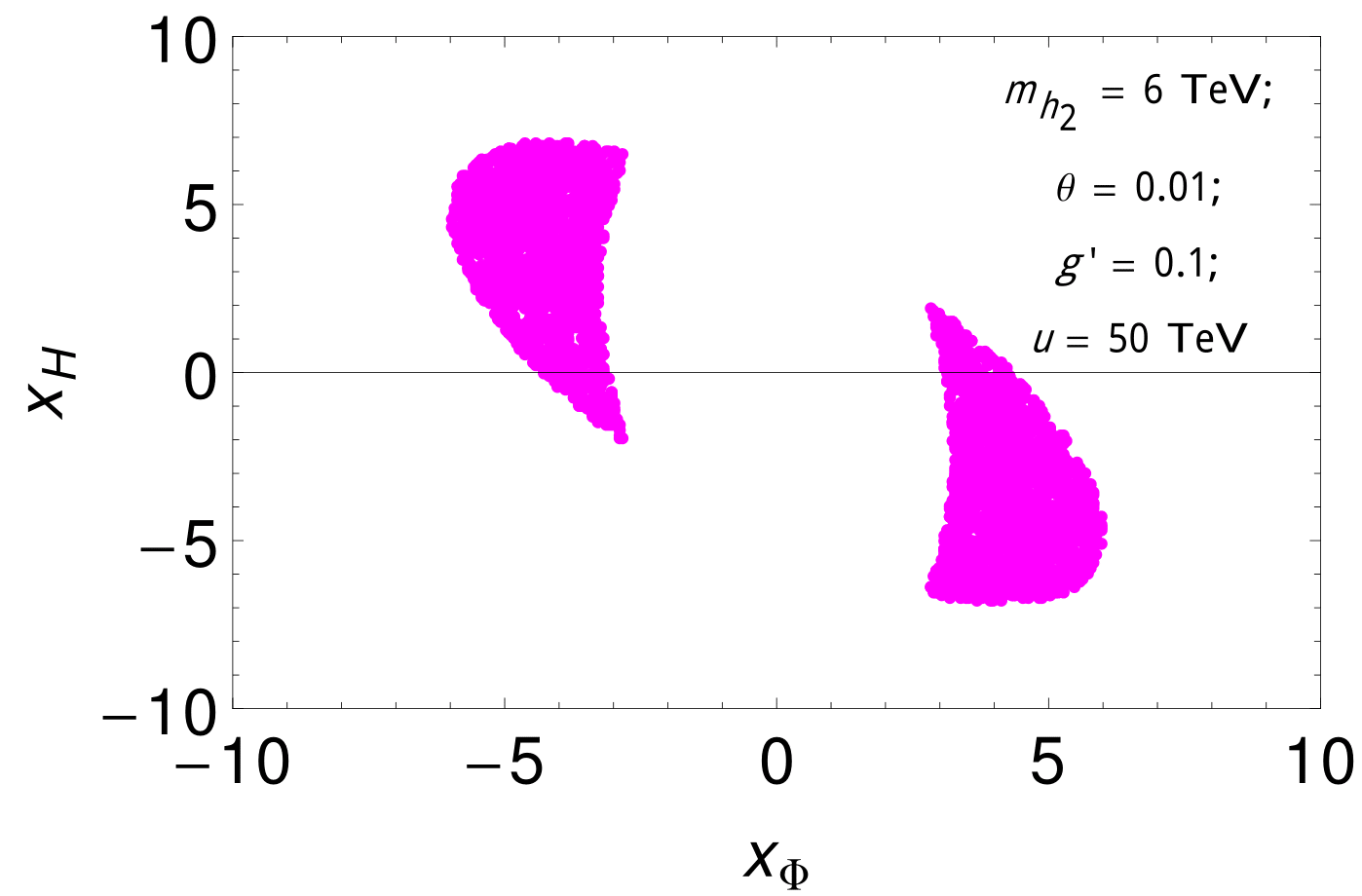}
        \caption{\centering $m_{h_2} = 6$ TeV} \label{xhxphi6}
    \end{subfigure}
    ~ 
    \begin{subfigure}[h]{0.49\textwidth}
        \includegraphics[width=\textwidth]{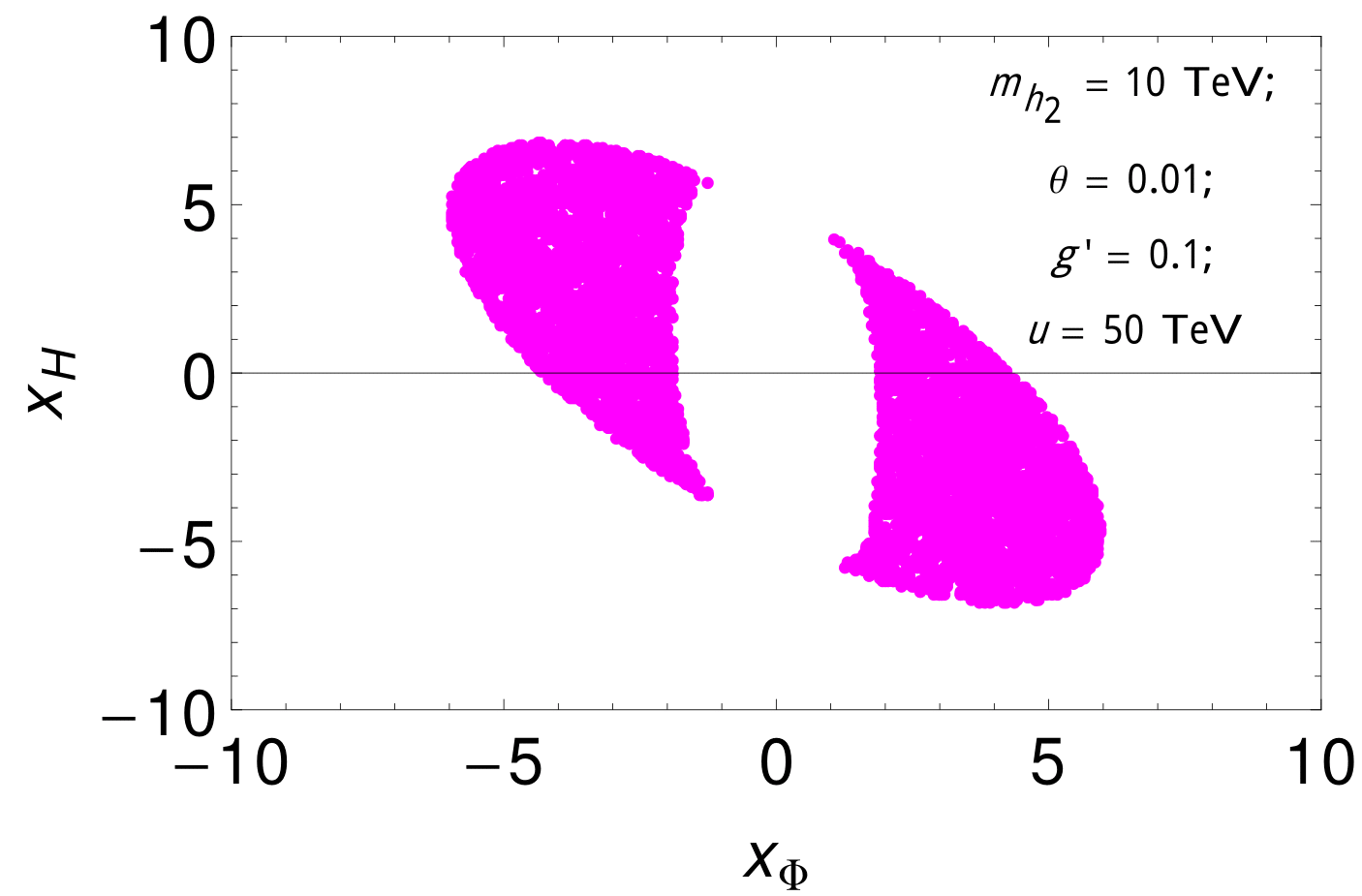}
        \caption{ \centering $m_{h_2} = 10 $ TeV} \label{xhxphi10}
            \end{subfigure}

      \caption{Regions in the $x_\Phi-x_H$ plane allowed by both vacuum stability and perturbativity up to $M_{Planck}$. We have taken the mass of the extra scalar to be 6 TeV (10 TeV) in the left (right) panel. For the neutrino Yukawa couplings, we have used BM-I from the Table \ref{benchmark} and we have fixed $\theta = 0.01$, $g'=0.1$ and $y^{33}_{NS} = 0.5$ for both the plots.}\label{xhxphi}
\end{figure}

In Fig.\ref{xhxphi}, we have presented the regions in the $x_\Phi-x_H$ plane allowed by both vacuum stability (absolute stability) and perturbativity up to $M_{Planck}$ for fixed values of $m_{h_2}$, $\theta$ and $g'$. For the neutrino Yukawa couplings, we have used the BM-I in Table \ref{benchmark} and we have taken and $y^{33}_{NS} = 0.5$. The mass of the extra scalar have been taken to be 6 TeV (10 TeV) in the left (right) panel and the values of $\theta $ and $g'$ are taken to be 0.01 and 0.1   respectively for both the plots. 
From these two figures, we can see that increasing the scalar mass will allow more values of $x_\Phi$ for a given value of $x_H$. In fact, one can see that the allowed values for $x_\Phi$ lie in the ranges  $\approx \pm 3$ to $ \pm 6$ and $\approx \pm 1$ to $ \pm 6$ for the figures in the left and the right panels respectively. Also, $x_H$ lies in the range $\approx -7$ to $7 $ for both the cases with the considered values of the parameters. This can be understood from Eq.\ref{scalarcouplings} which shows that higher value of $m_{h_2}$ implies higher value of the scalar couplings which in turn favors stability.

\begin{figure}[h!]
    \centering
    \begin{subfigure}[h]{0.49\textwidth}
        \includegraphics[width=\textwidth]{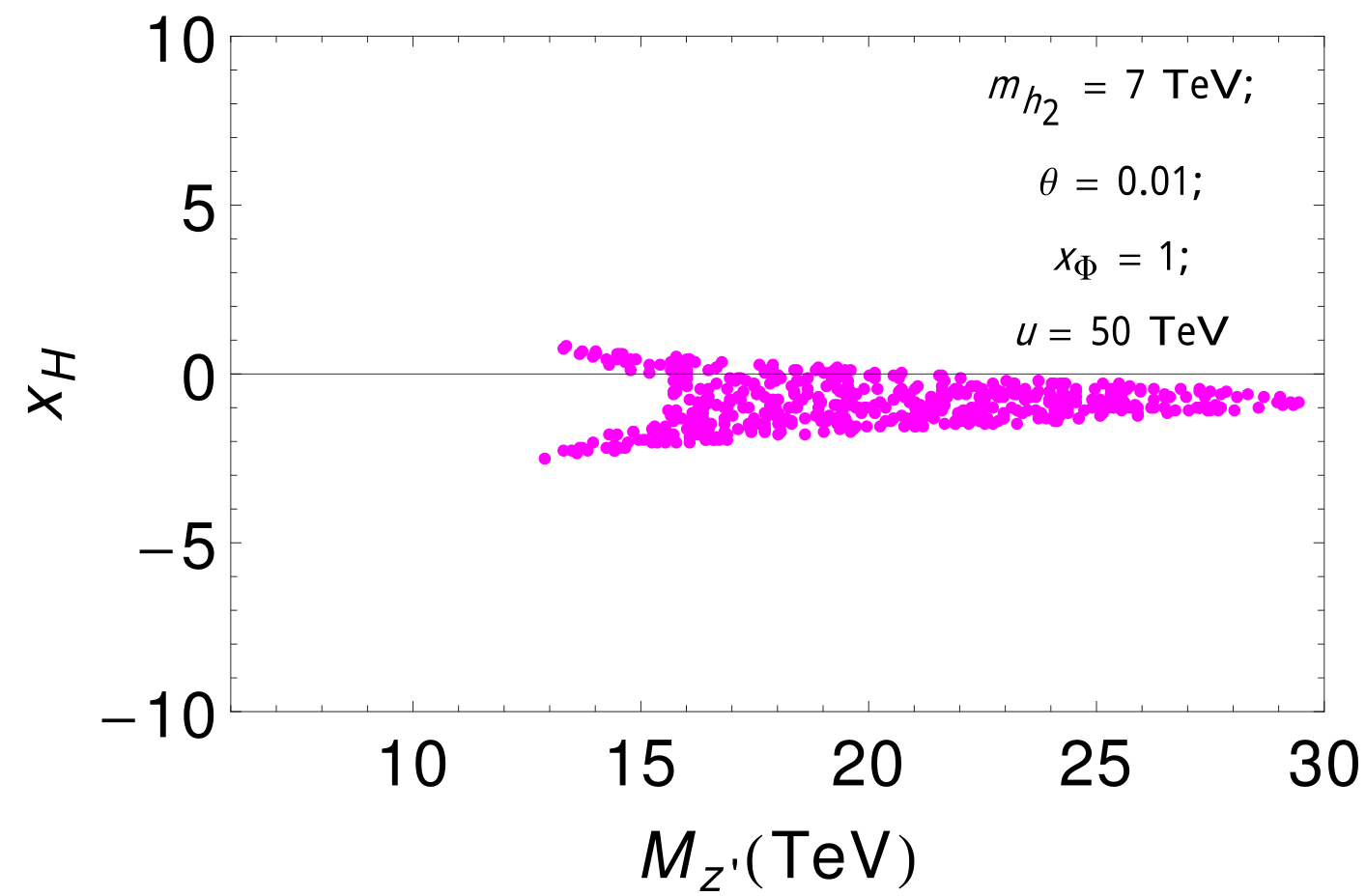}
        \caption{\centering $m_{h_2} = 7$ TeV} \label{mzxh7}
    \end{subfigure}
    ~ 
    \begin{subfigure}[h]{0.49\textwidth}
        \includegraphics[width=\textwidth]{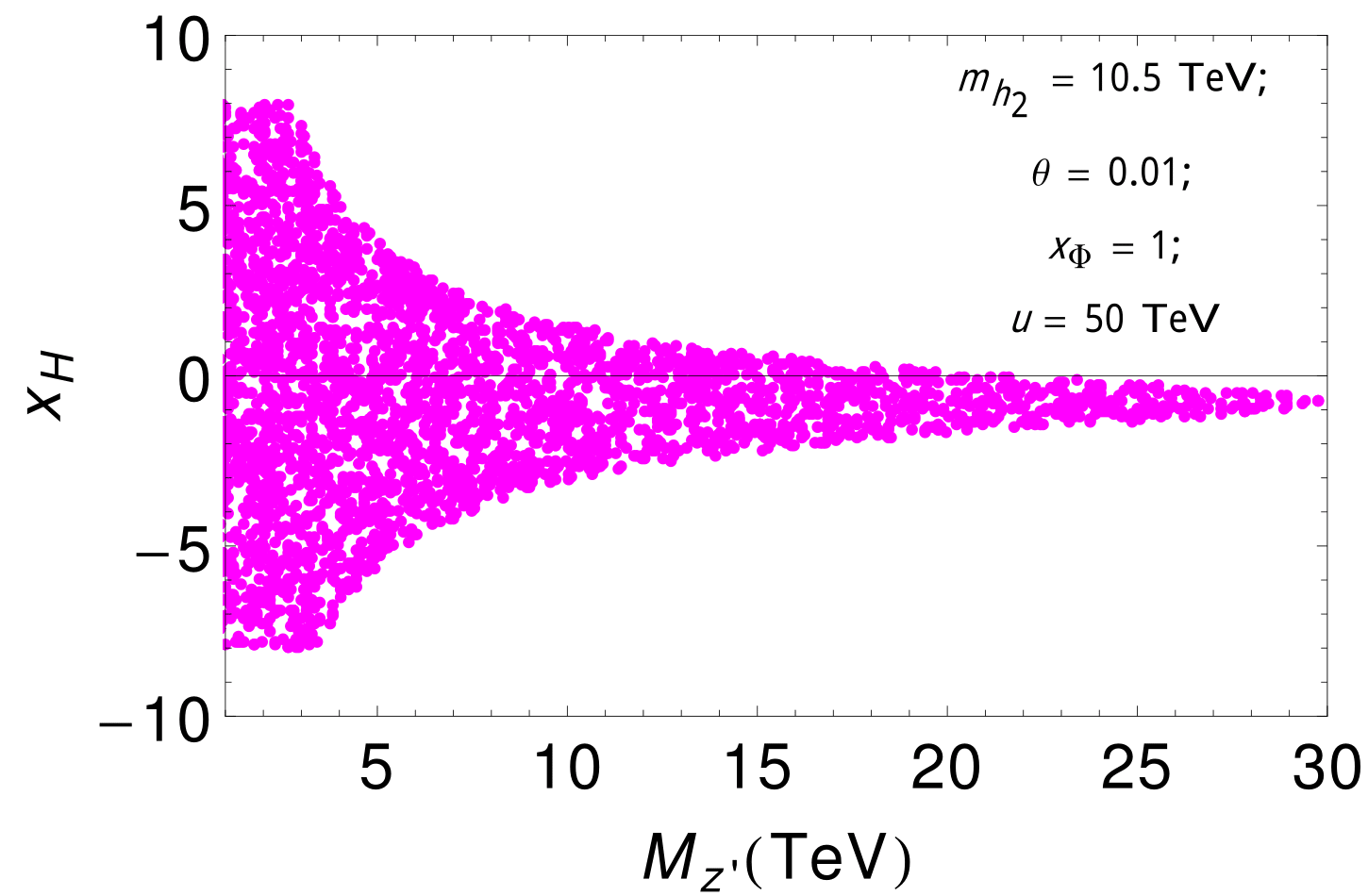}
        \caption{ \centering $m_{h_2} = 10.5$ TeV} \label{mzxh10}
            \end{subfigure}

      \caption{Regions in the $M_{Z'}-x_H$ plane allowed by both vacuum stability and perturbativity bounds up to $M_{Planck}$. We have taken the mass of the extra scalar to be 7 TeV (10.5 TeV) in the left (right) panel. For the neutrino Yukawa couplings, we have used BM-I from the Table \ref{benchmark} and we have fixed $\theta = 0.01$, $x_\Phi=1$ and and $y^{33}_{NS} = 0.5$ for both the plots.}\label{mzxh}
\end{figure}

Fig.\ref{mzxh}, displays the regions allowed by both vacuum stability and perturbativity up to $M_{Planck}$  in the $M_{Z'}-x_H$ plane for fixed values of $m_{h_2}$, $\theta$ and $x_\Phi$. Here also, we have used the BM-I in Table \ref{benchmark} for the neutrino Yukawa couplings and we have taken and $y^{33}_{NS} = 0.5$. The mass of the extra scalar have been taken to be 7 and 10.5 TeV in the left and the right panels respectively and the values of $\theta $ and $x_\Phi$ are taken to be 0.01 and 1 for both the plots. Also, we have varied $g'$ from 0 to 1 keeping $u$ fixed at 50 TeV and $x_H$ in the range -8 to 8. The corresponding values of $M_{Z'}$ have been calculated using,
\be  M_{Z'}=  \sqrt{(x_\Phi g^\prime u)^2 + (\frac{x_H}{2} g^{\prime} v_{\rm SM})^2}. \label{Zmass} \ee 
From these figures, we can see that lower values of $M_{Z'}$ allow large values of $x_H$ (or, equivalently lower values of $g^\prime$). From these figures, one can see that for a lower scalar mass, the lower values of $M_Z'$ (or equivalently, lower values of $g'$) are disfavored. For $m_{h_2} = 7$ TeV, values of $M_{Z'}$ less than 12 TeV
are disallowed and a very small range of $x_H$ is allowed whereas for $m_{h_2}= 10.5$ TeV, values of $M_{Z'}$ as low as 1 TeV are allowed and correspondingly, $x_H$ is allowed from $-8$ to $8$.

\section{Dark matter scenario}
In this section we discuss dark matter physics in our model with respect to the constraints from relic density and  direct detection experiments. As mentioned earlier, the third generations of $N_R$ and $S_L$  $(N_R^3, S_L^3)$ are odd under the $Z_2$ parity  in the general $U(1)^\prime$ inverse seesaw model that we consider.
 This ensures the stability of $N_R^3$ and $S_L^3$ which is required for these to be potential DM candidates. 
As a result the relevant interactions in the Lagrangian can be written as
\bea
-\mathcal{L}_{mass}^{2}\supset y_{NS}^{33} \overline{N_R^3} S_L^3 \Phi + M_S^{33} \overline{S_L^{3c}} S_L^3.
\label{mass11}
\eea
Note that $N_R^3$ can not couple to the SM Higgs and lepton doublets due to the $Z_2$ symmetry.
After the symmetry breaking we have $\langle \Phi \rangle =\frac{u}{\sqrt{2}}$ and the mass matrix can be written as, 
\bea
M_{N^3 S^3}=
\begin{pmatrix}
0&M_{NS}^{33}\\
M_{NS}^{33}&M_{S}^{33}
\end{pmatrix}
\eea
where $M_{NS}^{33}=\frac{y_{NS}^{33}u}{\sqrt{2}}$.  
Now rotating the basis we can write the physical eigenstates as 
\bea
\begin{pmatrix}
N_R^{3c}\\
S_L^3
\end{pmatrix}
=
\begin{pmatrix}
\cos\overline{\theta}&\sin\overline{\theta}\\
-\sin\overline{\theta}&\cos\overline{\theta}
\end{pmatrix}
\begin{pmatrix}
\psi_{1}\\
\psi_{2}
\end{pmatrix}
\eea
where $\tan 2 \overline{\theta}= \left| \frac{2M_{NS}^{33}}{-M_{S}^{33}} \right|=\sqrt{2}\frac{y_{NS}^{33} u}{M_{S}^{33}}.$ 
Note that $\psi_{1}$ and $\psi_2$ are Majorana fermions. The mass eigenvalues are obtained as
\bea
m_{\psi_1, \psi_2} = \frac{1}{2} \sqrt{(M_S^{33})^2 + 4 (M_{NS}^{33})^2} \mp \frac{1}{2} M_S^{33},
\eea
where we take $m_{\psi_1} < m_{\psi_2}$. 
Thus $\psi_1$ is the lightest $Z_2$ odd particle and our DM candidate.
Putting $\psi_1$ and $\psi_2$ back into Eq.~\ref{mass11} along with the physical mass eigenstates of $h$ and $\phi$ we write the interaction among $Z_2$ odd fermion and scalars as,
\bea
-\mathcal{L}\supset y_{NS}^{33}\Big(-\sin\theta \cos\overline{\theta} \cos\overline{\theta}~h_1+ \cos\theta \sin\overline{\theta} \sin\overline{\theta}~h_2\Big)\Big(-\overline{\psi_1^c} \psi_1+ \overline{\psi_2^c} \psi_2\Big).
\label{DMint1}
\eea
Then the DM candidate can annihilate through the scalar portal (Fig. \ref{DM1}a), where interactions between $h_2$ and SM particles are induced by scalar mixing (See Eq.\ref{Higgsmixing}) and these couplings are equal to the SM Higgs couplings times $\sin \theta$.
In addition, the DM can annihilate to the SM particles via $Z'$ exchange (Fig. \ref{DM1}c) where the gauge interactions are given by,
\bea
\mathcal{L}\supset - \frac{ x_\Phi g' }{2} Z'_\mu \left( \cos^2 \overline{\theta} \bar \psi_1 \gamma^\mu \gamma_5 \psi_1 + \sin^2 \overline{\theta} \bar \psi_2 \gamma^\mu \gamma_5 \psi_2  
- 2 \cos \overline{\theta} \sin \overline{\theta} \bar \psi_1 \gamma^\mu \gamma_5 \psi_2 \right).
\eea
 Furthermore, DM can annihilate into $Z'Z'$ mode via scalar portal where the relevant scalar-$Z'Z'$ interaction is given by
 \begin{equation}
 \mathcal{L} \supset \frac{M_{Z'}^2}{u} \cos\theta ~h_2 Z' Z' - \frac{M_{Z'}^2}{u} \sin \theta ~h_1 Z' Z'.
 \label{Eq:scalarZpZp}
 \end{equation}

\subsection{Relic density}

\begin{figure}
\begin{center}
\includegraphics[scale=0.75]{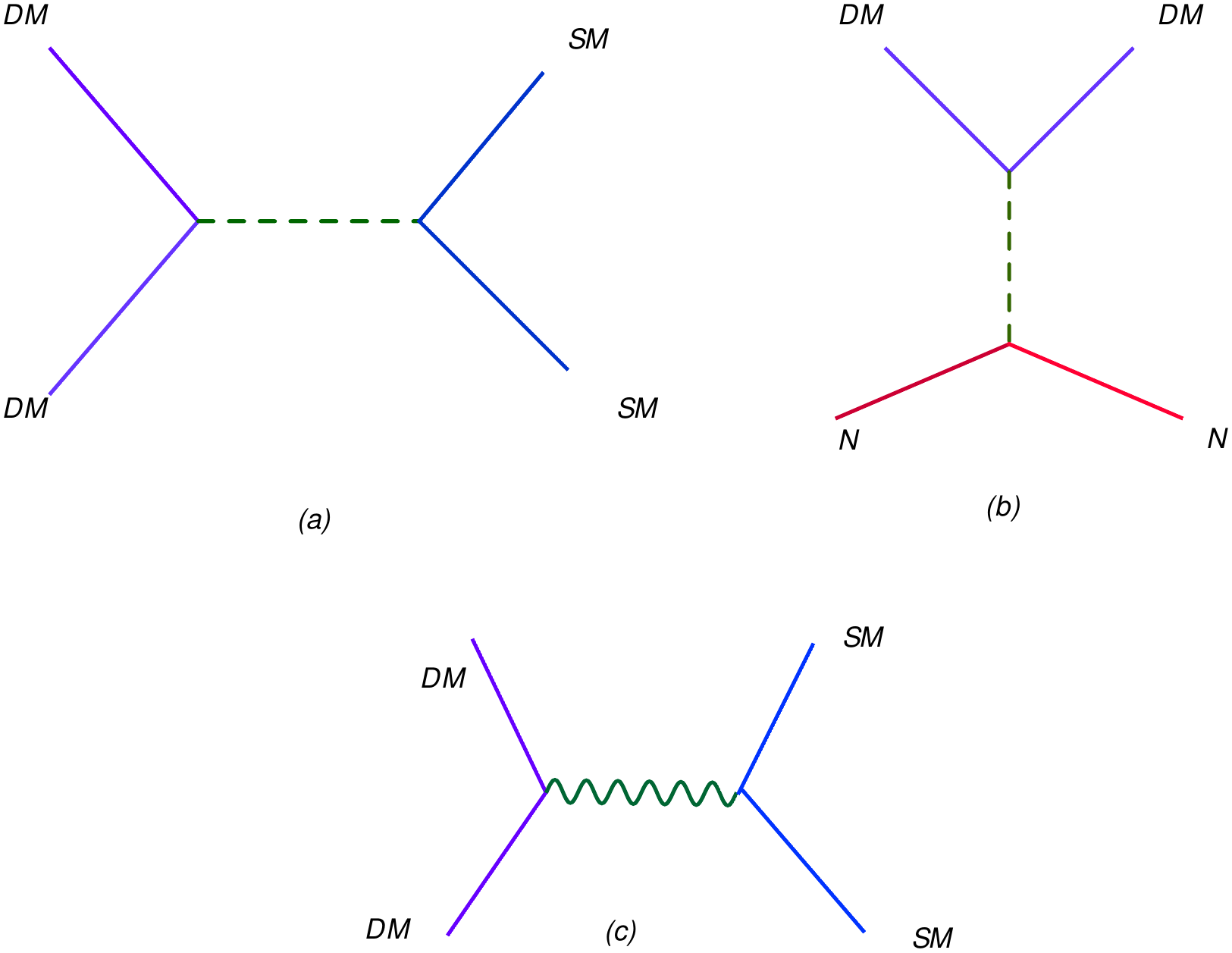}
\end{center}
\caption{(a) Scalar mediated DM annihilation (b) Direct detection and (c) $Z^{\prime}$ mediated DM annihilation.}
\label{DM1}
\end{figure}
\begin{figure}
\begin{center}
\includegraphics[scale=0.2]{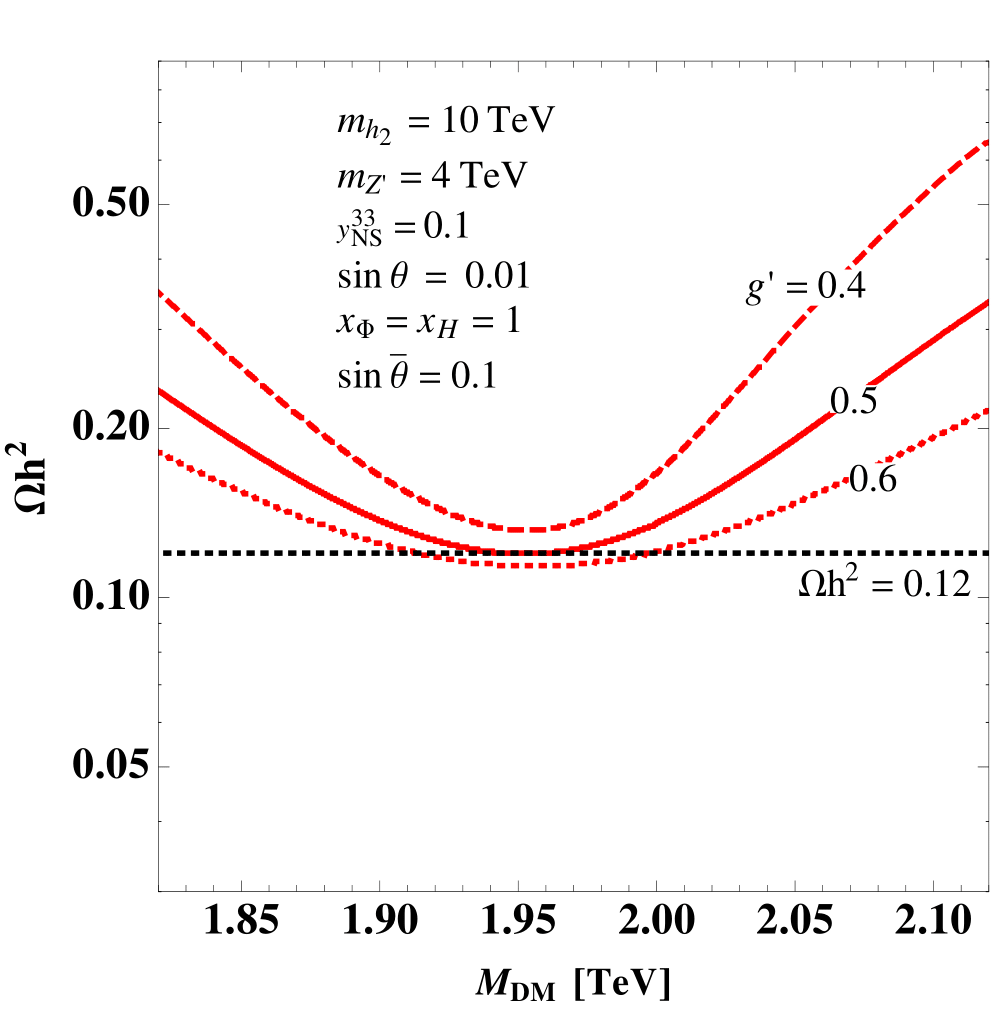}
\end{center}
\caption{Relic abundance as a function of DM mass for different values of $g'$. All the other parameters have been fixed as given in the plot.}
\label{Relic1}
\end{figure}

Here we analyze the relic density of our DM candidate.
The DM candidate $\psi_1$ annihilate into the SM particles via processes induced by $Z'$ and scalar boson interactions as shown in Fig.~\ref{DM1}.
Then we estimate the relic density using {\it micrOMEGAs 4.3.5}~\cite{Belanger:2014vza} implementing the relevant interactions.
Firstly we focus on the parameter space where the $Z'$ mediated process dominates for DM annihilation.
For illustration, in Fig.~\ref{Relic1}, we show the relic density as a function of DM mass ($M_{DM} \equiv m_{\psi_1}$) for $m_{Z'} = 4$ TeV, fixing the other parameters as indicated in the plot. 
The plot indicates that the required gauge coupling is $g' \gtrsim 0.5$ but it is excluded by the LHC data as we will see later.
Note that in this case, the value of $g'$ that gives the correct relic density depends on the choice of $x_H$ and $x_\Phi$ since the interaction strength of $Z'$ with the other particles is a product of $g'$ and a linear combination of $x_H$ and $x_\Phi$. If we increase $x_H$ and $x_\Phi$, then the value of $g'$ that can give the correct relic density can be lowered. However, for smaller values of $g'$, the LHC constraints imply much lower values of $M_Z'$ where the $Z'$ exchange is not a dominant process. We also find that the $Z'$ mediated process cannot provide sufficient annihilation cross section to explain the observed relic density if DM is heavier than $\sim 3$ TeV, complying with the  requirement that the gauge coupling satisfy $(x_{q,d,u,l,e,\nu,\Phi}) g', ~  (x_{H}/2)g'   < \sqrt{4 \pi}$ for perturbativity. This tendency comes from the fact that the annihilation cross section is P-wave suppressed since our DM is Majorana fermion.

\begin{figure}
\begin{center}
\includegraphics[scale=0.2]{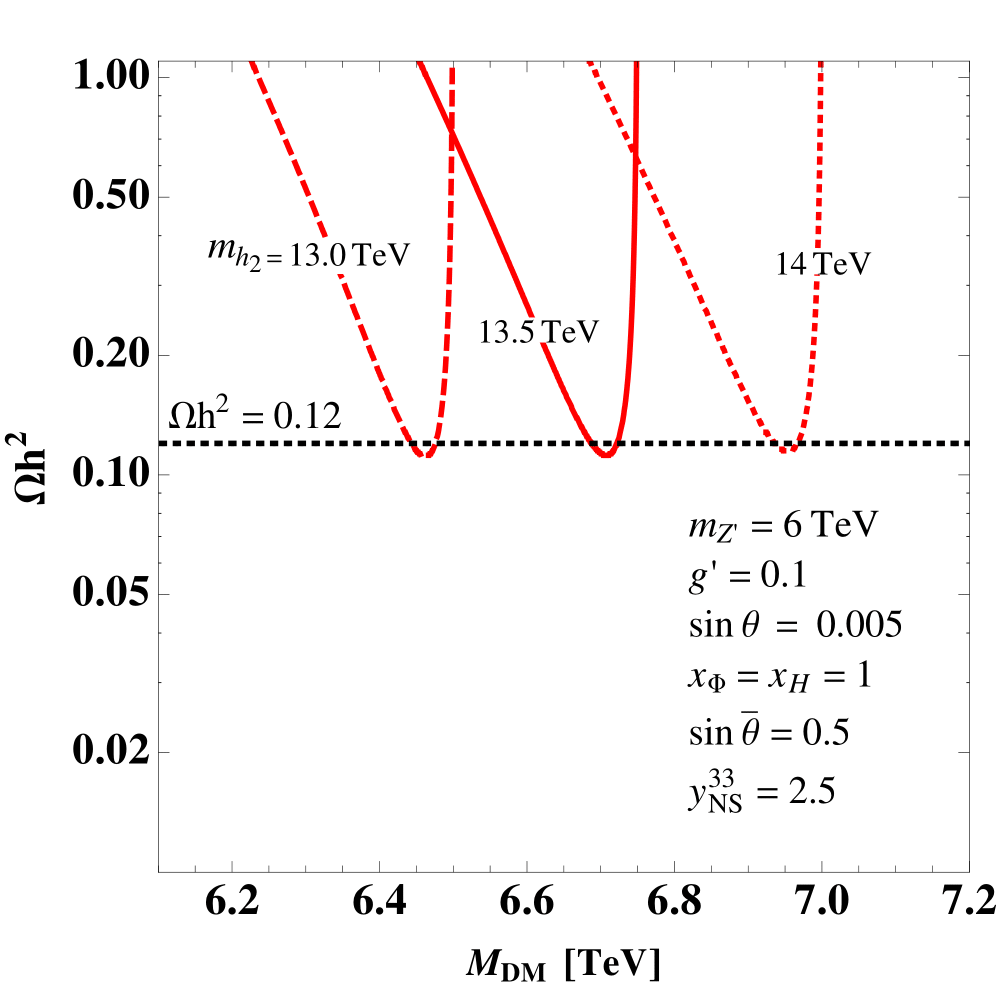} \quad
\includegraphics[scale=0.2]{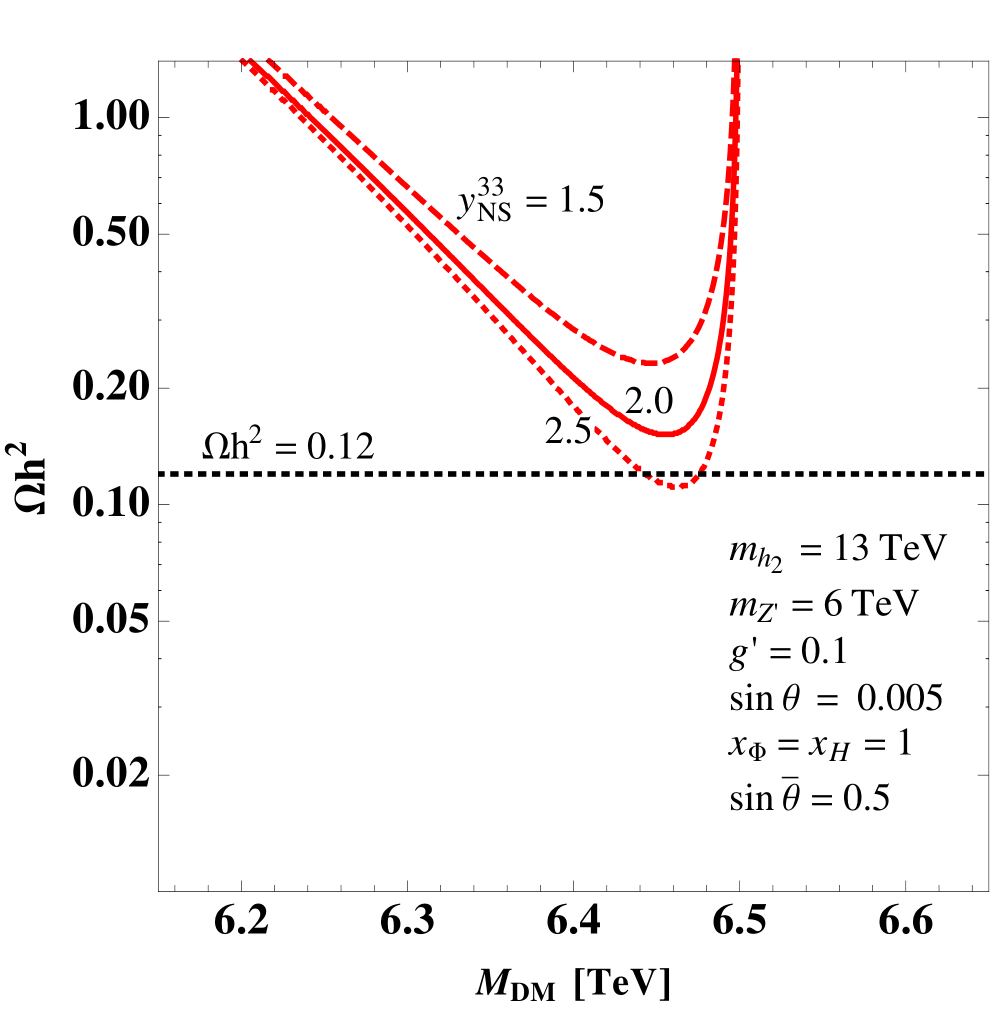}
\end{center}
\caption{Relic abundance as a function of DM mass : (a) For different values of $y_{NS}^{33}$ and fixed $m_{h_2} = 13$ TeV ; (b) For different values of $m_{h_2}$ and fixed $y_{NS}^{33} = 2.5$.}
\label{Relic2}
\end{figure}

We will now focus on the contribution of $h_2$ exchange process to the relic density of DM. For illustrating the effect of this process, we show the relic density as a function of DM mass for different values of $y_{NS}^{33}$ and $m_{h_2}$ in Fig.~\ref{Relic2}. In the left panel, we have fixed $y_{NS}^{33} = 2.5$ and plotted the relic density as a function of $M_{DM}$ for three different values of $m_{h_2}$, keeping all the other parameters fixed. Similarly, we have taken $m_{h_2} = 13$ TeV in the right plot and plotted the relic density for three different values of $y_{NS}^{33}$.
We find that the observed relic density can be realized for $y_{NS}^{33} \gtrsim 2$ when $m_{h_2}= 13$ TeV. In addition, $m_{h_2} \sim 2 M_{DM}$ is preferred to enhance the annihilation cross section which implies that $m_{h_2}$ mass is around $\mathcal{O}(10)$ TeV in our model. Note that such a heavy mass scale for $h_2$ is also preferred in stabilizing the scalar potential as we already discussed in the previous section.


 We perform a parameter scan and search for the allowed regions which can explain the relic density of DM. Firstly, we perform parameter scan in the following ranges focusing on the scalar exchange process,
\begin{align}
& M_{DM} \in [1.0, 10.0] \ {\rm TeV}, \quad m_{h_2} \in [1.8 M_{DM}, 2.2 M_{DM}], \quad y_{NS}^{33} \in [0.2, 3.0], \quad \sin \theta \in [0.001, 0.02], \nonumber \\
& x_H \in [-5, 5], \quad x_\Phi \in [-5, 5], \quad \sin \overline{\theta} \in [0.2, 0.7], \quad m_{Z'} = 5 \ {\rm TeV}, \quad g' = 0.01. \label{eq:scan1}
\end{align}

\begin{figure}
\begin{center}
\includegraphics[scale=0.21]{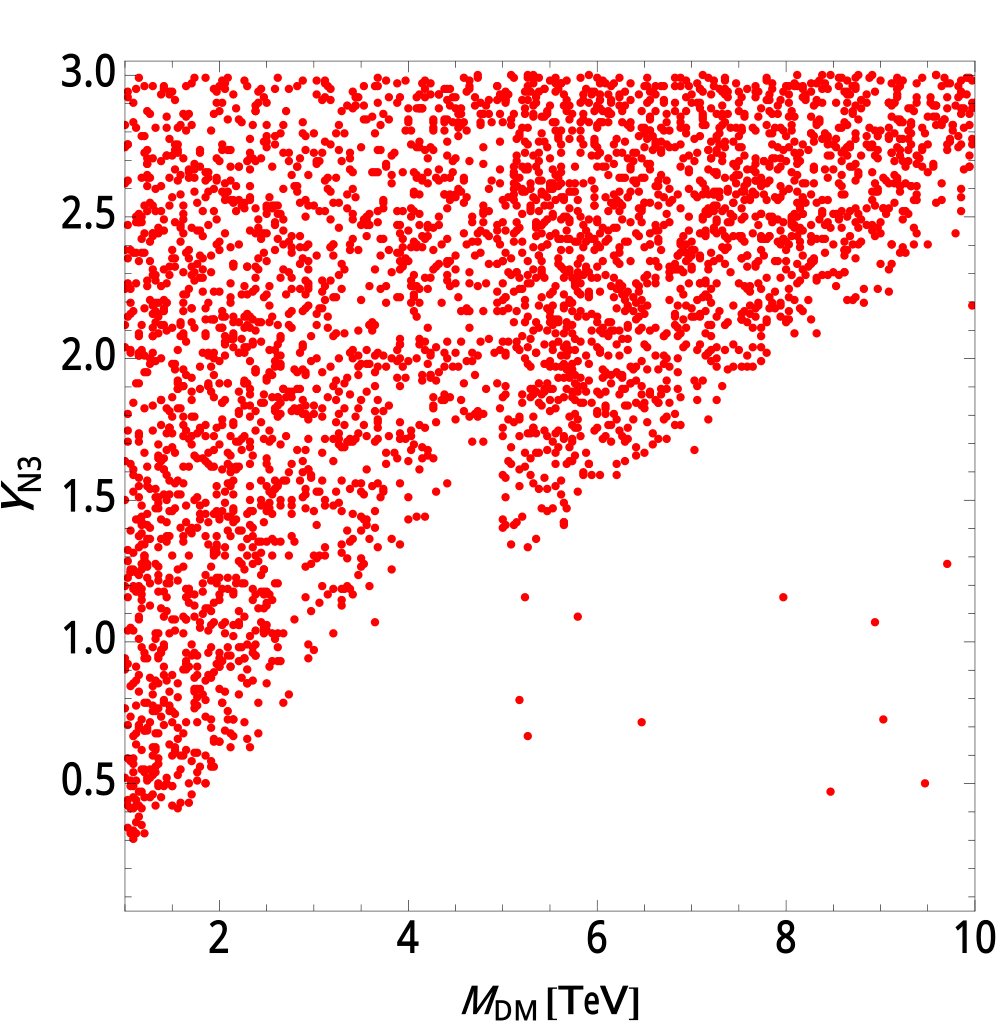} \quad
\includegraphics[scale=0.22]{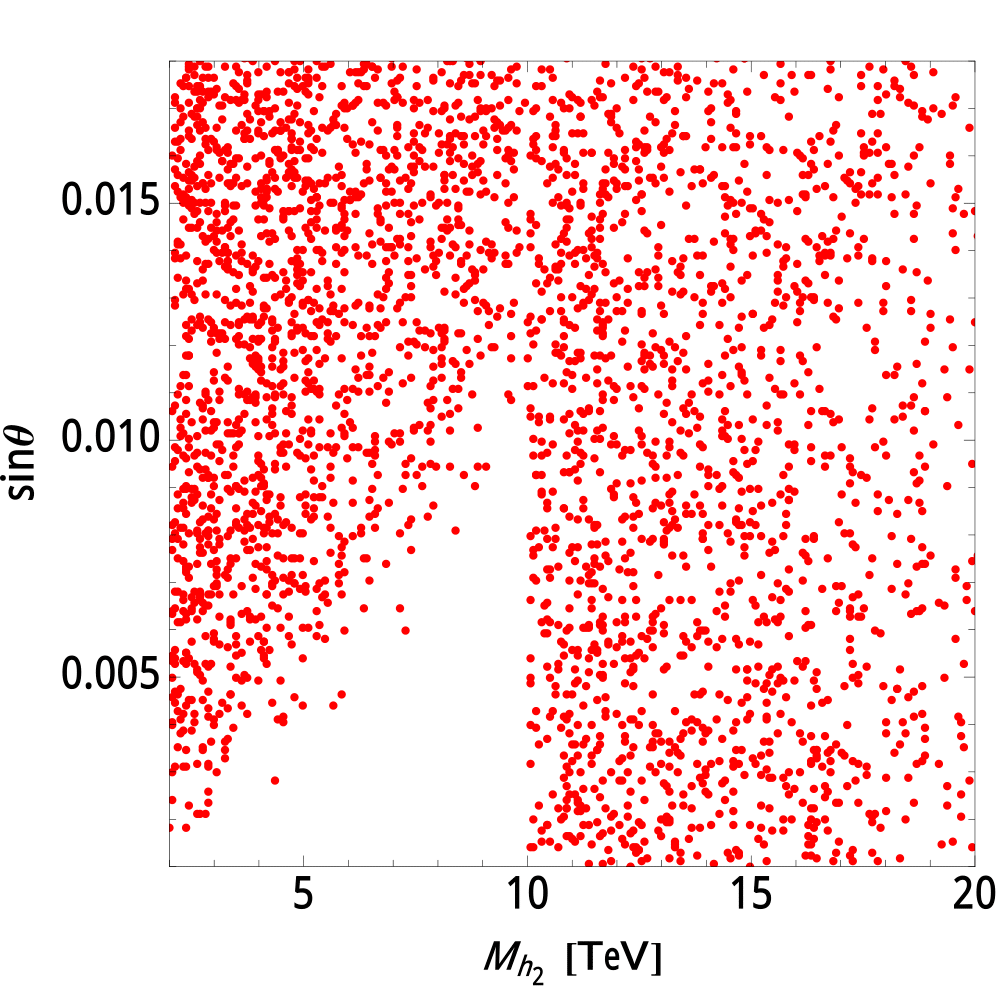}\\
\caption{ Parameter regions that give the correct relic density of DM in $M_{DM}$-$y_{NS}^{33}$ and $M_{h_2}$-$\sin \theta$ planes for scanning done in the ranges of parameters as given by Eq.~(\ref{eq:scan1}).}
\label{Param}
\end{center}
\end{figure}

We fixed $Z'$ mass and $g'$ for simplicity. Note that we chose $m_{h_2} \sim 2 M_{DM}$ since we can obtain the observed relic density in this region via $h_2$ exchange process as discussed above.
In Fig.~\ref{Param}, we show the allowed parameter space in $M_{DM}- y_{NS}^{33}$ and $m_{h_2}- \sin \theta$ planes that give the correct relic density of DM, $0.11 < \Omega h^2 < 0.13$, adopting the approximate range around the best fit value~\cite{Aghanim:2018eyx}. From the left panel of Fig. \ref{Param}, we can see that in general, for larger values of $M_{DM}$, the allowed values of $y_{NS}^{33}$ are large. But, a few points with smaller values of $y_{NS}^{33}$ are also obtained for $M_{DM} > M_{Z'}$ since $\psi_1 \psi_1 \to h_2 \to Z' Z'$ process is kinematically allowed there. In the right panel of Fig. \ref{Param}, we have shown the allowed parameter space in the $m_{h_2}- \sin \theta $ plane. From this plot, we can see that $\sin \theta$ can be small for $M_{DM} > M_{Z'}$ ($m_{h_2} \sim 2 M_{DM}$)  since $h_2 Z' Z'$ coupling is not suppressed by $\sin \theta$ as we can see from Eq.~(\ref{Eq:scalarZpZp}). However, we have some lower limit of $\sin \theta$ for $M_{DM} < m_{Z'}$ since here, $\psi_1 \psi_1 \to h_2 \to Z' Z'$ process is kinematically disallowed and the coupling of $h_2$ to the SM particles is suppressed by $\sin ~\theta$. 

\subsection{Direct detection}
Here we briefly discuss the constraints from the direct detection experiments estimating the DM-nucleon $(N)$ scattering in our model. 
Firstly note that the $Z'$ exchange process between DM and nucleon will not get stringent constraint 
since DM-$Z'$ interaction is via axial vector current due to the Majorana property of DM and provides spin-dependent operator for DM-nucleon interaction.
We thus focus on the scalar mediated processes for DM-nucleon scattering where
the corresponding Feynman diagram is given in Fig~\ref{DM1}b. 
In our case, the DM interacts with the nucleon through the scalar boson exchange $(h_1, h_2)$. The relevant interaction Lagrangian with the mixing effect is given by,

\bea
\mathcal{L} \supset C_{\psi_1 \psi_1 h_1} h_1 \overline{\psi_1^c} \psi_1+ C_{\psi_1 \psi_1 h_1} h_2 \overline{\psi_1^c} \psi_1 + C_{NNh_1} h_1 \overline{N} N+ C_{NNh_2} h_2 \overline{N} N,
\eea
where the effective couplings are,
\bea
&& C_{\psi_1 \psi_1 h_1} = \sin\overline{\theta} \cos\overline{\theta} \cos\theta \frac{y_{NS}^{33}}{\sqrt{2}}, \quad C_{\psi_1 \psi_1 h_2} = -\sin\overline{\theta} \cos\overline{\theta} \sin\theta \frac{y_{NS}^{33}}{\sqrt{2}}, \\
&& C_{NNh_1} = \sin\theta g_{hNN}, \quad C_{NNh_2} = \cos\theta g_{h NN}.
\eea
Hence the effective Lagrangian can be written as, 
\bea
&& \mathcal{L}_{eff}= G_{h} \overline{\psi_{1}}\psi_{1} \overline{N}N, \\
&&  G_{h}=\Big[\frac{C_{\psi_1 \psi_1 h_{1}} C_{h_1 NN}}{m_{h_{1}}^2}+\frac{C_{\psi_2 \psi_2 h_{2}} C_{h_2 NN}}{m_{h_{2}}^2}\Big]
 \eea
where $m_{h_1}$ and $m_{h_{2}}$ are the SM and BSM Higgs masses. The corresponding cross section of Fig.~\ref{DM1}b in the non-relativistic limit can be calculated as,
\bea
\sigma=g_{hNN}^2 \frac{M_{DM}^2 M_{N}^2}{16\pi (M_{DM}^2+M_{N}^2)^2} (y_{NS}^{33} \sin2\overline{\theta} \sin2\theta)^2 \Big(\frac{1}{m_{h_1}^2}-\frac{1}{m_{h_2}^2}\Big)^2,
\eea  
\begin{figure}
\begin{center}
\includegraphics[scale=0.275]{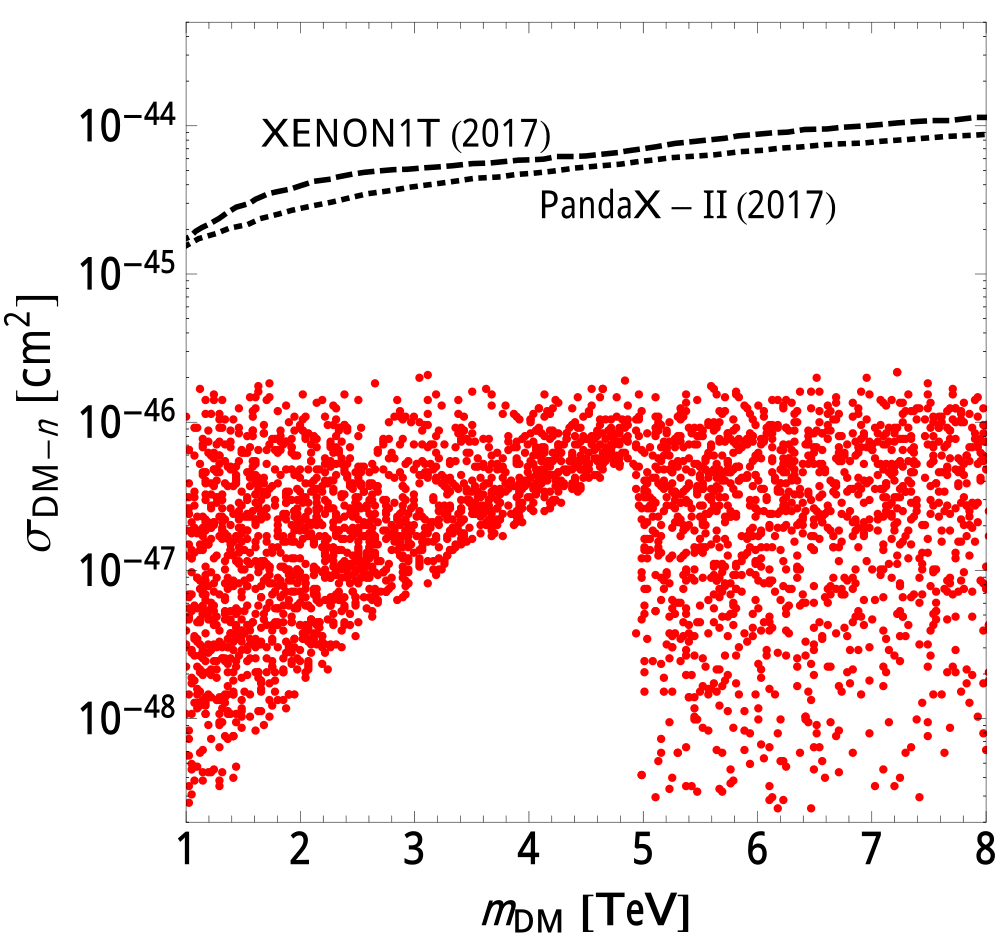}
\caption{Nucleon-DM scattering cross section as a function of DM mass for parameters that give the correct relic density. The current upper bounds from PANDAX-II \cite{Cui:2017nnn} (black dotted line) and XENON-1t \cite{Aprile:2017iyp} (back dashed line) are also shown.}
\label{DD1}
\end{center}
\end{figure}
where, $M_{DM}$ and $M_N$ are the DM and nucleon masses respectively. The effective coupling can be written as $g_{hNN}=\frac{f_{N} M_N}{v\sqrt{2}}$ where we apply $f_N=0.287$ for neutron~\cite{Belanger:2013oya}~\footnote{$f_N$ for proton has similar value and we here just use $f_N$ in estimating the cross section.}
 and $v=246$ GeV. 
We then estimate the cross sections applying allowed parameter sets obtained in previous subsection and the results are shown in Fig.~\ref{DD1}. The black dotted and dashed lines show the current upper bounds from PANDAX-II \cite{Cui:2017nnn} and XENON-1t \cite{Aprile:2017iyp} respectively.
 We find that our parameter region is allowed by the direct detection constraints since the cross section is suppressed by small $\sin \theta$ which is also preferred by the constraints from vacuum stability. 
 The cross section will be further explored by the future direct detection experiments like XENON 1t, PandaX, etc.

\section{Bounds on the $M_Z^\prime-g^\prime$ plane}

\begin{figure}
\begin{center}
\includegraphics[scale=0.21]{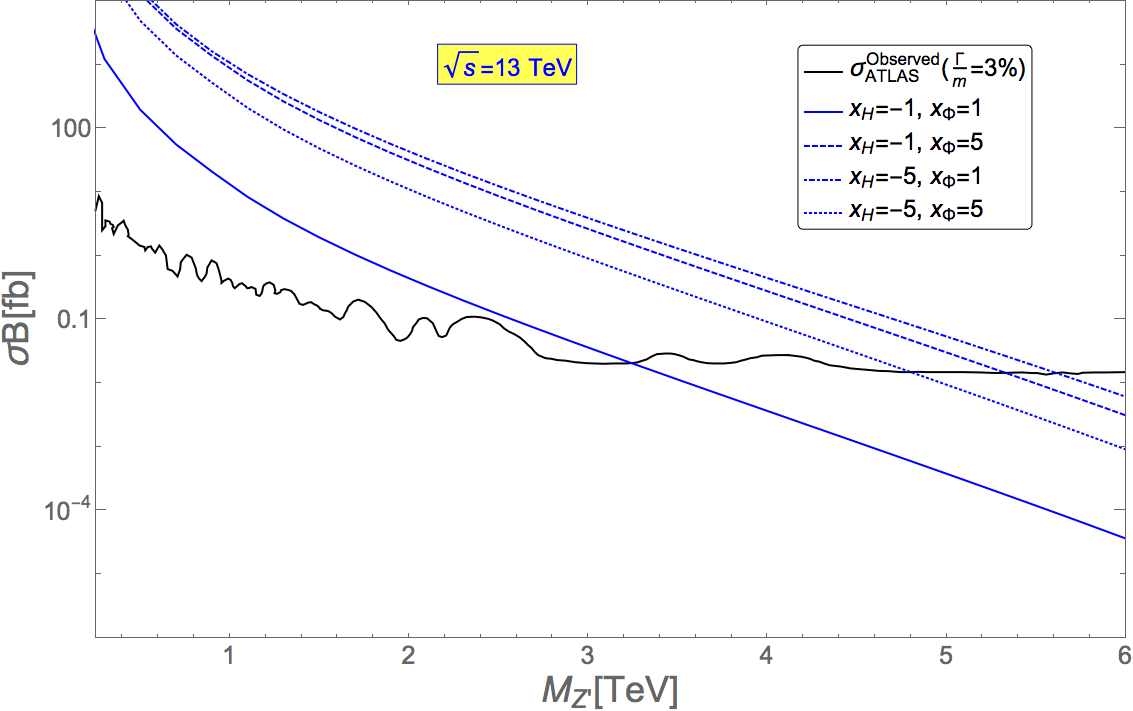} 
\includegraphics[scale=0.21]{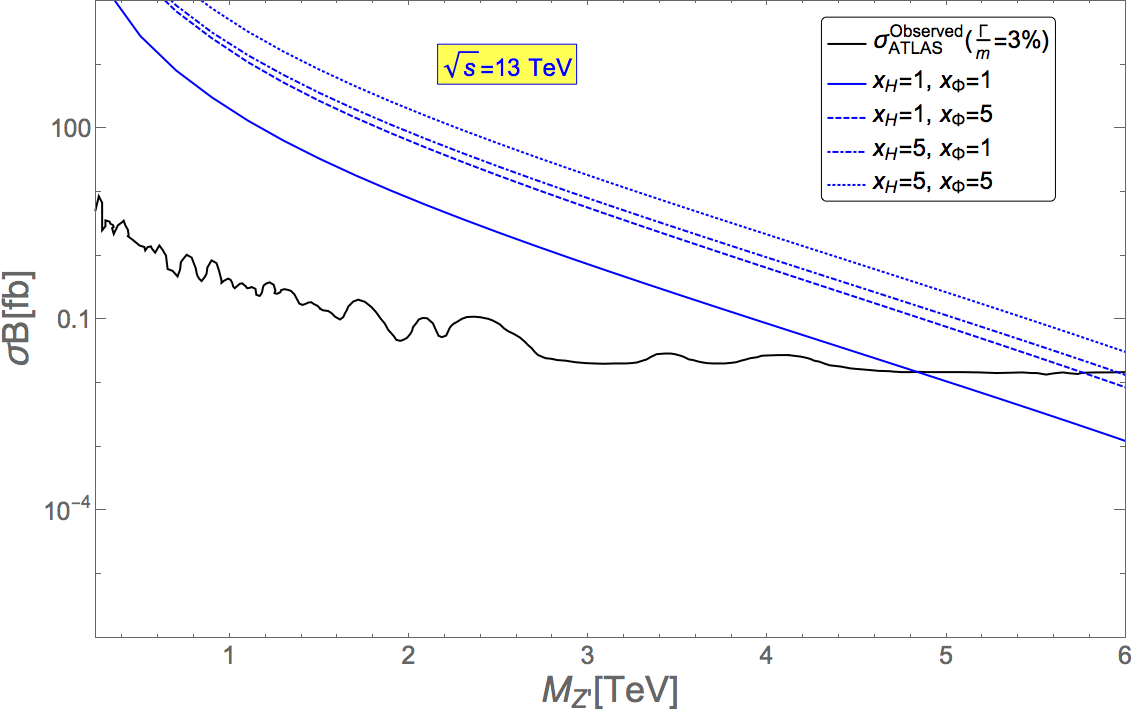}
\caption{Comparison between the ATLAS  \cite{Aad:2019fac} (black solid line) result and model cross sections (blue lines) for the different values of $x_H$ and $x_\Phi$. The model cross sections are produced with $g_{\rm Model}=0.05$. The left and right panels correspond to $x_H <0$ and $x_H >0$ respectively and we have considered $x_{\Phi} >0$ for both the cases.  }
\label{ATLAS1} 
\end{center}
\end{figure}
\begin{figure}
\begin{center}
\includegraphics[scale=0.3]{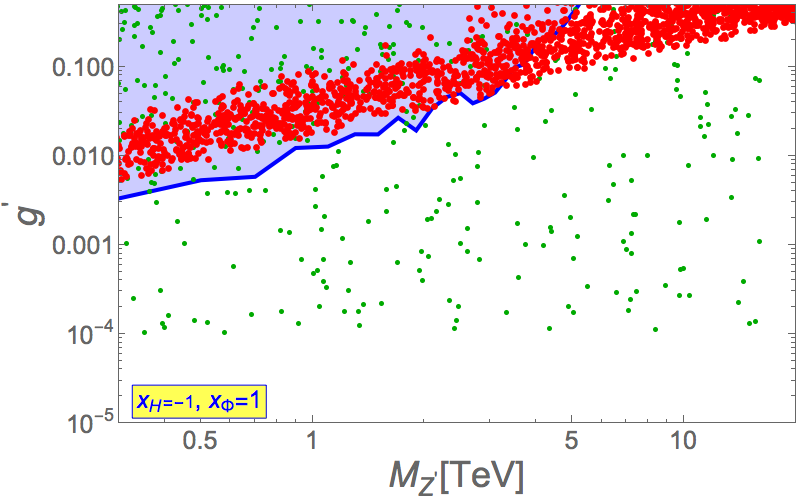} 
\includegraphics[scale=0.3]{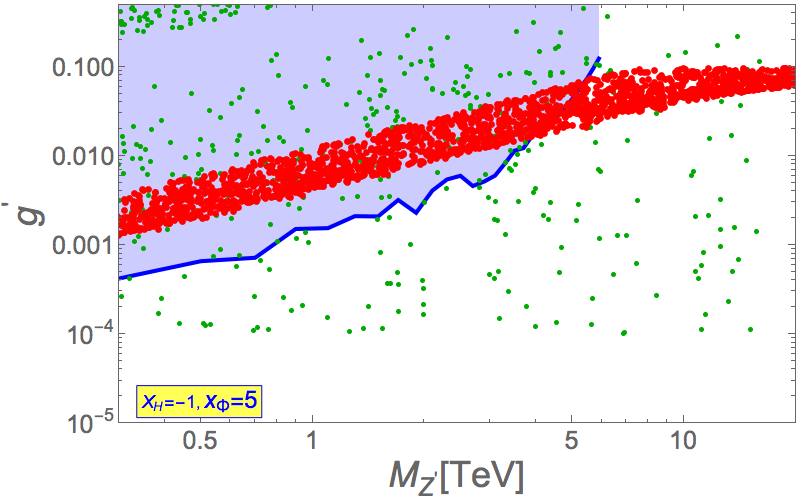}\\
\includegraphics[scale=0.3]{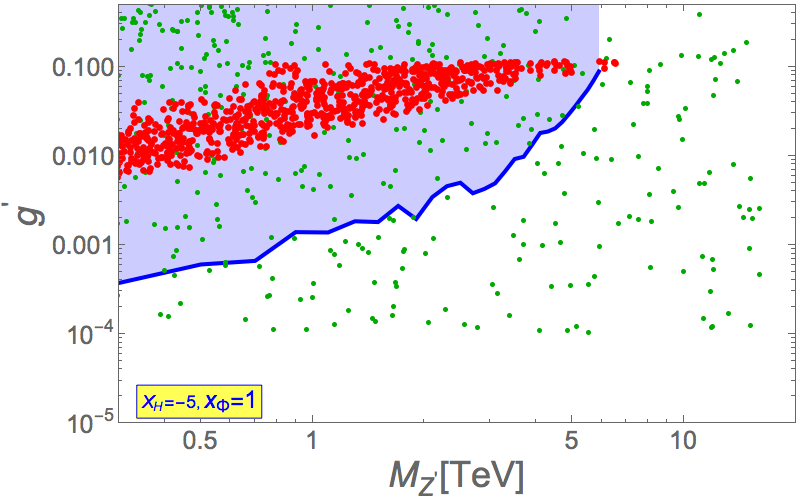} 
\includegraphics[scale=0.3]{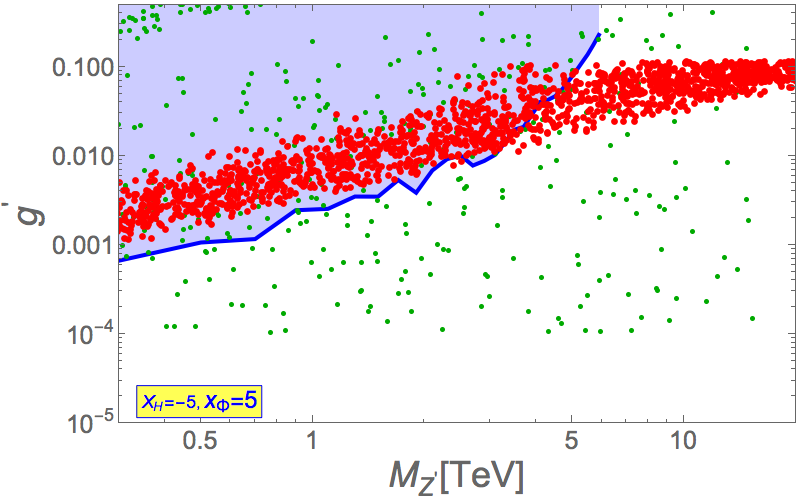}
\caption{Allowed parameter space combining the bounds obtained on $g^\prime$ as a function of $M_Z^\prime$ from vacuum stability and perturbativity (red dots), DM constraints (green dots) and collider (region below the blue solid line). The blue shaded regions  are ruled out by the recent ATLAS search \cite{Aad:2019fac} at $139$ fb $^{-1}$ luminosity.}
\label{Com1}
\end{center}
\end{figure}
\begin{figure}
\begin{center}
\includegraphics[scale=0.3]{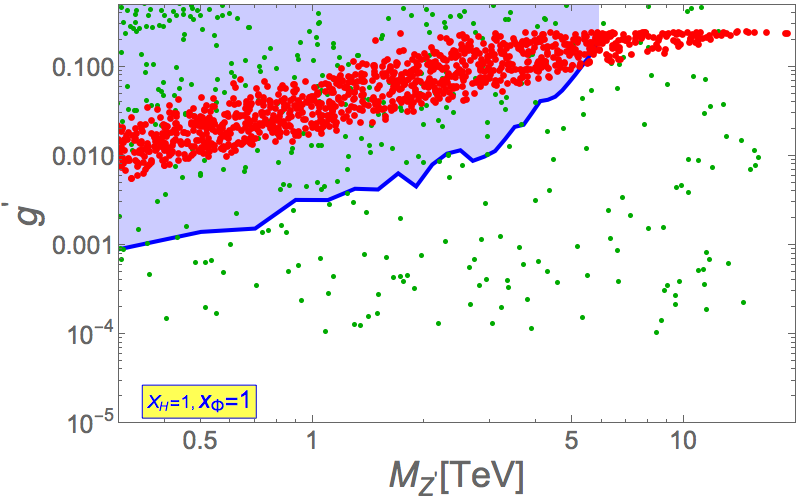} 
\includegraphics[scale=0.3]{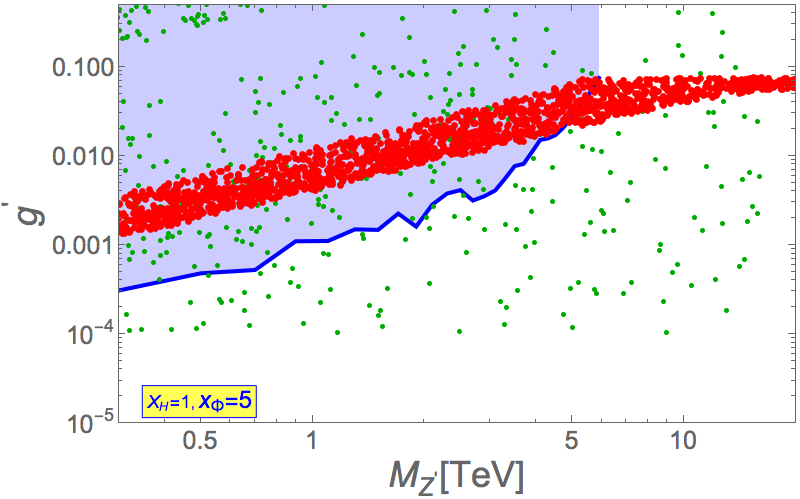}\\
\includegraphics[scale=0.3]{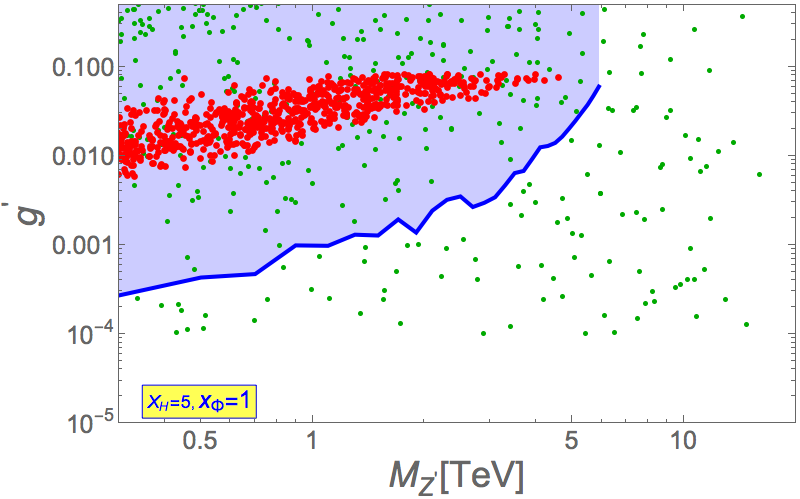} 
\includegraphics[scale=0.3]{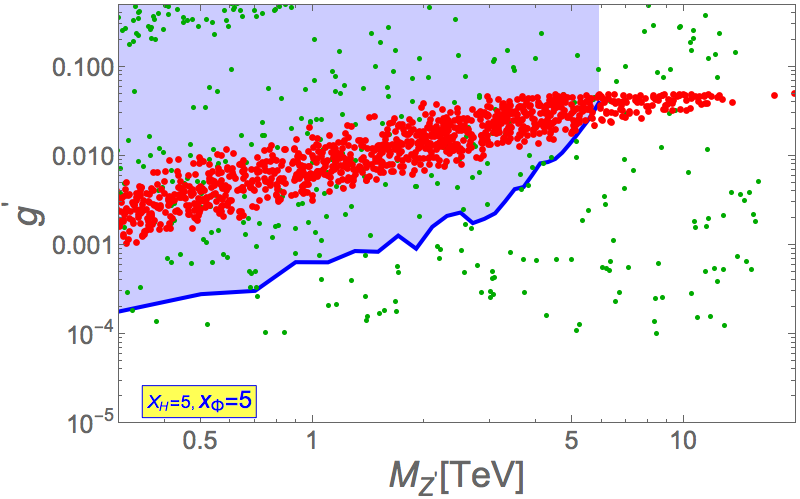}
\caption{Allowed parameter space combining the bounds obtained on $g^\prime$ as a function of $M_Z^\prime$ from vacuum stability and perturbativity (red dots), DM constraints (green dots) and collider (region below the blue solid line). The blue shaded regions are ruled out by the recent ATLAS search \cite{Aad:2019fac} at $139$ fb$^{-1}$ luminosity.}
\label{Com2}
\end{center}
\end{figure}

In this section, we consider the production of $Z^\prime$ from the proton proton collision at the LHC and its decay into different types of leptons.
We first calculate the $Z^\prime$ production cross section at the LHC from protons followed by the decay into lepton, $pp\to Z^\prime\to \ell^+\ell^-$ with $\ell=e,~\mu$.
In our analysis we calculate the cross section combining the electron and muon final states. We compare our cross section with the latest ATLAS search \cite{Aad:2019fac} for the heavy $Z^\prime$ resonance. 
Since we are considering $U(1)'$ models with extra $Z'$, the ATLAS results can be compared directly with our results. 
Atlas analysis has considered  
different models like SSM and $Z^\prime_\psi$ \cite{Langacker:2008yv} where the $Z^{\prime}$ decays into $e$ and $\mu$.
Conservatively considering these limits for our case we first produce the $Z^\prime$ $(300$ GeV $\leq M_Z^\prime \leq 6$ TeV$)$ at the $13$ TeV LHC followed by the decay into the dilepton mode and finally compare with the cross sections in our model. To calculate the bounds on the $g^\prime$, we calculate the model cross section, $\sigma_{\rm Model}$, for the process $pp \to Z^\prime \to 2e,~2\mu$, with a $U(1)'$ coupling constant $g_{\rm Model}$ at the LHC at the $13 $TeV center of mass energy. Then we compare this with the observed ATLAS bound $(\sigma_{\rm ATLAS}^{\rm Observed})$ for $\frac{\Gamma}{m}=3\%$ which has been studied for the SSM. The corresponding cross sections are plotted in Fig.~\ref{ATLAS1} for different choices of $x_H$ and $x_\Phi$. Thus, the value of $g'$ corresponding to a given $M_{Z'}$ is given as,

\bea
g^\prime =\sqrt{\frac{\sigma_{\rm ATLAS}^{\rm Observed}}{\Big(\frac{\sigma_{\rm Model}}{g_{\rm Model}^2}\Big)}} ,
\label{bound}
\eea
since the cross section varies with the square of the $U(1)'$ coupling $(g_{\rm Model}^2)$.

In this analysis we consider several choices of the $x_H$ and $x_{\Phi}$ to calculate the bounds in the $M_Z^{\prime}-g^\prime$ plane. These correspond to two scenarios : (1) $x_H$ is negative and $x_{\Phi}$ is positive for which the results are shown in Fig.~\ref{Com1} and (2) both $x_H$ and $x_{\Phi}$ are positive and the corresponding constraints in the $M_Z^{\prime}-g^\prime$ plane are shown in Fig.~\ref{Com2}. The interaction of the $Z^\prime$ with the fermions via the covariant derivative will depend on the $x_H$ and $x_{\phi}$ values and is given by the Lagrangian,

\bea
-L_{int}\supset \overline{f_L} \gamma^\mu g^\prime Q_x Z_\mu^\prime f_L + \overline{f_R} \gamma^\mu g^\prime Q_x^\prime Z_\mu^\prime f_R.
\label{Zpint}
\eea

Here, $f_L$ and $f_R$ are the left handed and right handed fermions and $Q_x$ and $Q_x^\prime$ are the corresponding charges under the $U(1)^\prime$ gauge group.
These charges are linear combinations of $x_H$ and $x_\Phi$ and will appear in the $C_V$ and $C_A$ coefficients of the $Z^\prime$ interactions. The $Z^\prime$ interaction with the colored fermions will contain the color factor $N_c=3$ in the interaction whereas $N_c=1$ for the uncolored fermions. The bounds from the collider for various models are shown by the blue solid lines in Figs.~\ref{Com1} and \ref{Com2}. The blue shaded regions in these figures are ruled out by the current LHC data obtained from the ATLAS experiment 
\cite{Aad:2019fac} at $139$ fb$^{-1}$ luminosity.

In these figures, we have also given the bounds from vacuum stability, perturbativity and relic density for purposes of comparison.
For finding the regions that
are allowed by vacuum stability and perturbativity, we have done a scanning in the following ranges of parameters,

\begin{align}
& g' \in [0.0001, 1.0], \quad u \in [0.3, 100] \ {\rm TeV} \quad m_{h_2} \in [2.0, 16] \ {\rm TeV}, \quad y^{33}_{NS} \in [0.2, 2.5] \label{eq:scan2},
\end{align}
with $\theta = 0.01$. For $Y_\nu$ and $(y_{NS})_{2 \times 2}$, we have used BM-I from the Table \ref{benchmark} and we have scaled $y_{NS}$ according to the variation in $u$. The values of $M_{Z'}$ have been calculated using Eq.\ref{Zmass} and the allowed regions are shown by the red points in  Figs.~\ref{Com1} and \ref{Com2}. It can be seen from these figures that the bulk of the parameter 
space allowed by vacuum stability lies in the region disfavoured by the ATLAS results. Regions beyond $M_{Z'} > 5 TeV$ that is not explored by ATLAS are 
seen to be allowed by vacuum stability and perturbativity constraints. 
Future ATLAS results will be able to explore this region. 

Similarly, to find out the points that can give the correct DM relic density, we have performed a scanning of parameters in the ranges, 
\begin{align}
& g' \in [0.0001, 1.0], \quad m_{Z'} \in [0.1, 16] \ {\rm TeV} \quad m_{h_2} \in [2.0, 16] \ {\rm TeV}, \nonumber \\ & \quad y^{33}_{NS} \in [0.2, 2.5], M_{DM} \in [1.0, 10.0] {\rm TeV}.\label{eq:scan2}
\end{align}

Here also, we have fixed $\theta = 0.01$. The green dots in  Figs.~\ref{Com1} and \ref{Com2} correspond to the values that give the correct DM relic density.
The constraints coming from this is seen to be less stringent than the 
combined constraints from vacuum stability, perturbativity and ATLAS 
analysis.

\section{Concluding Remarks}
In this paper we have studied the inverse seesaw model in a class of  general $U(1)$ extensions of the SM. We have studied the parameter spaces in various planes that are allowed by vacuum stability and perturbativity as well as consistent with the low energy neutrino data. In addition, this model has a prospective DM candidate resulting from the stabilization of the third generations of the $SU(2)_L$ singlet neutral fermions using the odd parity under the discrete $Z_2$ symmetry. Comparing the $Z^{\prime}$ production and its decay into the dilepton mode at the LHC with the current ATLAS results, we find the bounds on the $U(1)^\prime$ coupling constant with respect to the $Z^\prime$ mass. Finally, combining all the constraints, we obtain the resultant allowed parameter space which can be probed in the future experiments.

\section*{Acknowledgments}
This work of A. D. is supported by the Japan Society for the Promotion of Science (JSPS) Postdoctoral Fellowship for Research in Japan. 

\begin{appendices}   

\section{One-loop RG Equations}

\be \begin{split}
\beta_{g_1} =  \frac{1}{6} &\Big(41  g_1^3 + g_1^2  g_{1p1}  (78  x_H + 64  x_\Phi) + 
       g_{11p}  g_{1p1}  (39  g_{11p}  x_H + 41  g'  x_H^2 + 32  g_{11p}  x_\Phi \\
        & + 64  g'  x_H  x_\Phi + 66  g'  x_\Phi^2) + 
       g_1  (41  g_{11p}^2 + g_{11p}  g'  (39  x_H + 32  x_\Phi) \\ & + 
          g_{1p1}^2  (41  x_H^2 + 64  x_H  x_\Phi + 66  x_\Phi^2))\Big)  
          \end{split} \ee

\be \beta_{g_2} = \frac{(-19  g_2^3)}{6 } \ee

\be \beta{g_3} =   (-7  g_3^3)  \ee

\be \begin{split} \beta_{g'} =  \frac{1}{6} & \Big(41  g_{11p}^2  g' + 
       g_{11p}  (41  g1  g_{1p1} + (2  g'^2 + g_{1p1}^2)  (39  x_H + 32  x_\Phi))  \\
       & + g'  (g1  g_{1p1}  (39  x_H + 32  x_\Phi) + (g'^2 + g_{1p1}^2)  (41  x_H^2 + 
             64  x_H  x_\Phi + 66  x_\Phi^2))\Big)  \end{split} \ee

\be \begin{split} \beta_{g_{1p1}} =   \frac{1}{6}& \Big(41  g_1^2  g_{1p1} + 
       g1  (41  g_{11p}  g' + (g'^2 + 2  g_{1p1}^2)  (39  x_H + 32  x_\Phi)) + \\ &
       g_{1p1}  (g_{11p}  
           g'  (39  x_H + 32  x_\Phi) + (g'^2 + g_{1p1}^2)  (41  x_H^2 + 
             64  x_H  x_\Phi + 66  x_\Phi^2))\Big) \end{split} \ee

\be \begin{split} \beta_{g_{11p}} =  \frac{1}{6} & \Big(g_1^2  (41  g_{11p} + 39  g'  x_H + 32  g'  x_\Phi) + 
       g1  g_{1p1}  (39  g_{11p}  x_H + 41  g'  x_H^2 + 32  g_{11p}  x_\Phi  \\ &+ 
          64  g'  x_H  x_\Phi + 66  g'  x_\Phi^2) + 
       g_{11p}  (41  g_{11p}^2 + g_{11p}  g'  (78  x_H + 64  x_\Phi) \\ & + 
          g'^2  (41  x_H^2 + 64  x_H  x_\Phi + 66  x_\Phi^2))\Big)   \end{split} \ee

\be \begin{split} \beta_{\lambda_1} =  \frac{1}{8} & \Big(3  g_1^4 + 6  g_1^2  g_{11p}^2 + 3  g_{11p}^4 + 6  g_1^2  g_2^2 + 
       6  g_{11p}^2  g_2^2 + 9  g_2^4 - 24  g_1^2  \lambda_1 - 24  g_{11p}^2  \lambda_1 - 
       72  g_2^2  \lambda_1 \\ & + 192  \lambda_1^2 + 8  \lambda_3^2 - 12  g_1^2  g_{11p}  g'  x_H - 
       12  g_{11p}^3  g'  x_H - 12  g_1^3  g_{1p1}  x_H - 12  g_1  g_{11p}^2  g_{1p1}  x_H \\& - 
       12  g_{11p}  g'  g_2^2  x_H   - 12  g_1  g_{1p1}  g_2^2  x_H + 48  g_{11p}  g'  \lambda_1  x_H + 
       48  g_1  g_{1p1}  \lambda_1  x_H + 6  g_1^2  g'^2  x_H^2   \\ & + 18  g_{11p}^2  g'^2  x_H^2 + 
       24  g_1  g_{11p}  g'  g_{1p1}  x_H^2 +  18  g_1^2  g_{1p1}^2  x_H^2 + 
       6  g_{11p}^2  g_{1p1}^2  x_H^2 + 6  g'^2  g_2^2  x_H^2  \\& + 
       6  g_{1p1}^2  g_2^2  x_H^2  - 24  g'^2  \lambda_1  x_H^2 - 24  g_{1p1}^2  \lambda_1  x_H^2  - 
       12  g_{11p}  g'^3  x_H^3 - 12  g_1  g'^2  g_{1p1}  x_H^3 \\ & - 
       12  g_{11p}  g'  g_{1p1}^2  x_H^3  - 12  g_1  g_{1p1}^3  x_H^3 + 3  g'^4  x_H^4  + 
       6  g'^2  g_{1p1}^2  x_H^4 + 3  g_{1p1}^4  x_H^4 + 96  \lambda_1  y_t^2 \\ & + 
       32  \lambda_1  \textrm{Tr} [Y_\nu Y_\nu^\dag ] - 48  y_t^4 - 
       16  \textrm{Tr} [Y_\nu Y_\nu^\dag Y_\nu Y_\nu^\dag  ]\Big)  \end{split} \ee

\be \beta_{\lambda_2} =    \Big(10  \lambda_2^2 + \lambda_3^2 - 6  g'^2  \lambda_2  x_\Phi^2 - 6  g_{1p1}^2  \lambda_2  x_\Phi^2 + 
       3  g'^4  x_\Phi^4 + 2  \lambda_2  \textrm{Tr}[y_{NS} y_{NS}^\dag] - 
       \textrm{Tr}[y_{NS} y_{NS}^\dag y_{NS} y_{NS}^\dag] \Big) \ee

\be \begin{split} \beta_{\lambda_3} =   \frac{1}{2} & \Big(-3  g_1^2  \lambda_3 - 3  g_{11p}^2  \lambda_3 - 9  g_2^2  \lambda_3 + 24  \lambda_1  \lambda_3 + 16  \lambda_2  \lambda_3 + 
       8  \lambda_3^2 + 6  g_{11p}  g'  \lambda_3  x_H   + 6  g1  g_{1p1}  \lambda_3  x_H \\ & - 
       3  g'^2  \lambda_3  x_H^2 - 3  g_{1p1}^2  \lambda_3  x_H^2 + 6  g_{11p}^2  g'^2  x_\Phi^2 - 
       12  g'^2  \lambda_3  x_\Phi^2 - 12  g_{1p1}^2  \lambda_3  x_\Phi^2 - 
       12  g_{11p}  g'^3  x_H  x_\Phi^2  \\ & + 6  g'^4  x_H^2  x_\Phi^2   + 12  \lambda_3  y_t^2 + 
       4  \lambda_3  \textrm{Tr} [Y_\nu Y_\nu^\dag ] + 
       4  \lambda_3  \textrm{Tr}[y_{NS} y_{NS}^\dag] - 
       8  \textrm{Tr}[y_{NS} y_{NS}^\dag Y_\nu Y_\nu^\dag]\Big) \end{split} \ee

\be \begin{split} \beta_{y_t^{(1)}} =   \frac{1}{12} & \Big(-( \Big( 17  g_1^2 + 17  g_{11p}^2 + 27  g_2^2 + 96  g_3^2 + 34  g_{11p}  g'  x_H + 
            34  g1  g_{1p1}  x_H + 17  g'^2  x_H^2 \\ &  + 17  g_{1p1}^2  x_H^2 + 
            20  g_{11p}  g'  x_\Phi + 20  g1  g_{1p1}  x_\Phi + 20  g'^2  x_H  x_\Phi + 
            20  g_{1p1}^2  x_H  x_\Phi + 8  g'^2  x_\Phi^2 \\ & + 8  g_{1p1}^2  x_\Phi^2 - 
            36  y_t^2  - 12  \textrm{Tr} [Y_\nu Y_\nu^\dag ]\Big)  y_t) + 
       18  (y_t^3)\Big) \end{split} \ee

\be \beta_{y_{NS}^{(1)}} =   \Big( \Big(-3  (g'^2 + g_{1p1}^2)  x_\Phi^2 + \textrm{Tr}[y_{NS} y_{NS}^\dag]\Big)  y_{NS} + y_{NS} y_{NS}^\dag y_{NS} + 
     Y_\nu^T Y_\nu^* y_{NS} \Big)   \ee

\be \begin{split} \beta_{Y_\nu^{(1)}} =   \frac{1}{4} & \Big(-(\Big(3  g_1^2 + 3  g_{11p}^2 + 9  g_2^2 + 6  g_{11p}  g'  x_H + 
            6  g1  g_{1p1}  x_H + 3  g'^2  x_H^2 + 3  g_{1p1}^2  x_H^2  \\ & + 
            12  g_{11p}  g'  x_\Phi + 12  g1  g_{1p1}  x_\Phi + 12  g'^2  x_H  x_\Phi + 
            12  g_{1p1}^2  x_H  x_\Phi + 24  g'^2  x_\Phi^2 \\& + 24  g_{1p1}^2  x_\Phi^2  - 
            12  y_t^2 - 4  \textrm{Tr} [Y_\nu Y_\nu^\dag ]\Big)  Y\nu) + 
       2  (3  Y\nu Y_\nu^\dag Y_\nu + 
          Y_\nu y_{NS}^* y_{NS}^T)\Big)  \end{split} \ee

\section{Two-loop RG Equations}

\be   \begin{split} \beta_{g_1}^{(2)} = &  \frac{1}{18}\Big (199 g_1^5 + 398 g_1^3 g_{11 p}^2 + 
   199 g_1 g_{11 p}^4 + 81 g_1^3 g_2^2 + 
   81 g_1 g_{11 p}^2 g_2^2 + 264 g_1^3 g_3^2  + 
   264 g_1 g_{11 p}^2 g_3^2  + 543 g_1^3 g_{11 p} g' x_H \\& + 
   543 g_1 g_{11 p}^3 g' x_H + 724 g_1^4 g_{1 p1} x_H + 
   905 g_1^2 g_{11 p}^2 g_{1 p1} x_H  + 
   181 g_{11 p}^4 g_{1 p1} x_H  + 27 g_1 g_{11 p} g' g_2^2 x_H \\&  + 
   54 g_1^2 g_{1 p1} g_2^2 x_H  + 
   27 g_{11 p}^2 g_{1 p1} g_2^2 x_H + 
   264 g_1 g_{11 p} g' g_3^2 x_H + 
   528 g_1^2 g_{1 p1} g_3^2 x_H  + 
   264 g_{11 p}^2 g_{1 p1} g_3^2 x_H \\& + 199 g_1^3 g'^2 x_H^2 + 
   597 g_1 g_{11 p}^2 g'^2 x_H^2  + 
   1393 g_1^2 g_{11 p} g' g_{1 p1} x_H^2 + 
   597 g_{11 p}^3 g' g_{1 p1} x_H^2 + 
   1194 g_1^3 g_{1 p1}^2 x_H^2 \\& + 
   796 g_1 g_{11 p}^2 g_{1 p1}^2 x_H^2 + 
   81 g_{11 p} g' g_{1 p1} g_2^2 x_H^2 + 
   81 g_1 g_{1 p1}^2 g_2^2 x_H^2 + 
   264 g_{11 p} g' g_{1 p1} g_3^2 x_H^2  + 
   264 g_1 g_{1 p1}^2 g_3^2 x_H^2 \\&  + 
   181 g_1 g_{11 p} g'^3 x_H^3 + 362 g_1^2 g'^2 g_{1 p1} x_H^3 + 
   543 g_{11 p}^2 g'^2 g_{1 p1} x_H^3 + 
   905 g_1 g_{11 p} g' g_{1 p1}^2 x_H^3  + 
   724 g_1^2 g_{1 p1}^3 x_H^3  \\& + 181 g_{11 p}^2 g_{1 p1}^3 x_H^3 + 
   199 g_{11 p} g'^3 g_{1 p1} x_H^4 + 
   199 g_1 g'^2 g_{1 p1}^2 x_H^4 + 
   199 g_{11 p} g' g_{1 p1}^3 x_H^4 + 199 g_1 g_{1 p1}^4 x_H^4   \\& + 
   492 g_1^3 g_{11 p} g' x_\Phi + 492 g_1 g_{11 p}^3 g' x_\Phi + 
   656 g_1^4 g_{1 p1} x_\Phi + 
   820 g_1^2 g_{11 p}^2 g_{1 p1} x_\Phi  + 
   164 g_{11 p}^4 g_{1 p1} x_\Phi \\& + 
   108 g_1 g_{11 p} g' g_2^2 x_\Phi + 
   216 g_1^2 g_{1 p1} g_2^2 x_\Phi + 
   108 g_{11 p}^2 g_{1 p1} g_2^2 x_\Phi + 
   96 g_1 g_{11 p} g' g_3^2 x_\Phi  + 
   192 g_1^2 g_{1 p1} g_3^2 x_\Phi  \\& + 
   96 g_{11 p}^2 g_{1 p1} g_3^2 x_\Phi + 
   328 g_1^3 g'^2 x_H x_\Phi + 
   984 g_1 g_{11 p}^2 g'^2 x_H x_\Phi + 
   2296 g_1^2 g_{11 p} g' g_{1 p1} x_H x_\Phi  + 
   984 g_{11 p}^3 g' g_{1 p1} x_H x_\Phi  \\&   + 
   1968 g_1^3 g_{1 p1}^2 x_H x_\Phi + 
   1312 g_1 g_{11 p}^2 g_{1 p1}^2 x_H x_\Phi  + 
   216 g_{11 p} g' g_{1 p1} g_2^2 x_H x_\Phi + 
   216 g_1 g_{1 p1}^2 g_2^2 x_H x_\Phi \\& + 
   192 g_{11 p} g' g_{1 p1} g_3^2 x_H x_\Phi  + 
   192 g_1 g_{1 p1}^2 g_3^2 x_H x_\Phi + 
   492 g_1 g_{11 p} g'^3 x_H^2 x_\Phi + 
   984 g_1^2 g'^2 g_{1 p1} x_H^2 x_\Phi \\& + 
   1476 g_{11 p}^2 g'^2 g_{1 p1} x_H^2 x_\Phi  + 
   2460 g_1 g_{11 p} g' g_{1 p1}^2 x_H^2 x_\Phi  + 
   1968 g_1^2 g_{1 p1}^3 x_H^2 x_\Phi + 
   492 g_{11 p}^2 g_{1 p1}^3 x_H^2 x_\Phi \\& + 
   656 g_{11 p} g'^3 g_{1 p1} x_H^3 x_\Phi + 
   656 g_1 g'^2 g_{1 p1}^2 x_H^3 x_\Phi  + 
   656 g_{11 p} g' g_{1 p1}^3 x_H^3 x_\Phi  + 
   656 g_1 g_{1 p1}^4 x_H^3 x_\Phi + 184 g_1^3 g'^2 x_\Phi^2 \\& + 
   552 g_1 g_{11 p}^2 g'^2 x_\Phi^2 + 
   1288 g_1^2 g_{11 p} g' g_{1 p1} x_\Phi^2 + 
   552 g_{11 p}^3 g' g_{1 p1} x_\Phi^2  + 
   1104 g_1^3 g_{1 p1}^2 x_\Phi^2 + 
   736 g_1 g_{11 p}^2 g_{1 p1}^2 x_\Phi^2 \\& + 
   216 g_{11 p} g' g_{1 p1} g_2^2 x_\Phi^2 + 
   216 g_1 g_{1 p1}^2 g_2^2 x_\Phi^2 + 
   192 g_{11 p} g' g_{1 p1} g_3^2 x_\Phi^2  + 
   192 g_1 g_{1 p1}^2 g_3^2 x_\Phi^2 + 
   552 g_1 g_{11 p} g'^3 x_H x_\Phi^2 \\& + 
   1104 g_1^2 g'^2 g_{1 p1} x_H x_\Phi^2 + 
   1656 g_{11 p}^2 g'^2 g_{1 p1} x_H x_\Phi^2 + 
   2760 g_1 g_{11 p} g' g_{1 p1}^2 x_H x_\Phi^2  + 
   2208 g_1^2 g_{1 p1}^3 x_H x_\Phi^2 \\& + 
   552 g_{11 p}^2 g_{1 p1}^3 x_H x_\Phi^2 + 
   1104 g_{11 p} g'^3 g_{1 p1} x_H^2 x_\Phi^2 + 
   1104 g_1 g'^2 g_{1 p1}^2 x_H^2 x_\Phi^2 + 
   1104 g_{11 p} g' g_{1 p1}^3 x_H^2 x_\Phi^2 \\& + 
   1104 g_1 g_{1 p1}^4 x_H^2 x_\Phi^2 + 
   224 g_1 g_{11 p} g'^3 x_\Phi^3 + 
   448 g_1^2 g'^2 g_{1 p1} x_\Phi^3 + 
   672 g_{11 p}^2 g'^2 g_{1 p1} x_\Phi^3 + 
   1120 g_1 g_{11 p} g' g_{1 p1}^2 x_\Phi^3 \\& + 
   896 g_1^2 g_{1 p1}^3 x_\Phi^3 + 
   224 g_{11 p}^2 g_{1 p1}^3 x_\Phi^3 + 
   896 g_{11 p} g'^3 g_{1 p1} x_H x_\Phi^3 + 
   896 g_1 g'^2 g_{1 p1}^2 x_H x_\Phi^3 + 
   896 g_{11 p} g' g_{1 p1}^3 x_H x_\Phi^3 \\& + 
   896 g_1 g_{1 p1}^4 x_H x_\Phi^3 + 
   520 g_{11 p} g'^3 g_{1 p1} x_\Phi^4 + 
   520 g_1 g'^2 g_{1 p1}^2 x_\Phi^4 + 
   520 g_{11 p} g' g_{1 p1}^3 x_\Phi^4 + 
   520 g_1 g_{1 p1}^4 x_\Phi^4 - 51 g_1^3 y_t^2 \\& - 
   51 g_1 g_{11 p}^2 y_t^2 - 51 g_1 g_{11 p} g' x_H y_t^2 - 
   102 g_1^2 g_{1 p1} x_H y_t^2 - 
   51 g_{11 p}^2 g_{1 p1} x_H y_t^2 - 
   51 g_{11 p} g' g_{1 p1} x_H^2 y_t^2  \\& - 
   51 g_1 g_{1 p1}^2 x_H^2 y_t^2 - 
   30 g_1 g_{11 p} g' x_\Phi y_t^2 - 
   60 g_1^2 g_{1 p1} x_\Phi y_t^2 - 
   30 g_{11 p}^2 g_{1 p1} x_\Phi y_t^2 - 
   60 g_{11 p} g' g_{1 p1} x_H x_\Phi y_t^2 \end{split} \nonumber \ee \be \begin{split} \,\,\,\,\,\,\,\,\,\, &- 
   60 g_1 g_{1 p1}^2 x_H x_\Phi y_t^2 - 
   24 g_{11 p} g' g_{1 p1} x_\Phi^2 y_t^2 - 
   24 g_1 g_{1 p1}^2 x_\Phi^2 y_t^2  + \textrm{Tr}[Y_\nu Y_\nu^\dag]\Big(- 
   9 g_1^3  - 
   9 g_1 g_{11 p}^2  - 
   9 g_1 g_{11 p} g' x_H \\& - 
   18 g_1^2 g_{1 p1} x_H  - 
   9 g_{11 p}^2 g_{1 p1} x_H  - 
   9 g_{11 p} g' g_{1 p1} x_H^2   - 
   9 g_1 g_{1 p1}^2 x_H^2  - 
   18 g_1 g_{11 p} g' x_\Phi  - 
   36 g_1^2 g_{1 p1} x_\Phi \\& - 
   18 g_{11 p}^2 g_{1 p1} x_\Phi   - 
   36 g_{11 p} g' g_{1 p1} x_H x_\Phi  - 
   36 g_1 g_{1 p1}^2 x_H x_\Phi    - 
   72 g_{11 p} g' g_{1 p1} x_\Phi^2  - 
   72 g_1 g_{1 p1}^2 x_\Phi^2  \Big) \\& - 
   18 g_{11 p} g' g_{1 p1} x_\Phi^2 \textrm{Tr}[y_{NS} y_{NS}^\dag] - 
   18 g_1 g_{1 p1}^2 x_\Phi^2 \textrm{Tr}[y_{NS} y_{NS}^\dag] \Big) \end{split}  \ee

\be  \begin{split} \beta_{g_2}^{(2)} = & \frac{g_2^3}{6} \Big (9 g_1^2 + 9 g_{11 p}^2 + 35 g_2^2 + 
    72 g_3^2 + 6 g_{11 p} g' x_H + 6 g_1 g_{1 p1} x_H + 
    9 g'^2 x_H^2 + 9 g_{1 p1}^2 x_H^2 + 24 g_{11 p} g' x_\Phi \\& + 
    24 g_1 g_{1 p1} x_\Phi + 24 g'^2 x_H x_\Phi + 
    24 g_{1 p1}^2 x_H x_\Phi + 24 g'^2 x_\Phi^2 + 
    24 g_{1 p1}^2 x_\Phi^2 - 9 y_t^2 - 
    3 \textrm{Tr}[Y_\nu Y_\nu^\dag]\Big) \end{split} \ee

   \be \begin{split} \beta_{g_3}^{(2)} = & \frac{g_3^3}{6} \Big (11 g_1^2 + 11 g_{11 p}^2 + 27 g_2^2 - 
   156 g_3^2 + 22 g_{11 p} g' x_H + 22 g_1 g_{1 p1} x_H + 
   11 g'^2 x_H^2 + 11 g_{1 p1}^2 x_H^2  \\& + 8 g_{11 p} g' x_\Phi + 
   8 g_1 g_{1 p1} x_\Phi + 8 g'^2 x_H x_\Phi + 
   8 g_{1 p1}^2 x_H x_\Phi + 8 g'^2 x_\Phi^2 + 
   8 g_{1 p1}^2 x_\Phi^2 - 12 y_t^2 \Big) \end{split}  \ee

   \be \begin{split}   \beta_{g'}^{(2)} = & \frac {1} {18}\Big (199 g_1^2 g_{11 p}^2 g' + 
     199 g_{11 p}^4 g' + 199 g_1^3 g_{11 p} g_{1 p1} + 
     199 g_1 g_{11 p}^3 g_{1 p1} + 81 g_{11 p}^2 g' g_2^2 + 
     81 g_1 g_{11 p} g_{1 p1} g_2^2 \\& + 264 g_{11 p}^2 g' g_3^2 + 
     264 g_1 g_{11 p} g_{1 p1} g_3^2 + 
     362 g_1^2 g_{11 p} g'^2 x_H + 724 g_{11 p}^3 g'^2 x_H + 
     181 g_1^3 g' g_{1 p1} x_H \\& + 
     905 g_1 g_{11 p}^2 g' g_{1 p1} x_H  + 
     543 g_1^2 g_{11 p} g_{1 p1}^2 x_H + 
     181 g_{11 p}^3 g_{1 p1}^2 x_H + 54 g_{11 p} g'^2 g_2^2 x_H + 
     27 g_1 g' g_{1 p1} g_2^2 x_H \\& + 
     27 g_{11 p} g_{1 p1}^2 g_2^2 x_H  + 
     528 g_{11 p} g'^2 g_3^2 x_H + 
     264 g_1 g' g_{1 p1} g_3^2 x_H + 
     264 g_{11 p} g_{1 p1}^2 g_3^2 x_H + 199 g_1^2 g'^3 x_H^2 \\& + 
     1194 g_{11 p}^2 g'^3 x_H^2  + 
     1393 g_1 g_{11 p} g'^2 g_{1 p1} x_H^2 + 
     597 g_1^2 g' g_{1 p1}^2 x_H^2 + 
     796 g_{11 p}^2 g' g_{1 p1}^2 x_H^2 + 
     597 g_1 g_{11 p} g_{1 p1}^3 x_H^2 \\& + 81 g'^3 g_2^2 x_H^2 + 
     81 g' g_{1 p1}^2 g_2^2 x_H^2 + 264 g'^3 g_3^2 x_H^2 + 
     264 g' g_{1 p1}^2 g_3^2 x_H^2 + 724 g_{11 p} g'^4 x_H^3 \\& + 
     543 g_1 g'^3 g_{1 p1} x_H^3 + 
     905 g_{11 p} g'^2 g_{1 p1}^2 x_H^3 + 
     543 g_1 g' g_{1 p1}^3 x_H^3 + 
     181 g_{11 p} g_{1 p1}^4 x_H^3 + 199 g'^5 x_H^4 \\& + 
     398 g'^3 g_{1 p1}^2 x_H^4 + 199 g' g_{1 p1}^4 x_H^4 + 
     328 g_1^2 g_{11 p} g'^2 x_\Phi + 656 g_{11 p}^3 g'^2 x_\Phi + 
     164 g_1^3 g' g_{1 p1} x_\Phi \\& + 
     820 g_1 g_{11 p}^2 g' g_{1 p1} x_\Phi + 
     492 g_1^2 g_{11 p} g_{1 p1}^2 x_\Phi + 
     164 g_{11 p}^3 g_{1 p1}^2 x_\Phi + 
     216 g_{11 p} g'^2 g_2^2 x_\Phi + 
     108 g_1 g' g_{1 p1} g_2^2 x_\Phi \\& + 
     108 g_{11 p} g_{1 p1}^2 g_2^2 x_\Phi + 
     192 g_{11 p} g'^2 g_3^2 x_\Phi + 
     96 g_1 g' g_{1 p1} g_3^2 x_\Phi + 
     96 g_{11 p} g_{1 p1}^2 g_3^2 x_\Phi + 
     328 g_1^2 g'^3 x_H x_\Phi  \\& + 1968 g_{11 p}^2 g'^3 x_H x_\Phi  + 
     2296 g_1 g_{11 p} g'^2 g_{1 p1} x_H x_\Phi + 
     984 g_1^2 g' g_{1 p1}^2 x_H x_\Phi + 
     1312 g_{11 p}^2 g' g_{1 p1}^2 x_H x_\Phi \\& + 
     984 g_1 g_{11 p} g_{1 p1}^3 x_H x_\Phi   + 
     216 g'^3 g_2^2 x_H x_\Phi + 
     216 g' g_{1 p1}^2 g_2^2 x_H x_\Phi  + 
     192 g'^3 g_3^2 x_H x_\Phi + 
     192 g' g_{1 p1}^2 g_3^2 x_H x_\Phi  \end{split} \nonumber \ee \be \begin{split} \,\,\,\,\,\,\,\,\,\, & + 
     1968 g_{11 p} g'^4 x_H^2 x_\Phi  + 
     1476 g_1 g'^3 g_{1 p1} x_H^2 x_\Phi + 
     2460 g_{11 p} g'^2 g_{1 p1}^2 x_H^2 x_\Phi + 
     1476 g_1 g' g_{1 p1}^3 x_H^2 x_\Phi \\& + 
     492 g_{11 p} g_{1 p1}^4 x_H^2 x_\Phi  + 656 g'^5 x_H^3 x_\Phi + 
     1312 g'^3 g_{1 p1}^2 x_H^3 x_\Phi + 
     656 g' g_{1 p1}^4 x_H^3 x_\Phi + 184 g_1^2 g'^3 x_\Phi^2 \\& + 
     1104 g_{11 p}^2 g'^3 x_\Phi^2  + 
     1288 g_1 g_{11 p} g'^2 g_{1 p1} x_\Phi^2  + 
     552 g_1^2 g' g_{1 p1}^2 x_\Phi^2 + 
     736 g_{11 p}^2 g' g_{1 p1}^2 x_\Phi^2 + 
     552 g_1 g_{11 p} g_{1 p1}^3 x_\Phi^2 \\& + 
     216 g'^3 g_2^2 x_\Phi^2 + 216 g' g_{1 p1}^2 g_2^2 x_\Phi^2  + 
     192 g'^3 g_3^2 x_\Phi^2 + 192 g' g_{1 p1}^2 g_3^2 x_\Phi^2 + 
     2208 g_{11 p} g'^4 x_H x_\Phi^2 + 
     1104 g'^5 x_H^2 x_\Phi^2 \\& + 
     1656 g_1 g'^3 g_{1 p1} x_H x_\Phi^2 + 
     2760 g_{11 p} g'^2 g_{1 p1}^2 x_H x_\Phi^2 + 
     1656 g_1 g' g_{1 p1}^3 x_H x_\Phi^2 + 
     552 g_{11 p} g_{1 p1}^4 x_H x_\Phi^2  \\& + 
     2208 g'^3 g_{1 p1}^2 x_H^2 x_\Phi^2 + 
     1104 g' g_{1 p1}^4 x_H^2 x_\Phi^2 + 
     896 g_{11 p} g'^4 x_\Phi^3 + 672 g_1 g'^3 g_{1 p1} x_\Phi^3  + 
     1120 g_{11 p} g'^2 g_{1 p1}^2 x_\Phi^3 \\& + 
     672 g_1 g' g_{1 p1}^3 x_\Phi^3 + 
     224 g_{11 p} g_{1 p1}^4 x_\Phi^3 + 896 g'^5 x_H x_\Phi^3 + 
     1792 g'^3 g_{1 p1}^2 x_H x_\Phi^3 + 
     896 g' g_{1 p1}^4 x_H x_\Phi^3 \\& + 520 g'^5 x_\Phi^4 + 
     1040 g'^3 g_{1 p1}^2 x_\Phi^4 + 520 g' g_{1 p1}^4 x_\Phi^4 - 
     51 g_{11 p}^2 g'  y_t^2 - 51 g_1 g_{11 p} g_{1 p1}  y_t^2 \\& - 
     102 g_{11 p} g'^2 x_H  y_t^2 - 
     51 g_1 g' g_{1 p1} x_H  y_t^2 - 
     51 g_{11 p} g_{1 p1}^2 x_H  y_t^2 - 51 g'^3 x_H^2  y_t^2 - 
     51 g' g_{1 p1}^2 x_H^2  y_t^2 \\& - 
     60 g_{11 p} g'^2 x_\Phi  y_t^2 - 
     30 g_1 g' g_{1 p1} x_\Phi  y_t^2 - 
     30 g_{11 p} g_{1 p1}^2 x_\Phi  y_t^2 - 
     60 g'^3 x_H x_\Phi  y_t^2 - 
     60 g' g_{1 p1}^2 x_H x_\Phi  y_t^2 \\& - 24 g'^3 x_\Phi^2  y_t^2 - 
     24 g' g_{1 p1}^2 x_\Phi^2  y_t^2 - 
     9 g_{11 p}^2 g'  \textrm{Tr}[Y_\nu Y_\nu^\dag] - 
     9 g_1 g_{11 p} g_{1 p1}  \textrm{Tr}[Y_\nu Y_\nu^\dag] \\& - 
     18 g_{11 p} g'^2 x_H  \textrm{Tr}[Y_\nu Y_\nu^\dag] - 
     9 g_1 g' g_{1 p1} x_H  \textrm{Tr}[Y_\nu Y_\nu^\dag] - 
     9 g_{11 p} g_{1 p1}^2 
      x_H  \textrm{Tr}[Y_\nu Y_\nu^\dag] - 
     9 g'^3 x_H^2  \textrm{Tr}[Y_\nu Y_\nu^\dag] \\& - 
     9 g' g_{1 p1}^2 x_H^2  \textrm{Tr}[Y_\nu Y_\nu^\dag] - 
     36 g_{11 p} g'^2 x_\Phi  \textrm{Tr}[Y_\nu Y_\nu^\dag] - 
     18 g_1 g' g_{1 p1} 
      x_\Phi  \textrm{Tr}[Y_\nu Y_\nu^\dag] - 
     18 g_{11 p} g_{1 p1}^2 
      x_\Phi  \textrm{Tr}[Y_\nu Y_\nu^\dag] \\& - 
     36 g'^3 x_H x_\Phi  \textrm{Tr}[Y_\nu Y_\nu^\dag] - 
     36 g' g_{1 p1}^2 x_H 
      x_\Phi  \textrm{Tr}[Y_\nu Y_\nu^\dag] - 
     72 g'^3 x_\Phi^2  \textrm{Tr}[Y_\nu Y_\nu^\dag] - 
     72 g' g_{1 p1}^2 x_\Phi^2  \textrm{Tr}[Y_\nu Y_\nu^\dag] \\& - 
     18 g'^3 x_\Phi^2  \textrm{Tr}[y_{NS} y_{NS}^\dag] - 
     18 g' g_{1 p1}^2 x_\Phi^2  \textrm{Tr}[y_{NS} y_{NS}^\dag]\Big) \end{split}  \ee

   \be \begin{split} \beta_{g_{1p1}}^{(2)} =   \frac {1}{18}\Big ( & 199 g_1^3 g_{11 p} g' + 
   199 g_1 g_{11 p}^3 g' + 199 g_1^4 g_{1 p1} + 
   199 g_1^2 g_{11 p}^2 g_{1 p1} + 81 g_1 g_{11 p} g' g_2^2 \\& + 
   81 g_1^2 g_{1 p1} g_2^2 + 264 g_1 g_{11 p} g' g_3^2 + 
   264 g_1^2 g_{1 p1} g_3^2 + 181 g_1^3 g'^2 x_H + 
   543 g_1 g_{11 p}^2 g'^2 x_H \\& + 
   905 g_1^2 g_{11 p} g' g_{1 p1} x_H + 
   181 g_{11 p}^3 g' g_{1 p1} x_H + 724 g_1^3 g_{1 p1}^2 x_H + 
   362 g_1 g_{11 p}^2 g_{1 p1}^2 x_H + 27 g_1 g'^2 g_2^2 x_H  \\& + 
   27 g_{11 p} g' g_{1 p1} g_2^2 x_H + 
   54 g_1 g_{1 p1}^2 g_2^2 x_H + 264 g_1 g'^2 g_3^2 x_H + 
   264 g_{11 p} g' g_{1 p1} g_3^2 x_H + 
   528 g_1 g_{1 p1}^2 g_3^2 x_H  \\& + 597 g_1 g_{11 p} g'^3 x_H^2 + 
   796 g_1^2 g'^2 g_{1 p1} x_H^2 + 
   597 g_{11 p}^2 g'^2 g_{1 p1} x_H^2 + 
   1393 g_1 g_{11 p} g' g_{1 p1}^2 x_H^2 + 
   1194 g_1^2 g_{1 p1}^3 x_H^2 \\& + 
   199 g_{11 p}^2 g_{1 p1}^3 x_H^2 + 
   81 g'^2 g_{1 p1} g_2^2 x_H^2 + 81 g_{1 p1}^3 g_2^2 x_H^2 + 
   264 g'^2 g_{1 p1} g_3^2 x_H^2 + 264 g_{1 p1}^3 g_3^2 x_H^2 + 
   181 g_1 g'^4 x_H^3 \\& + 543 g_{11 p} g'^3 g_{1 p1} x_H^3 + 
   905 g_1 g'^2 g_{1 p1}^2 x_H^3 + 
   543 g_{11 p} g' g_{1 p1}^3 x_H^3 + 724 g_1 g_{1 p1}^4 x_H^3 + 
   199 g'^4 g_{1 p1} x_H^4 \\& + 398 g'^2 g_{1 p1}^3 x_H^4 + 
   199 g_{1 p1}^5 x_H^4 + 164 g_1^3 g'^2 x_\Phi + 
   492 g_1 g_{11 p}^2 g'^2 x_\Phi + 
   820 g_1^2 g_{11 p} g' g_{1 p1} x_\Phi \end{split} \nonumber \ee \be \begin{split} \,\,\,\,\,\,\,\,\,\, & + 
   164 g_{11 p}^3 g' g_{1 p1} x_\Phi + 
   656 g_1^3 g_{1 p1}^2 x_\Phi + 
   328 g_1 g_{11 p}^2 g_{1 p1}^2 x_\Phi + 
   108 g_1 g'^2 g_2^2 x_\Phi + 
   108 g_{11 p} g' g_{1 p1} g_2^2 x_\Phi \\& + 
   216 g_1 g_{1 p1}^2 g_2^2 x_\Phi + 96 g_1 g'^2 g_3^2 x_\Phi + 
   96 g_{11 p} g' g_{1 p1} g_3^2 x_\Phi + 
   192 g_1 g_{1 p1}^2 g_3^2 x_\Phi + 
   984 g_1 g_{11 p} g'^3 x_H x_\Phi \\& + 
   1312 g_1^2 g'^2 g_{1 p1} x_H x_\Phi + 
   984 g_{11 p}^2 g'^2 g_{1 p1} x_H x_\Phi + 
   2296 g_1 g_{11 p} g' g_{1 p1}^2 x_H x_\Phi + 
   1968 g_1^2 g_{1 p1}^3 x_H x_\Phi \\& + 
   328 g_{11 p}^2 g_{1 p1}^3 x_H x_\Phi  + 
   216 g'^2 g_{1 p1} g_2^2 x_H x_\Phi + 
   216 g_{1 p1}^3 g_2^2 x_H x_\Phi + 
   192 g'^2 g_{1 p1} g_3^2 x_H x_\Phi \\& + 
   192 g_{1 p1}^3 g_3^2 x_H x_\Phi  + 492 g_1 g'^4 x_H^2 x_\Phi  + 
   1476 g_{11 p} g'^3 g_{1 p1} x_H^2 x_\Phi + 
   2460 g_1 g'^2 g_{1 p1}^2 x_H^2 x_\Phi \\& + 
   1476 g_{11 p} g' g_{1 p1}^3 x_H^2 x_\Phi + 
   1968 g_1 g_{1 p1}^4 x_H^2 x_\Phi + 
   656 g'^4 g_{1 p1} x_H^3 x_\Phi  + 
   1312 g'^2 g_{1 p1}^3 x_H^3 x_\Phi \\& + 
   656 g_{1 p1}^5 x_H^3 x_\Phi + 552 g_1 g_{11 p} g'^3 x_\Phi^2 + 
   736 g_1^2 g'^2 g_{1 p1} x_\Phi^2 + 
   552 g_{11 p}^2 g'^2 g_{1 p1} x_\Phi^2 \\& + 
   1288 g_1 g_{11 p} g' g_{1 p1}^2 x_\Phi^2 + 
   1104 g_1^2 g_{1 p1}^3 x_\Phi^2 + 
   184 g_{11 p}^2 g_{1 p1}^3 x_\Phi^2 + 
   216 g'^2 g_{1 p1} g_2^2 x_\Phi^2 + 
   216 g_{1 p1}^3 g_2^2 x_\Phi^2 \\& + 
   192 g'^2 g_{1 p1} g_3^2 x_\Phi^2 + 
   192 g_{1 p1}^3 g_3^2 x_\Phi^2 + 552 g_1 g'^4 x_H x_\Phi^2 + 
   1656 g_{11 p} g'^3 g_{1 p1} x_H x_\Phi^2 + 
   2760 g_1 g'^2 g_{1 p1}^2 x_H x_\Phi^2 \\&  + 
   1656 g_{11 p} g' g_{1 p1}^3 x_H x_\Phi^2 + 
   2208 g_1 g_{1 p1}^4 x_H x_\Phi^2 + 
   1104 g'^4 g_{1 p1} x_H^2 x_\Phi^2 + 
   2208 g'^2 g_{1 p1}^3 x_H^2 x_\Phi^2  + 
   1104 g_{1 p1}^5 x_H^2 x_\Phi^2 \\ & + 224 g_1 g'^4 x_\Phi^3 + 
   672 g_{11 p} g'^3 g_{1 p1} x_\Phi^3 + 
   1120 g_1 g'^2 g_{1 p1}^2 x_\Phi^3  + 
   672 g_{11 p} g' g_{1 p1}^3 x_\Phi^3 + 
   896 g_1 g_{1 p1}^4 x_\Phi^3  \\ & + 896 g'^4 g_{1 p1} x_H x_\Phi^3 + 
   1792 g'^2 g_{1 p1}^3 x_H x_\Phi^3 + 
   896 g_{1 p1}^5 x_H x_\Phi^3 + 520 g'^4 g_{1 p1} x_\Phi^4 + 
   1040 g'^2 g_{1 p1}^3 x_\Phi^4  \\ & + 520 g_{1 p1}^5 x_\Phi^4  - 
   51 g_1 g_{11 p} g' y_t^2 - 51 g_1^2 g_{1 p1} y_t^2 - 
   51 g_1 g'^2 x_H y_t^2 - 51 g_{11 p} g' g_{1 p1} x_H y_t^2  \\ & - 
   102 g_1 g_{1 p1}^2 x_H y_t^2 - 51 g'^2 g_{1 p1} x_H^2 y_t^2 - 
   51 g_{1 p1}^3 x_H^2 y_t^2 - 30 g_1 g'^2 x_\Phi y_t^2 - 
   30 g_{11 p} g' g_{1 p1} x_\Phi y_t^2  \\ & - 
   60 g_1 g_{1 p1}^2 x_\Phi y_t^2 - 
   60 g'^2 g_{1 p1} x_H x_\Phi y_t^2 - 
   60 g_{1 p1}^3 x_H x_\Phi y_t^2 - 
   24 g'^2 g_{1 p1} x_\Phi^2 y_t^2 - 24 g_{1 p1}^3 x_\Phi^2 y_t^2  \\ & - 
   9 g_1 g_{11 p} g' \textrm{Tr}[Y_\nu Y_\nu^\dag] - 
   9 g_1^2 g_{1 p1} \textrm{Tr}[Y_\nu Y_\nu^\dag] - 
   9 g_1 g'^2 x_H \textrm{Tr}[Y_\nu Y_\nu^\dag] - 
   9 g_{11 p} g' g_{1 p1} x_H \textrm{Tr}[Y_\nu Y_\nu^\dag] \\ & - 
   18 g_1 g_{1 p1}^2 x_H \textrm{Tr}[Y_\nu Y_\nu^\dag] - 
   9 g'^2 g_{1 p1} x_H^2 \textrm{Tr}[Y_\nu Y_\nu^\dag] - 
   9 g_{1 p1}^3 x_H^2 \textrm{Tr}[Y_\nu Y_\nu^\dag] - 
   18 g_1 g'^2 x_\Phi \textrm{Tr}[Y_\nu Y_\nu^\dag] \\ & - 
   18 g_{11 p} g' g_{1 p1} x_\Phi \textrm{Tr}[
     Y_\nu Y_\nu^\dag] - 
   36 g_1 g_{1 p1}^2 x_\Phi \textrm{Tr}[Y_\nu Y_\nu^\dag] - 
   36 g'^2 g_{1 p1} x_H x_\Phi \textrm{Tr}[Y_\nu Y_\nu^\dag] - 
   36 g_{1 p1}^3 x_H x_\Phi \textrm{Tr}[Y_\nu Y_\nu^\dag]  \\ & - 
   72 g'^2 g_{1 p1} x_\Phi^2 \textrm{Tr}[Y_\nu Y_\nu^\dag] - 
   72 g_{1 p1}^3 x_\Phi^2 \textrm{Tr}[Y_\nu Y_\nu^\dag] - 
   18 g'^2 g_{1 p1} x_\Phi^2 \textrm{Tr}[y_{NS} y_{NS}^\dag]  - 
   18 g_{1 p1}^3 x_\Phi^2 \textrm{Tr}[y_{NS} y_{NS}^\dag]\Big) \end{split} \ee

 \be \begin{split} \beta_{g_{11p}}^{(2)} =  & \frac {1}{18}\Big (199 g_1^4 g_{11 p} + 
     398 g_1^2 g_{11 p}^3 + 199 g_{11 p}^5 + 
     81 g_1^2 g_{11 p} g_2^2 + 81 g_{11 p}^3 g_2^2  \\& + 
     264 g_1^2 g_{11 p} g_3^2 + 264 g_{11 p}^3 g_3^2 + 
     181 g_1^4 g' x_H + 905 g_1^2 g_{11 p}^2 g' x_H + 
     724 g_{11 p}^4 g' x_H  \\& + 543 g_1^3 g_{11 p} g_{1 p1} x_H + 
     543 g_1 g_{11 p}^3 g_{1 p1} x_H + 27 g_1^2 g' g_2^2 x_H + 
     54 g_{11 p}^2 g' g_2^2 x_H + 
     27 g_1 g_{11 p} g_{1 p1} g_2^2 x_H \\& + 
     264 g_1^2 g' g_3^2 x_H + 528 g_{11 p}^2 g' g_3^2 x_H + 
     264 g_1 g_{11 p} g_{1 p1} g_3^2 x_H + 
     796 g_1^2 g_{11 p} g'^2 x_H^2 + 1194 g_{11 p}^3 g'^2 x_H^2  \end{split} \nonumber \ee \be \begin{split} \,\,\,\,\,\,\,\,\,\, & + 
     597 g_1^3 g' g_{1 p1} x_H^2 + 
     1393 g_1 g_{11 p}^2 g' g_{1 p1} x_H^2 + 
     597 g_1^2 g_{11 p} g_{1 p1}^2 x_H^2 + 
     199 g_{11 p}^3 g_{1 p1}^2 x_H^2 + 
     81 g_{11 p} g'^2 g_2^2 x_H^2  \\& + 
     81 g_1 g' g_{1 p1} g_2^2 x_H^2 + 
     264 g_{11 p} g'^2 g_3^2 x_H^2 + 
     264 g_1 g' g_{1 p1} g_3^2 x_H^2 + 181 g_1^2 g'^3 x_H^3 + 
     724 g_{11 p}^2 g'^3 x_H^3 \\& + 
     905 g_1 g_{11 p} g'^2 g_{1 p1} x_H^3 + 
     543 g_1^2 g' g_{1 p1}^2 x_H^3 + 
     362 g_{11 p}^2 g' g_{1 p1}^2 x_H^3 + 
     181 g_1 g_{11 p} g_{1 p1}^3 x_H^3 + 
     199 g_{11 p} g'^4 x_H^4 \\& + 199 g_1 g'^3 g_{1 p1} x_H^4 + 
     199 g_{11 p} g'^2 g_{1 p1}^2 x_H^4 + 
     199 g_1 g' g_{1 p1}^3 x_H^4 + 164 g_1^4 g' x_\Phi + 
     820 g_1^2 g_{11 p}^2 g' x_\Phi \\& + 656 g_{11 p}^4 g' x_\Phi + 
     492 g_1^3 g_{11 p} g_{1 p1} x_\Phi + 
     492 g_1 g_{11 p}^3 g_{1 p1} x_\Phi + 
     108 g_1^2 g' g_2^2 x_\Phi + 216 g_{11 p}^2 g' g_2^2 x_\Phi \\&+ 
     108 g_1 g_{11 p} g_{1 p1} g_2^2 x_\Phi  + 
     96 g_1^2 g' g_3^2 x_\Phi + 192 g_{11 p}^2 g' g_3^2 x_\Phi + 
     96 g_1 g_{11 p} g_{1 p1} g_3^2 x_\Phi + 
     1312 g_1^2 g_{11 p} g'^2 x_H x_\Phi \\& + 
     1968 g_{11 p}^3 g'^2 x_H x_\Phi + 
     984 g_1^3 g' g_{1 p1} x_H x_\Phi + 
     2296 g_1 g_{11 p}^2 g' g_{1 p1} x_H x_\Phi + 
     984 g_1^2 g_{11 p} g_{1 p1}^2 x_H x_\Phi \\& + 
     328 g_{11 p}^3 g_{1 p1}^2 x_H x_\Phi  + 
     216 g_{11 p} g'^2 g_2^2 x_H x_\Phi + 
     216 g_1 g' g_{1 p1} g_2^2 x_H x_\Phi + 
     192 g_{11 p} g'^2 g_3^2 x_H x_\Phi \\& + 
     192 g_1 g' g_{1 p1} g_3^2 x_H x_\Phi + 
     492 g_1^2 g'^3 x_H^2 x_\Phi + 
     1968 g_{11 p}^2 g'^3 x_H^2 x_\Phi + 
     2460 g_1 g_{11 p} g'^2 g_{1 p1} x_H^2 x_\Phi \\& + 
     1476 g_1^2 g' g_{1 p1}^2 x_H^2 x_\Phi + 
     984 g_{11 p}^2 g' g_{1 p1}^2 x_H^2 x_\Phi + 
     492 g_1 g_{11 p} g_{1 p1}^3 x_H^2 x_\Phi + 
     656 g_{11 p} g'^4 x_H^3 x_\Phi \\& + 
     656 g_1 g'^3 g_{1 p1} x_H^3 x_\Phi + 
     656 g_{11 p} g'^2 g_{1 p1}^2 x_H^3 x_\Phi + 
     656 g_1 g' g_{1 p1}^3 x_H^3 x_\Phi + 
     736 g_1^2 g_{11 p} g'^2 x_\Phi^2 \\& + 
     1104 g_{11 p}^3 g'^2 x_\Phi^2 + 
     552 g_1^3 g' g_{1 p1} x_\Phi^2 + 
     1288 g_1 g_{11 p}^2 g' g_{1 p1} x_\Phi^2 + 
     552 g_1^2 g_{11 p} g_{1 p1}^2 x_\Phi^2 + 
     184 g_{11 p}^3 g_{1 p1}^2 x_\Phi^2 \\& + 
     216 g_{11 p} g'^2 g_2^2 x_\Phi^2 + 
     216 g_1 g' g_{1 p1} g_2^2 x_\Phi^2 + 
     192 g_{11 p} g'^2 g_3^2 x_\Phi^2 + 
     192 g_1 g' g_{1 p1} g_3^2 x_\Phi^2 + 
     552 g_1^2 g'^3 x_H x_\Phi^2 \\& + 
     2208 g_{11 p}^2 g'^3 x_H x_\Phi^2 + 
     2760 g_1 g_{11 p} g'^2 g_{1 p1} x_H x_\Phi^2 + 
     1656 g_1^2 g' g_{1 p1}^2 x_H x_\Phi^2 + 
     1104 g_{11 p}^2 g' g_{1 p1}^2 x_H x_\Phi^2 \\& + 
     552 g_1 g_{11 p} g_{1 p1}^3 x_H x_\Phi^2 + 
     1104 g_{11 p} g'^4 x_H^2 x_\Phi^2 + 
     1104 g_1 g'^3 g_{1 p1} x_H^2 x_\Phi^2 + 
     1104 g_{11 p} g'^2 g_{1 p1}^2 x_H^2 x_\Phi^2 \\& + 
     1104 g_1 g' g_{1 p1}^3 x_H^2 x_\Phi^2 + 
     224 g_1^2 g'^3 x_\Phi^3 + 896 g_{11 p}^2 g'^3 x_\Phi^3 + 
     1120 g_1 g_{11 p} g'^2 g_{1 p1} x_\Phi^3 + 
     672 g_1^2 g' g_{1 p1}^2 x_\Phi^3 \\& + 
     448 g_{11 p}^2 g' g_{1 p1}^2 x_\Phi^3 + 
     224 g_1 g_{11 p} g_{1 p1}^3 x_\Phi^3 + 
     896 g_{11 p} g'^4 x_H x_\Phi^3 + 
     896 g_1 g'^3 g_{1 p1} x_H x_\Phi^3 + 
     896 g_{11 p} g'^2 g_{1 p1}^2 x_H x_\Phi^3 \\& + 
     896 g_1 g' g_{1 p1}^3 x_H x_\Phi^3 + 
     520 g_{11 p} g'^4 x_\Phi^4 + 520 g_1 g'^3 g_{1 p1} x_\Phi^4 + 
     520 g_{11 p} g'^2 g_{1 p1}^2 x_\Phi^4 + 
     520 g_1 g' g_{1 p1}^3 x_\Phi^4 \\& - 51 g_1^2 g_{11 p}  y_t^2 - 
     51 g_{11 p}^3  y_t^2 - 51 g_1^2 g' x_H  y_t^2 - 
     102 g_{11 p}^2 g' x_H  y_t^2 - 
     51 g_1 g_{11 p} g_{1 p1} x_H  y_t^2 \\& - 
     51 g_{11 p} g'^2 x_H^2  y_t^2 - 
     51 g_1 g' g_{1 p1} x_H^2  y_t^2 - 30 g_1^2 g' x_\Phi  y_t^2 - 
     60 g_{11 p}^2 g' x_\Phi  y_t^2 - 
     30 g_1 g_{11 p} g_{1 p1} x_\Phi  y_t^2 \\& - 
     60 g_{11 p} g'^2 x_H x_\Phi  y_t^2 - 
     60 g_1 g' g_{1 p1} x_H x_\Phi  y_t^2 - 
     24 g_{11 p} g'^2 x_\Phi^2  y_t^2 - 
     24 g_1 g' g_{1 p1} x_\Phi^2  y_t^2 - 
     9 g_1^2 g_{11 p}  \textrm{Tr}[Y_\nu Y_\nu^\dag] \\& - 
     9 g_{11 p}^3  \textrm{Tr}[Y_\nu Y_\nu^\dag] - 
     9 g_1^2 g' x_H  \textrm{Tr}[Y_\nu Y_\nu^\dag] - 
     18 g_{11 p}^2 g' x_H  \textrm{Tr}[Y_\nu Y_\nu^\dag] - 
     9 g_1 g_{11 p} g_{1 p1} 
      x_H  \textrm{Tr}[Y_\nu Y_\nu^\dag] \\& - 
     9 g_{11 p} g'^2 x_H^2  \textrm{Tr}[Y_\nu Y_\nu^\dag] - 
     9 g_1 g' g_{1 p1} x_H^2  \textrm{Tr}[Y_\nu Y_\nu^\dag] - 
     18 g_1^2 g' x_\Phi  \textrm{Tr}[Y_\nu Y_\nu^\dag] - 
     36 g_{11 p}^2 g' x_\Phi  \textrm{Tr}[Y_\nu Y_\nu^\dag] \\& - 
     18 g_1 g_{11 p} g_{1 p1} 
      x_\Phi  \textrm{Tr}[Y_\nu Y_\nu^\dag] - 
     36 g_{11 p} g'^2 x_H 
      x_\Phi  \textrm{Tr}[Y_\nu Y_\nu^\dag] - 
     36 g_1 g' g_{1 p1} x_H 
      x_\Phi  \textrm{Tr}[Y_\nu Y_\nu^\dag] - 
     72 g_{11 p} g'^2 x_\Phi^2  \textrm{Tr}[Y_\nu Y_\nu^\dag] \\& - 
     72 g_1 g' g_{1 p1} 
      x_\Phi^2  \textrm{Tr}[Y_\nu Y_\nu^\dag] - 
     18 g_{11 p} g'^2 x_\Phi^2  \textrm{Tr}[y_{NS} y_{NS}^\dag] - 
     18 g_1 g' g_{1 p1} 
      x_\Phi^2  \textrm{Tr}[y_{NS} y_{NS}^\dag] \Big) \end{split} \ee

\be \begin{split} \beta_{\lambda_1}^{(2)} = & \frac {1}{48}\Big (-379 g_1^6 - 
     469 g_1^4 g_{11 p}^2 - 469 g_1^2 g_{11 p}^4 - 
     379 g_{11 p}^6 - 559 g_1^4 g_2^2 - 
     450 g_1^2 g_{11 p}^2 g_2^2 - 559 g_{11 p}^4 g_2^2 \\&- 
     289 g_1^2 g_2^4 - 289 g_{11 p}^2 g_2^4 + 915 g_2^6 + 
     1258 g_1^4 \lambda_1 + 828 g_1^2 g_{11 p}^2 \lambda_1 + 
     1258 g_{11 p}^4 \lambda_1 + 468 g_1^2 g_2^2 \lambda_1  + 
     468 g_{11 p}^2 g_2^2 \lambda_1 - 438 g_2^4 \lambda_1  \\& + 
     1728 g_1^2 \lambda_1^2   + 1728 g_{11 p}^2 \lambda_1^2 + 
     5184 g_2^2 \lambda_1^2 - 14976 \lambda_1^3 - 
     480 \lambda_1 \lambda_3^2 - 192 \lambda_3^3  \\& + 
     938 g_1^4 g_{11 p} g' x_H + 596 g_1^2 g_{11 p}^3 g' x_H   + 
     994 g_{11 p}^5 g' x_H + 994 g_1^5 g_{1 p1} x_H + 
     596 g_1^3 g_{11 p}^2 g_{1 p1} x_H  \\& + 
     938 g_1 g_{11 p}^4 g_{1 p1} x_H + 
     900 g_1^2 g_{11 p} g' g_2^2 x_H  + 
     956 g_{11 p}^3 g' g_2^2 x_H + 
     956 g_1^3 g_{1 p1} g_2^2 x_H + 
     900 g_1 g_{11 p}^2 g_{1 p1} g_2^2 x_H  \\& + 
     578 g_{11 p} g' g_2^4 x_H + 578 g_1 g_{1 p1} g_2^4 x_H - 
     1656 g_1^2 g_{11 p} g' \lambda_1 x_H - 
     1832 g_{11 p}^3 g' \lambda_1 x_H - 
     1832 g_1^3 g_{1 p1} \lambda_1 x_H  \\& - 
     1656 g_1 g_{11 p}^2 g_{1 p1} \lambda_1 x_H - 
     936 g_{11 p} g' g_2^2 \lambda_1 x_H - 
     936 g_1 g_{1 p1} g_2^2 \lambda_1 x_H - 
     3456 g_{11 p} g' \lambda_1^2 x_H  - 
     3456 g_1 g_{1 p1} \lambda_1^2 x_H  \\& - 469 g_1^4 g'^2 x_H^2 - 
     254 g_1^2 g_{11 p}^2 g'^2 x_H^2 - 565 g_{11 p}^4 g'^2 x_H^2 - 
     1192 g_1^3 g_{11 p} g' g_{1 p1} x_H^2 - 
     1192 g_1 g_{11 p}^3 g' g_{1 p1} x_H^2  \\& - 
     565 g_1^4 g_{1 p1}^2 x_H^2 - 
     254 g_1^2 g_{11 p}^2 g_{1 p1}^2 x_H^2 - 
     469 g_{11 p}^4 g_{1 p1}^2 x_H^2 - 
     450 g_1^2 g'^2 g_2^2 x_H^2 - 
     794 g_{11 p}^2 g'^2 g_2^2 x_H^2  \\& - 
     1800 g_1 g_{11 p} g' g_{1 p1} g_2^2 x_H^2 - 
     794 g_1^2 g_{1 p1}^2 g_2^2 x_H^2 - 
     450 g_{11 p}^2 g_{1 p1}^2 g_2^2 x_H^2 - 
     289 g'^2 g_2^4 x_H^2 - 289 g_{1 p1}^2 g_2^4 x_H^2 \\& + 
     828 g_1^2 g'^2 \lambda_1 x_H^2   + 
     1148 g_{11 p}^2 g'^2 \lambda_1 x_H^2 + 
     3312 g_1 g_{11 p} g' g_{1 p1} \lambda_1 x_H^2 + 
     1148 g_1^2 g_{1 p1}^2 \lambda_1 x_H^2 + 
     828 g_{11 p}^2 g_{1 p1}^2 \lambda_1 x_H^2 \\& + 
     468 g'^2 g_2^2 \lambda_1 x_H^2   + 
     468 g_{1 p1}^2 g_2^2 \lambda_1 x_H^2 + 
     1728 g'^2 \lambda_1^2 x_H^2 + 
     1728 g_{1 p1}^2 \lambda_1^2 x_H^2  + 
     596 g_1^2 g_{11 p} g'^3 x_H^3 \\& - 100 g_{11 p}^3 g'^3 x_H^3  + 
     596 g_1^3 g'^2 g_{1 p1} x_H^3 + 
     508 g_1 g_{11 p}^2 g'^2 g_{1 p1} x_H^3 + 
     508 g_1^2 g_{11 p} g' g_{1 p1}^2 x_H^3 + 
     596 g_{11 p}^3 g' g_{1 p1}^2 x_H^3 \\& - 
     100 g_1^3 g_{1 p1}^3 x_H^3   + 
     596 g_1 g_{11 p}^2 g_{1 p1}^3 x_H^3 + 
     956 g_{11 p} g'^3 g_2^2 x_H^3 + 
     900 g_1 g'^2 g_{1 p1} g_2^2 x_H^3  + 
     900 g_{11 p} g' g_{1 p1}^2 g_2^2 x_H^3 \\& + 
     956 g_1 g_{1 p1}^3 g_2^2 x_H^3   - 
     1832 g_{11 p} g'^3 \lambda_1 x_H^3 - 
     1656 g_1 g'^2 g_{1 p1} \lambda_1 x_H^3 - 
     1656 g_{11 p} g' g_{1 p1}^2 \lambda_1 x_H^3  \\& - 
     1832 g_1 g_{1 p1}^3 \lambda_1 x_H^3 - 469 g_1^2 g'^4 x_H^4  - 
     565 g_{11 p}^2 g'^4 x_H^4 - 
     1192 g_1 g_{11 p} g'^3 g_{1 p1} x_H^4 - 
     254 g_1^2 g'^2 g_{1 p1}^2 x_H^4 \\& - 
     254 g_{11 p}^2 g'^2 g_{1 p1}^2 x_H^4 - 
     1192 g_1 g_{11 p} g' g_{1 p1}^3 x_H^4 - 
     565 g_1^2 g_{1 p1}^4 x_H^4 - 
     469 g_{11 p}^2 g_{1 p1}^4 x_H^4 - 559 g'^4 g_2^2 x_H^4 \\& - 
     450 g'^2 g_{1 p1}^2 g_2^2 x_H^4 - 
     559 g_{1 p1}^4 g_2^2 x_H^4 + 1258 g'^4 \lambda_1 x_H^4 + 
     828 g'^2 g_{1 p1}^2 \lambda_1 x_H^4 + 
     1258 g_{1 p1}^4 \lambda_1 x_H^4 \\& + 994 g_{11 p} g'^5 x_H^5 + 
     938 g_1 g'^4 g_{1 p1} x_H^5 + 
     596 g_{11 p} g'^3 g_{1 p1}^2 x_H^5  + 
     596 g_1 g'^2 g_{1 p1}^3 x_H^5 + 
     938 g_{11 p} g' g_{1 p1}^4 x_H^5 \\& + 
     994 g_1 g_{1 p1}^5 x_H^5 - 379 g'^6 x_H^6 - 
     469 g'^4 g_{1 p1}^2 x_H^6 - 469 g'^2 g_{1 p1}^4 x_H^6 - 
     379 g_{1 p1}^6 x_H^6 \\&- 512 g_1^2 g_{11 p}^3 g' x_\Phi - 
     512 g_{11 p}^5 g' x_\Phi - 512 g_1^5 g_{1 p1} x_\Phi - 
     512 g_1^3 g_{11 p}^2 g_{1 p1} x_\Phi - 
     512 g_{11 p}^3 g' g_2^2 x_\Phi \\& - 
     512 g_1^3 g_{1 p1} g_2^2 x_\Phi + 
     1280 g_{11 p}^3 g' \lambda_1 x_\Phi + 
     1280 g_1^3 g_{1 p1} \lambda_1 x_\Phi + 
     512 g_1^2 g_{11 p}^2 g'^2 x_H x_\Phi  + 
     1536 g_{11 p}^4 g'^2 x_H x_\Phi \\& + 
     1024 g_1^3 g_{11 p} g' g_{1 p1} x_H x_\Phi + 
     1024 g_1 g_{11 p}^3 g' g_{1 p1} x_H x_\Phi + 
     1536 g_1^4 g_{1 p1}^2 x_H x_\Phi  + 
     512 g_1^2 g_{11 p}^2 g_{1 p1}^2 x_H x_\Phi  \\& +
     512 g_{11 p}^2 g'^2 g_2^2 x_H x_\Phi  + 
     512 g_1^2 g_{1 p1}^2 g_2^2 x_H x_\Phi - 
     1280 g_{11 p}^2 g'^2 \lambda_1 x_H x_\Phi - 
     1280 g_1^2 g_{1 p1}^2 \lambda_1 x_H x_\Phi \\&+ 
     512 g_1^2 g_{11 p} g'^3 x_H^2 x_\Phi  - 
     1024 g_{11 p}^3 g'^3 x_H^2 x_\Phi - 
     512 g_1^3 g'^2 g_{1 p1} x_H^2 x_\Phi - 
     1024 g_1 g_{11 p}^2 g'^2 g_{1 p1} x_H^2 x_\Phi \end{split} \nonumber \ee \be \begin{split} \,\,\,\,\,\,\,\,\,\, & - 
     1024 g_1^2 g_{11 p} g' g_{1 p1}^2 x_H^2 x_\Phi - 
     512 g_{11 p}^3 g' g_{1 p1}^2 x_H^2 x_\Phi - 
     1024 g_1^3 g_{1 p1}^3 x_H^2 x_\Phi + 
     512 g_1 g_{11 p}^2 g_{1 p1}^3 x_H^2 x_\Phi \\& + 
     512 g_{11 p} g'^3 g_2^2 x_H^2 x_\Phi + 
     512 g_1 g_{1 p1}^3 g_2^2 x_H^2 x_\Phi  - 
     1280 g_{11 p} g'^3 \lambda_1 x_H^2 x_\Phi - 
     1280 g_1 g_{1 p1}^3 \lambda_1 x_H^2 x_\Phi - 
     512 g_1^2 g'^4 x_H^3 x_\Phi \\& - 
     1024 g_{11 p}^2 g'^4 x_H^3 x_\Phi  - 
     1024 g_1 g_{11 p} g'^3 g_{1 p1} x_H^3 x_\Phi + 
     512 g_1^2 g'^2 g_{1 p1}^2 x_H^3 x_\Phi + 
     512 g_{11 p}^2 g'^2 g_{1 p1}^2 x_H^3 x_\Phi \\& - 
     1024 g_1 g_{11 p} g' g_{1 p1}^3 x_H^3 x_\Phi  - 
     1024 g_1^2 g_{1 p1}^4 x_H^3 x_\Phi - 
     512 g_{11 p}^2 g_{1 p1}^4 x_H^3 x_\Phi  - 
     512 g'^4 g_2^2 x_H^3 x_\Phi - 
     512 g_{1 p1}^4 g_2^2 x_H^3 x_\Phi \\& + 
     1280 g'^4 \lambda_1 x_H^3 x_\Phi  + 
     1280 g_{1 p1}^4 \lambda_1 x_H^3 x_\Phi + 
     1536 g_{11 p} g'^5 x_H^4 x_\Phi + 
     1024 g_1 g'^4 g_{1 p1} x_H^4 x_\Phi + 
     512 g_{11 p} g'^3 g_{1 p1}^2 x_H^4 x_\Phi \\& + 
     512 g_1 g'^2 g_{1 p1}^3 x_H^4 x_\Phi  + 
     1024 g_{11 p} g' g_{1 p1}^4 x_H^4 x_\Phi + 
     1536 g_1 g_{1 p1}^5 x_H^4 x_\Phi - 512 g'^6 x_H^5 x_\Phi - 
     512 g'^4 g_{1 p1}^2 x_H^5 x_\Phi \\& - 
     512 g'^2 g_{1 p1}^4 x_H^5 x_\Phi  - 
     512 g_{1 p1}^6 x_H^5 x_\Phi - 
     540 g_1^2 g_{11 p}^2 g'^2 x_\Phi^2 - 
     540 g_{11 p}^4 g'^2 x_\Phi^2 - 
     540 g_1^4 g_{1 p1}^2 x_\Phi^2 \\& - 
     540 g_1^2 g_{11 p}^2 g_{1 p1}^2 x_\Phi^2  - 
     540 g_{11 p}^2 g'^2 g_2^2 x_\Phi^2 - 
     540 g_1^2 g_{1 p1}^2 g_2^2 x_\Phi^2 + 
     1368 g_{11 p}^2 g'^2 \lambda_1 x_\Phi^2  \\& + 
     1368 g_1^2 g_{1 p1}^2 \lambda_1 x_\Phi^2 + 
     240 g_{11 p}^2 g'^2 \lambda_3 x_\Phi^2  + 
     384 g'^2 \lambda_3^2 x_\Phi^2 + 
     384 g_{1 p1}^2 \lambda_3^2 x_\Phi^2 \\& + 
     1080 g_1^2 g_{11 p} g'^3 x_H x_\Phi^2 + 
     2160 g_{11 p}^3 g'^3 x_H x_\Phi^2 + 
     1080 g_1 g_{11 p}^2 g'^2 g_{1 p1} x_H x_\Phi^2  + 
     1080 g_1^2 g_{11 p} g' g_{1 p1}^2 x_H x_\Phi^2 \\& + 
     2160 g_1^3 g_{1 p1}^3 x_H x_\Phi^2 + 
     1080 g_1 g_{11 p}^2 g_{1 p1}^3 x_H x_\Phi^2 + 
     1080 g_{11 p} g'^3 g_2^2 x_H x_\Phi^2 + 
     1080 g_1 g_{1 p1}^3 g_2^2 x_H x_\Phi^2 \\& - 
     2736 g_{11 p} g'^3 \lambda_1 x_H x_\Phi^2  - 
     2736 g_1 g_{1 p1}^3 \lambda_1 x_H x_\Phi^2 - 
     480 g_{11 p} g'^3 \lambda_3 x_H x_\Phi^2 - 
     540 g_1^2 g'^4 x_H^2 x_\Phi^2 \\& - 
     3240 g_{11 p}^2 g'^4 x_H^2 x_\Phi^2  - 
     2160 g_1 g_{11 p} g'^3 g_{1 p1} x_H^2 x_\Phi^2 - 
     540 g_1^2 g'^2 g_{1 p1}^2 x_H^2 x_\Phi^2 - 
     540 g_{11 p}^2 g'^2 g_{1 p1}^2 x_H^2 x_\Phi^2 \\& - 
     2160 g_1 g_{11 p} g' g_{1 p1}^3 x_H^2 x_\Phi^2 - 
     3240 g_1^2 g_{1 p1}^4 x_H^2 x_\Phi^2  - 
     540 g_{11 p}^2 g_{1 p1}^4 x_H^2 x_\Phi^2 - 
     540 g'^4 g_2^2 x_H^2 x_\Phi^2 \\& - 
     540 g_{1 p1}^4 g_2^2 x_H^2 x_\Phi^2 + 
     1368 g'^4 \lambda_1 x_H^2 x_\Phi^2 + 
     1368 g_{1 p1}^4 \lambda_1 x_H^2 x_\Phi^2 + 
     240 g'^4 \lambda_3 x_H^2 x_\Phi^2 + 
     2160 g_{11 p} g'^5 x_H^3 x_\Phi^2 \\& + 
     1080 g_1 g'^4 g_{1 p1} x_H^3 x_\Phi^2 + 
     1080 g_{11 p} g'^3 g_{1 p1}^2 x_H^3 x_\Phi^2  + 
     1080 g_1 g'^2 g_{1 p1}^3 x_H^3 x_\Phi^2 + 
     1080 g_{11 p} g' g_{1 p1}^4 x_H^3 x_\Phi^2 \\& + 
     2160 g_1 g_{1 p1}^5 x_H^3 x_\Phi^2 - 540 g'^6 x_H^4 x_\Phi^2 - 
     540 g'^4 g_{1 p1}^2 x_H^4 x_\Phi^2  - 
     540 g'^2 g_{1 p1}^4 x_H^4 x_\Phi^2  - 
     540 g_{1 p1}^6 x_H^4 x_\Phi^2 \\& + y_t^2 \Big( - 228 g_1^4   - 
     456 g_1^2 g_{11 p}^2 - 228 g_{11 p}^4  + 
     504 g_1^2 g_2^2  + 504 g_{11 p}^2 g_2^2  - 
     108 g_2^4  + 680 g_1^2 \lambda_1  + 
     680 g_{11 p}^2 \lambda_1 \\& + 
     1080 g_2^2 \lambda_1  + 3840 g_3^2 \lambda_1   - 
     6912 \lambda_1^2  + 
     1360 g_{11 p} g' \lambda_1 x_H  + 
     1360 g_1 g_{1 p1} \lambda_1 x_H  + 
     456 g_1^2 g'^2 x_H^2  + 456 g_{11 p}^2 g'^2 x_H^2 \\& + 
     456 g_1^2 g_{1 p1}^2 x_H^2  + 
     456 g_{11 p}^2 g_{1 p1}^2 x_H^2  - 
     504 g'^2 g_2^2 x_H^2  - 
     504 g_{1 p1}^2 g_2^2 x_H^2  + 
     680 g'^2 \lambda_1 x_H^2  + 
     680 g_{1 p1}^2 \lambda_1 x_H^2 \\& - 228 g'^4 x_H^4  - 
     456 g'^2 g_{1 p1}^2 x_H^4  - 228 g_{1 p1}^4 x_H^4  - 
     480 g_1^2 g_{11 p} g' x_\Phi  - 
     480 g_{11 p}^3 g' x_\Phi  - 
     480 g_1^3 g_{1 p1} x_\Phi \\& - 
     480 g_1 g_{11 p}^2 g_{1 p1} x_\Phi  + 
     288 g_{11 p} g' g_2^2 x_\Phi  + 
     288 g_1 g_{1 p1} g_2^2 x_\Phi  + 
     800 g_{11 p} g' \lambda_1 x_\Phi   + 
     800 g_1 g_{1 p1} \lambda_1 x_\Phi  + 
     480 g_1^2 g'^2 x_H x_\Phi \\& + 
     480 g_{11 p}^2 g'^2 x_H x_\Phi  + 
     480 g_1^2 g_{1 p1}^2 x_H x_\Phi  + 
     480 g_{11 p}^2 g_{1 p1}^2 x_H x_\Phi  - 
     288 g'^2 g_2^2 x_H x_\Phi - 
     288 g_{1 p1}^2 g_2^2 x_H x_\Phi \\& + 
     800 g'^2 \lambda_1 x_H x_\Phi   + 
     800 g_{1 p1}^2 \lambda_1 x_H x_\Phi  + 
     480 g_{11 p} g'^3 x_H^2 x_\Phi  + 
     480 g_1 g'^2 g_{1 p1} x_H^2 x_\Phi \\&  + 
     480 g_{11 p} g' g_{1 p1}^2 x_H^2 x_\Phi   + 
     480 g_1 g_{1 p1}^3 x_H^2 x_\Phi  - 
     480 g'^4 x_H^3 x_\Phi  - 
     960 g'^2 g_{1 p1}^2 x_H^3 x_\Phi  - 
     480 g_{1 p1}^4 x_H^3 x_\Phi \\& - 
     192 g_{11 p}^2 g'^2 x_\Phi^2  - 
     384 g_1 g_{11 p} g' g_{1 p1} x_\Phi^2  - 
     192 g_1^2 g_{1 p1}^2 x_\Phi^2  + 
     320 g'^2 \lambda_1 x_\Phi^2  + 
     320 g_{1 p1}^2 \lambda_1 x_\Phi^2  \end{split} \nonumber \ee \be \begin{split} \,\,\,\,\,\,\,\,\,\, & + 
     384 g_{11 p} g'^3 x_H x_\Phi^2  + 
     384 g_1 g'^2 g_{1 p1} x_H x_\Phi^2  + 
     384 g_{11 p} g' g_{1 p1}^2 x_H x_\Phi^2  + 
     384 g_1 g_{1 p1}^3 x_H x_\Phi^2  \\& - 
     192 g'^4 x_H^2 x_\Phi^2  - 
     384 g'^2 g_{1 p1}^2 x_H^2 x_\Phi^2  - 
     192 g_{1 p1}^4 x_H^2 x_\Phi^2  \Big) \\& 
     + \textrm{Tr}[Y_\nu Y_\nu^\dag] \Big(- 
     12 g_1^4  - 
     24 g_1^2 g_{11 p}^2  - 
     12 g_{11 p}^4  - 
     24 g_1^2 g_2^2  - 
     24 g_{11 p}^2 g_2^2  - 
     36 g_2^4  + 
     120 g_1^2 \lambda_1  + 
     120 g_{11 p}^2 \lambda_1 \\& + 
     360 g_2^2 \lambda_1  - 
     2304 \lambda_1^2 + 
     240 g_{11 p} g' \lambda_1 x_H  + 
     240 g_1 g_{1 p1} \lambda_1 x_H  + 
     24 g_1^2 g'^2 x_H^2  + 
     24 g_{11 p}^2 g'^2 x_H^2  + 
     24 g_1^2 g_{1 p1}^2 x_H^2  \\& + 
     24 g_{11 p}^2 g_{1 p1}^2 x_H^2  + 
     24 g'^2 g_2^2 x_H^2  + 
     24 g_{1 p1}^2 g_2^2 x_H^2 + 
     120 g'^2 \lambda_1 x_H^2  + 
     120 g_{1 p1}^2 \lambda_1 x_H^2  - 
     12 g'^4 x_H^4  - 
     24 g'^2 g_{1 p1}^2 x_H^4 \\&  - 
     12 g_{1 p1}^4 x_H^4  - 
     288 g_1^2 g_{11 p} g' x_\Phi  - 
     288 g_{11 p}^3 g' x_\Phi  - 
     288 g_1^3 g_{1 p1} x_\Phi  - 
     288 g_1 g_{11 p}^2 g_{1 p1} x_\Phi \\& - 
     288 g_{11 p} g' g_2^2 x_\Phi  - 
     288 g_1 g_{1 p1} g_2^2 x_\Phi  + 
     480 g_{11 p} g' \lambda_1 x_\Phi  + 
     480 g_1 g_{1 p1} \lambda_1 x_\Phi  + 
     288 g_1^2 g'^2 x_H x_\Phi \\& + 
     288 g_{11 p}^2 g'^2 x_H x_\Phi  + 
     288 g_1^2 g_{1 p1}^2 x_H x_\Phi  + 
     288 g_{11 p}^2 g_{1 p1}^2 x_H x_\Phi  + 
     288 g'^2 g_2^2 x_H x_\Phi \\& + 
     288 g_{1 p1}^2 g_2^2 x_H x_\Phi  + 
     480 g'^2 \lambda_1 x_H x_\Phi + 
     480 g_{1 p1}^2 \lambda_1 x_H x_\Phi  + 
     288 g_{11 p} g'^3 x_H^2 x_\Phi \\& + 
     288 g_1 g'^2 g_{1 p1} x_H^2 x_\Phi + 
     288 g_{11 p} g' g_{1 p1}^2 x_H^2 x_\Phi  + 
     288 g_1 g_{1 p1}^3 x_H^2 x_\Phi  - 
     288 g'^4 x_H^3 x_\Phi - 
     576 g'^2 g_{1 p1}^2 x_H^3 x_\Phi \\&  - 
     288 g_{1 p1}^4 x_H^3 x_\Phi  - 
     576 g_{11 p}^2 g'^2 x_\Phi^2 - 
     1152 g_1 g_{11 p} g' g_{1 p1} x_\Phi^2  - 
     576 g_1^2 g_{1 p1}^2 x_\Phi^2 + 
     960 g'^2 \lambda_1 x_\Phi^2  \\& + 
     960 g_{1 p1}^2 \lambda_1 x_\Phi^2 + 
     1152 g_{11 p} g'^3 x_H x_\Phi^2  + 
     1152 g_1 g'^2 g_{1 p1} x_H x_\Phi^2  + 
     1152 g_{11 p} g' g_{1 p1}^2 x_H x_\Phi^2 + 
     1152 g_1 g_{1 p1}^3 x_H x_\Phi^2 \\& - 
     576 g'^4 x_H^2 x_\Phi^2  - 
     1152 g'^2 g_{1 p1}^2 x_H^2 x_\Phi^2  - 
     576 g_{1 p1}^4 x_H^2 x_\Phi^2  \Big) \\&- 
     96 \lambda_3^2 \textrm{Tr}[y_{NS} y_{NS}^\dag] - 128 g_1^2 y_t^4 - 
     128 g_{11 p}^2 y_t^4 - 1536 g_3^2 y_t^4 - 
     144 \lambda_1 y_t^4 - 832 g_{11 p} g' x_H y_t^4 \\& - 
     832 g_1 g_{1 p1} x_H y_t^4 - 128 g'^2 x_H^2 y_t^4 - 
     128 g_{1 p1}^2 x_H^2 y_t^4 - 320 g_{11 p} g' x_\Phi y_t^4 - 
     320 g_1 g_{1 p1} x_\Phi y_t^4 \\& - 320 g'^2 x_H x_\Phi y_t^4 - 
     320 g_{1 p1}^2 x_H x_\Phi y_t^4 - 128 g'^2 x_\Phi^2 y_t^4 - 
     128 g_{1 p1}^2 x_\Phi^2 y_t^4  - 
     48 \lambda_1 \textrm{Tr}[
       Y_\nu Y_\nu^\dag Y_\nu Y_\nu^\dag] \\& - 
     192 g_{11 p} g' x_H \textrm{Tr}[
       Y_\nu Y_\nu^\dag Y_\nu Y_\nu^\dag] - 
     192 g_1 g_{1 p1} x_H \textrm{Tr}[
       Y_\nu Y_\nu^\dag Y_\nu Y_\nu^\dag] - 
     192 g_{11 p} g' x_\Phi \textrm{Tr}[
       Y_\nu Y_\nu^\dag Y_\nu Y_\nu^\dag]\\& - 
     192 g_1 g_{1 p1} x_\Phi \textrm{Tr}[
       Y_\nu Y_\nu^\dag Y_\nu Y_\nu^\dag] - 
     192 g'^2 x_H x_\Phi \textrm{Tr}[
       Y_\nu Y_\nu^\dag Y_\nu Y_\nu^\dag] - 
     192 g_{1 p1}^2 x_H x_\Phi \textrm{Tr}[
       Y_\nu Y_\nu^\dag Y_\nu Y_\nu^\dag] \\& - 
     384 g'^2 x_\Phi^2 \textrm{Tr}[
       Y_\nu Y_\nu^\dag Y_\nu Y_\nu^\dag]  - 
     384 g_{1 p1}^2 x_\Phi^2 \textrm{Tr}[
       Y_\nu Y_\nu^\dag Y_\nu Y_\nu^\dag] 
       - 144 \lambda_1  \textrm {Tr}[y_{NS} y_{NS}^\dag Y_\nu^T Y_\nu^*] 
       + 1440 y_t^6 \\& + 
     480 \textrm{Tr}[Y_\nu Y_\nu^\dag Y_\nu Y_\nu^\dag Y_\nu Y_\nu^\dag] + 
     96  \textrm{Tr}[
        y_{NS} y_{NS}^\dag Y_\nu^T Y_\nu^* 
         Y_\nu^T Y_\nu^*] \Big) \end{split}  \ee

\be \begin{split} \beta_{\lambda_2}^{(2)} = & \frac {1}{3}\Big (-720 \lambda_2^3 + 
     12 g_1^2 \lambda_3^2 + 12 g_{11 p}^2 \lambda_3^2 + 
     36 g_2^2 \lambda_3^2  - 60 \lambda_2 \lambda_3^2 \\& - 
     24 \lambda_3^3 - 24 g_{11 p} g' \lambda_3^2 x_H - 
     24 g_1 g_{1 p1} \lambda_3^2 x_H + 
     12 g'^2 \lambda_3^2 x_H^2 + 12 g_{1 p1}^2 \lambda_3^2 x_H^2 \\& + 
     211 g_{11 p}^2 g'^2 \lambda_2 x_\Phi^2 + 
     211 g_1^2 g_{1 p1}^2 \lambda_2 x_\Phi^2 + 
     336 g'^2 \lambda_2^2 x_\Phi^2 + 
     336 g_{1 p1}^2 \lambda_2^2 x_\Phi^2 + 
     30 g_{11 p}^2 g'^2 \lambda_3 x_\Phi^2 \\& + 
     378 g_{11 p} g'^3 \lambda_2 x_H x_\Phi^2 + 
     378 g_1 g_{1 p1}^3 \lambda_2 x_H x_\Phi^2 - 
     60 g_{11 p} g'^3 \lambda_3 x_H x_\Phi^2 + 
     211 g'^4 \lambda_2 x_H^2 x_\Phi^2 + 
     211 g_{1 p1}^4 \lambda_2 x_H^2 x_\Phi^2 \\& + 
     30 g'^4 \lambda_3 x_H^2 x_\Phi^2 + 
     320 g_{11 p} g'^3 \lambda_2 x_\Phi^3 + 
     320 g_1 g_{1 p1}^3 \lambda_2 x_\Phi^3 + 
     320 g'^4 \lambda_2 x_H x_\Phi^3 + 
     320 g_{1 p1}^4 \lambda_2 x_H x_\Phi^3 \\& - 
     334 g_{11 p}^2 g'^4 x_\Phi^4 + 636 g'^4 \lambda_2 x_\Phi^4 + 
     36 g'^2 g_{1 p1}^2 \lambda_2 x_\Phi^4 + 
     360 g_{1 p1}^4 \lambda_2 x_\Phi^4 - 
     612 g_{11 p} g'^5 x_H x_\Phi^4 \\& - 334 g'^6 x_H^2 x_\Phi^4 - 
     512 g_{11 p} g'^5 x_\Phi^5 - 512 g'^6 x_H x_\Phi^5 - 
     720 g'^6 x_\Phi^6 - 180 g'^4 g_{1 p1}^2 x_\Phi^6 - 
     36 \lambda_3^2 y_t^2 \\& - 
     12 \lambda_3^2 \textrm{Tr}[Y_\nu Y_\nu^\dag] - 
     120 \lambda_2^2 \textrm{Tr}[y_{NS} y_{NS}^\dag] + 
     30 g'^2 \lambda_2 x_\Phi^2 \textrm{Tr}[y_{NS} y_{NS}^\dag] + 
     30 g_{1 p1}^2 \lambda_2 x_\Phi^2 \textrm{Tr}[y_{NS} y_{NS}^\dag] \\& - 
     12 g'^4 x_\Phi^4 \textrm{Tr}[y_{NS} y_{NS}^\dag] + 
     6 \lambda_2 \textrm{Tr}[y_{NS} y_{NS}^\dag y_{NS} y_{NS}^\dag ] - 
     18 \lambda_2  \textrm{Tr}[y_{NS} y_{NS}^\dag Y_\nu^T Y_\nu^*] + 
     24  \textrm{Tr}[y_{NS} y_{NS}^\dag y_{NS} y_{NS}^\dag y_{NS} y_{NS}^\dag] \\& + 
     12  \textrm{Tr}[
        y_{NS} y_{NS}^\dag Y_\nu^T Y_\nu^* y_{NS} y_{NS}^\dag] \Big) \end{split}  \ee

\be   \begin{split} \beta_{\lambda_3}^{(2)} = & (557 g_1^4 \lambda_3)/
   48 + (45 g_1^2 g_{11 p}^2 \lambda_3)/
   8 + (557 g_{11 p}^4 \lambda_3)/
   48 + (15 g_1^2 g_2^2 \lambda_3)/
   8   + (15 g_{11 p}^2 g_2^2 \lambda_3)/8 - (145 g_2^4 \lambda_3)/
   16 \\& + 24 g_1^2 \lambda_1 \lambda_3 + 
  24 g_{11 p}^2 \lambda_1 \lambda_3  + 
  72 g_2^2 \lambda_1 \lambda_3 - 60 \lambda_1^2 \lambda_3 - 
  40 \lambda_2^2 \lambda_3 + g_1^2 \lambda_3^2  + 
  g_{11 p}^2 \lambda_3^2 + 3 g_2^2 \lambda_3^2 \\& - 
  72 \lambda_1 \lambda_3^2 - 48 \lambda_2 \lambda_3^2 - 
  11 \lambda_3^3 - (45 g_1^2 g_{11 p} g' \lambda_3 x_H)/
   4 -  (157 g_{11 p}^3 g' \lambda_3 x_H)/
   12 - (157 g_1^3 g_{1 p1} \lambda_3 x_H)/
   12 \\& - (45 g_1 g_{11 p}^2 g_{1 p1} \lambda_3 x_H)/
   4 - (15 g_{11 p} g' g_2^2 \lambda_3 x_H)/
   4  - (15 g_1 g_{1 p1} g_2^2 \lambda_3 x_H)/4 - 
  48 g_{11 p} g' \lambda_1 \lambda_3 x_H  \\&- 
  48 g_1 g_{1 p1} \lambda_1 \lambda_3 x_H - 
  2 g_{11 p} g' \lambda_3^2 x_H \\& - 
  2 g_1 g_{1 p1} \lambda_3^2 x_H + (45 g_1^2 g'^2 \lambda_3
x_H^2)/8 + (71 g_{11 p}^2 g'^2 \lambda_3 x_H^2)/
   24 + (45 g_1 g_{11 p} g' g_{1 p1} \lambda_3 x_H^2)/
   2 \\& + (71 g_1^2 g_{1 p1}^2 \lambda_3 x_H^2)/
   24  + (45 g_{11 p}^2 g_{1 p1}^2 \lambda_3 x_H^2)/
   8 + (15 g'^2 g_2^2 \lambda_3 x_H^2)/
   8 + (15 g_{1 p1}^2 g_2^2 \lambda_3 x_H^2)/8 \\& + 
  24 g'^2 \lambda_1 \lambda_3 x_H^2 + 
  24 g_{1 p1}^2 \lambda_1 \lambda_3 x_H^2  + 
  g'^2 \lambda_3^2 x_H^2 + 
  g_{1 p1}^2 \lambda_3^2 x_H^2 \\& - (157 g_{11 p} g'^3 \lambda_3 
x_H^3)/12 - (45 g_1 g'^2 g_{1 p1} \lambda_3 x_H^3)/
   4 - (45 g_{11 p} g' g_{1 p1}^2 \lambda_3 x_H^3)/
   4 - (157 g_1 g_{1 p1}^3 \lambda_3 x_H^3)/
   12 \\& + (557 g'^4 \lambda_3 x_H^4)/
   48 + (45 g'^2 g_{1 p1}^2 \lambda_3 x_H^4)/
   8 + (557 g_{1 p1}^4 \lambda_3 x_H^4)/
   48 + (40 g_{11 p}^3 g' \lambda_3 x_\Phi)/
   3 \\& + (40 g_1^3 g_{1 p1} \lambda_3 x_\Phi)/
   3 - (40 g_{11 p}^2 g'^2 \lambda_3 x_H x_\Phi)/
   3 - (40 g_1^2 g_{1 p1}^2 \lambda_3 x_H x_\Phi)/
   3 - (40 g_{11 p} g'^3 \lambda_3 x_H^2 x_\Phi)/
   3 \\& - (40 g_1 g_{1 p1}^3 \lambda_3 x_H^2 x_\Phi)/
   3 + (40 g'^4 \lambda_3 x_H^3 x_\Phi)/
   3 + (40 g_{1 p1}^4 \lambda_3 x_H^3 x_\Phi)/
   3 - (15 g_1^2 g_{11 p}^2 g'^2 x_\Phi^2)/
   4 \\& - (713 g_{11 p}^4 g'^2 x_\Phi^2)/
   12 - (45 g_{11 p}^2 g'^2 g_2^2 x_\Phi^2)/4 + 
  30 g_{11 p}^2 g'^2 \lambda_1 x_\Phi^2 + 
  20 g_{11 p}^2 g'^2 \lambda_2 x_\Phi^2 \end{split} \nonumber \ee \be \begin{split} \,\,\,\,\,\,\,\,\,\, & + (617 g_{11 p}^2 g'^2 
\lambda_3 x_\Phi^2)/
   12 + (593 g_1^2 g_{1 p1}^2 \lambda_3 x_\Phi^2)/12 + 
  64 g'^2 \lambda_2 \lambda_3 x_\Phi^2 + 
  64 g_{1 p1}^2 \lambda_2 \lambda_3 x_\Phi^2 \\& + 
  4 g'^2 \lambda_3^2 x_\Phi^2 + 
  4 g_{1 p1}^2 \lambda_3^2 x_\Phi^2 + (15 g_1^2 g_{11 p} g'^3 x_H 
x_\Phi^2)/2 + (73 g_{11 p}^3 g'^3 x_H x_\Phi^2)/
   3 \\& + (15 g_1 g_{11 p}^2 g'^2 g_{1 p1} x_H x_\Phi^2)/
   2 + (45 g_{11 p} g'^3 g_2^2 x_H x_\Phi^2)/2 - 
  60 g_{11 p} g'^3 \lambda_1 x_H x_\Phi^2 - 
  40 g_{11 p} g'^3 \lambda_2 x_H x_\Phi^2 \\& + (61 g_{11 p} g'^3 
\lambda_3 x_H x_\Phi^2)/
   2 + (69 g_1 g_{1 p1}^3 \lambda_3 x_H x_\Phi^2)/
   2 - (15 g_1^2 g'^4 x_H^2 x_\Phi^2)/
   4 + (421 g_{11 p}^2 g'^4 x_H^2 x_\Phi^2)/6 \\& - 
  15 g_1 g_{11 p} g'^3 g_{1 p1} x_H^2 x_\Phi^2 - (15 g_{11 p}^2 
g'^2 g_{1 p1}^2 x_H^2 x_\Phi^2)/4 - (45 g'^4 g_2^2 x_H^2 x_\Phi^2)/
   4  \\& + 30 g'^4 \lambda_1 x_H^2 x_\Phi^2 + 
  20 g'^4 \lambda_2 x_H^2 x_\Phi^2 + (617 g'^4 \lambda_3 x_H^2 
x_\Phi^2)/12 + (593 g_{1 p1}^4 \lambda_3 x_H^2 x_\Phi^2)/
   12 \\&+ (73 g_{11 p} g'^5 x_H^3 x_\Phi^2)/
   3 + (15 g_1 g'^4 g_{1 p1} x_H^3 x_\Phi^2)/
   2 + (15 g_{11 p} g'^3 g_{1 p1}^2 x_H^3 x_\Phi^2)/
   2 - (713 g'^6 x_H^4 x_\Phi^2)/
   12 \\& - (15 g'^4 g_{1 p1}^2 x_H^4 x_\Phi^2)/
   4 - (256 g_{11 p}^3 g'^3 x_\Phi^3)/
   3 + (160 g_{11 p} g'^3 \lambda_3 x_\Phi^3)/
   3 + (160 g_1 g_{1 p1}^3 \lambda_3 x_\Phi^3)/
   3 + \\& (256 g_{11 p}^2 g'^4 x_H x_\Phi^3)/
   3 + (160 g'^4 \lambda_3 x_H x_\Phi^3)/
   3 + (160 g_{1 p1}^4 \lambda_3 x_H x_\Phi^3)/
   3 + (256 g_{11 p} g'^5 x_H^2 x_\Phi^3)/
   3 \\&- (256 g'^6 x_H^3 x_\Phi^3)/3 - 105 g_{11 p}^2 g'^4 x_\Phi^4 - 
  15 g_{11 p}^2 g'^2 g_{1 p1}^2 x_\Phi^4 + 
  82 g'^4 \lambda_3 x_\Phi^4 \\& + 
  6 g'^2 g_{1 p1}^2 \lambda_3 x_\Phi^4 + 
  60 g_{1 p1}^4 \lambda_3 x_\Phi^4 + 
  210 g_{11 p} g'^5 x_H x_\Phi^4 + 
  30 g_{11 p} g'^3 g_{1 p1}^2 x_H x_\Phi^4 \\& - 
  105 g'^6 x_H^2 x_\Phi^4 - 
  15 g'^4 g_{1 p1}^2 x_H^2 x_\Phi^4 + (85 g_1^2 \lambda_3 y_t^2)/
   12 + (85 g_{11 p}^2 \lambda_3 y_t^2)/
   12 \\&+ (45 g_2^2 \lambda_3 y_t^2)/4 + 40 g_3^2 \lambda_3 y_t^2 - 
  72 \lambda_1 \lambda_3 y_t^2 - 
  12 \lambda_3^2 y_t^2 + (85 g_{11 p} g' \lambda_3 x_H y_t^2)/
   6\\& + (85 g_1 g_{1 p1} \lambda_3 x_H y_t^2)/
   6 + (85 g'^2 \lambda_3 x_H^2 y_t^2)/
   12 + (85 g_{1 p1}^2 \lambda_3 x_H^2 y_t^2)/
   12 + (25 g_{11 p} g' \lambda_3 x_\Phi y_t^2)/
   3 \\&+ (25 g_1 g_{1 p1} \lambda_3 x_\Phi y_t^2)/
   3 + (25 g'^2 \lambda_3 x_H x_\Phi y_t^2)/
   3 + (25 g_{1 p1}^2 \lambda_3 x_H x_\Phi y_t^2)/3 - 
  19 g_{11 p}^2 g'^2 x_\Phi^2 y_t^2 \\& + (10 g'^2 \lambda_3 x_\Phi^2 
y_t^2)/3 + (10 g_{1 p1}^2 \lambda_3 x_\Phi^2 y_t^2)/3 - 
  38 g_{11 p} g'^3 x_H x_\Phi^2 y_t^2 - 
  19 g'^4 x_H^2 x_\Phi^2 y_t^2 \\& - 40 g_{11 p} g'^3 x_\Phi^3 y_t^2 - 
  40 g'^4 x_H x_\Phi^3 y_t^2 - 
  16 g'^4 x_\Phi^4 y_t^2 + (5 g_1^2 \lambda_3 \textrm{Tr}[
      Y_\nu Y_\nu^\dag])/
   4 \\& + (5 g_{11 p}^2 \lambda_3 \textrm{Tr}[Y_\nu Y_\nu^\dag])/
   4 + (15 g_2^2 \lambda_3 \textrm{Tr}[Y_\nu Y_\nu^\dag])/4 - 
  24 \lambda_1 \lambda_3 \textrm{Tr}[Y_\nu Y_\nu^\dag] \\& - 
  4 \lambda_3^2 \textrm{Tr}[
    Y_\nu Y_\nu^\dag] + (5 g_{11 p} g' \lambda_3 x_H \textrm {Tr}[Y_\nu Y_\nu^\dag])/
   2 \\& + (5 g_1 g_{1 p1} \lambda_3 x_H \textrm{Tr}[
      Y_\nu Y_\nu^\dag])/
   2 + (5 g'^2 \lambda_3 x_H^2 \textrm{Tr}[Y_\nu Y_\nu^\dag])/
   4 + (5 g_{1 p1}^2 \lambda_3 x_H^2 \textrm{Tr}[
      Y_\nu Y_\nu^\dag])/4 + 
  5 g_{11 p} g' \lambda_3 x_\Phi \textrm{Tr}[
    Y_\nu Y_\nu^\dag] \\& + 
  5 g_1 g_{1 p1} \lambda_3 x_\Phi \textrm{Tr}[
    Y_\nu Y_\nu^\dag] + 
  5 g'^2 \lambda_3 x_H x_\Phi \textrm{Tr}[Y_\nu Y_\nu^\dag] + 
  5 g_{1 p1}^2 \lambda_3 x_H x_\Phi \textrm{Tr}[
    Y_\nu Y_\nu^\dag] \\& - 
  g_{11 p}^2 g'^2 x_\Phi^2 \textrm{Tr}[Y_\nu Y_\nu^\dag] + 
  10 g'^2 \lambda_3 x_\Phi^2 \textrm{Tr}[Y_\nu Y_\nu^\dag] + 
  10 g_{1 p1}^2 \lambda_3 x_\Phi^2 \textrm{Tr}[
    Y_\nu Y_\nu^\dag] \\& - 
  2 g_{11 p} g'^3 x_H x_\Phi^2 \textrm{Tr}[Y_\nu Y_\nu^\dag] - 
  g'^4 x_H^2 x_\Phi^2 \textrm{Tr}[Y_\nu Y_\nu^\dag] - 
  24 g_{11 p} g'^3 x_\Phi^3 \textrm{Tr}[Y_\nu Y_\nu^\dag] \\& - 
  24 g'^4 x_H x_\Phi^3 \textrm{Tr}[Y_\nu Y_\nu^\dag] - 
  48 g'^4 x_\Phi^4 \textrm{Tr}[Y_\nu Y_\nu^\dag] - 
  16 \lambda_2 \lambda_3 \textrm{Tr}[y_{NS} y_{NS}^\dag] - 
  4 \lambda_3^2 \textrm{Tr}[y_{NS} y_{NS}^\dag] \\& - 
  g_{11 p}^2 g'^2 x_\Phi^2 \textrm{Tr}[y_{NS} y_{NS}^\dag] + 
  5 g'^2 \lambda_3 x_\Phi^2 \textrm{Tr}[y_{NS} y_{NS}^\dag] + 
  5 g_{1 p1}^2 \lambda_3 x_\Phi^2 \textrm{Tr}[y_{NS} y_{NS}^\dag] + 
  2 g_{11 p} g'^3 x_H x_\Phi^2 \textrm{Tr}[y_{NS} y_{NS}^\dag] \\& - 
  g'^4 x_H^2 x_\Phi^2 \textrm{Tr}[
    y_{NS} y_{NS}^\dag] - (27 \lambda_3 y_t^4)/
   2 - (9 \lambda_3 \textrm{Tr}[
      Y_\nu Y_\nu^\dag Y_\nu Y_\nu^\dag])/2 - 
  3 \lambda_3 \textrm{Tr}[
    y_{NS} y_{NS}^\dag y_{NS} y_{NS}^\dag ]  \end{split} \nonumber \ee \be \begin{split} \,\,\,\,\,\,\,\,\,\, & + (7 \lambda_3  \textrm{Tr}[
      y_{NS} y_{NS}^\dag Y_\nu^T Y_\nu^*])/2 \\&- 
  4 g_{11 p} g' x_H  \textrm{Tr}[
    y_{NS} y_{NS}^\dag Y_\nu^T Y_\nu^*] - 
  4 g_1 g_{1 p1} x_H  \textrm{Tr}[
    y_{NS} y_{NS}^\dag Y_\nu^T Y_\nu^*] - 
  4 g_{11 p} g' x_\Phi  \textrm{Tr}[
    y_{NS} y_{NS}^\dag Y_\nu^T Y_\nu^*] \\& - 
  4 g_1 g_{1 p1} x_\Phi  \textrm{Tr}[
    y_{NS} y_{NS}^\dag Y_\nu^T Y_\nu^*] - 
  4 g'^2 x_H x_\Phi  \textrm{Tr}[y_{NS} y_{NS}^\dag Y_\nu^T Y_\nu^*] - 
  4 g_{1 p1}^2 x_H x_\Phi  \textrm{Tr}[
    y_{NS} y_{NS}^\dag Y_\nu^T Y_\nu^*] \\&- 
  8 g'^2 x_\Phi^2  \textrm{Tr}[y_{NS} y_{NS}^\dag Y_\nu^T Y_\nu^*] - 
  8 g_{1 p1}^2 x_\Phi^2  \textrm{Tr}[
    y_{NS} y_{NS}^\dag Y_\nu^T Y_\nu^*] + 
  6 \textrm{Tr}[y_{NS} y_{NS}^\dag y_{NS} y_{NS}^\dag Y_\nu^T Y_\nu^*] \\&+  
  4  \textrm{Tr}[y_{NS} y_{NS}^\dag Y_\nu^T Y_\nu^* y_{NS} y_{NS}^\dag] + 
  14  \textrm{Tr}[
     y_{NS} y_{NS}^\dag Y_\nu^T Y_\nu^* Y_\nu^T Y_\nu^*] \end{split} \ee

\be \begin{split} \beta_{y_{NS}^{(2)}} = \frac {1} {24} & \Big (4 \Big (24  \lambda_2^2 + 
        6  \lambda_3^2 + 44 g_{11 p}^2 g'^2 x_\Phi^2 + 
        88 g_1 g_{11 p} g' g_{1 p1} x_\Phi^2 + 
        44 g_1^2 g_{1 p1}^2 x_\Phi^2 + 
        72 g_{11 p} g'^3 x_H x_\Phi^2 \\& + 
        72 g_1 g'^2 g_{1 p1} x_H x_\Phi^2 + 
        72 g_{11 p} g' g_{1 p1}^2 x_H x_\Phi^2 + 
        72 g_1 g_{1 p1}^3 x_H x_\Phi^2 + 44 g'^4 x_H^2 x_\Phi^2 + 
        88 g'^2 g_{1 p1}^2 x_H^2 x_\Phi^2 \\& + 
        44 g_{1 p1}^4 x_H^2 x_\Phi^2 + 64 g_{11 p} g'^3 x_\Phi^3 + 
        64 g_1 g'^2 g_{1 p1} x_\Phi^3 + 
        64 g_{11 p} g' g_{1 p1}^2 x_\Phi^3 + 
        64 g_1 g_{1 p1}^3 x_\Phi^3 \\& + 64 g'^4 x_H x_\Phi^3 + 
        128 g'^2 g_{1 p1}^2 x_H x_\Phi^3 + 
        64 g_{1 p1}^4 x_H x_\Phi^3 + 36 g'^4 x_\Phi^4 + 
        72 g'^2 g_{1 p1}^2 x_\Phi^4  \\& + 36 g_{1 p1}^4 x_\Phi^4 + 
        15 (g'^2 + g_{1 p1}^2) x_\Phi^2 \textrm {Tr}[y_{NS} y_{NS}^\dag] - 
        9 \textrm {Tr}[y_{NS} y_{NS}^\dag y_{NS} y_{NS}^\dag] - 
        9 \textrm {Tr}[y_{NS} y_{NS}^\dag Y_\nu^T Y_\nu^*]\Big) y_{NS}  \\& + 
     3 \Big (-4 (16  \lambda_2 - 26 (g'^2 + g_{1 p1}^2) x_\Phi^2 + 
            3 \textrm {Tr}[y_{NS} y_{NS}^\dag]) (y_{NS} y_{NS}^\dag y_{NS}) \\& + (17 g_1^2 + 17 g_{11 p}^2 + 51 g_2^2 - 32  \lambda_3 - 
            38 g_{11 p} g' x_H - 38 g_1 g_{1 p1} x_H + 
            17 g'^2 x_H^2 + 17 g_{1 p1}^2 x_H^2 \\& - 
            4 g_{11 p} g' x_\Phi - 4 g_1 g_{1 p1} x_\Phi - 
            4 g'^2 x_H x_\Phi - 4 g_{1 p1}^2 x_H x_\Phi - 
            32 g'^2 x_\Phi^2 - 32 g_{1 p1}^2 x_\Phi^2 - 36 y_t^2 \\& - 
            12 \textrm {Tr}[Y_\nu Y_\nu^\dag]) (Y_\nu^
             T Y_\nu^* y_{NS}) + 14 (y_{NS} y_{NS}^\dag y_{NS} y_{NS}^\dag y_{NS}) - 
         2 (y_{NS} y_{NS}^\dag Y_\nu^T Y_\nu^* y_{NS}) - 
         2 (Y_\nu^T Y_\nu^* Y_\nu^T Y_\nu^* y_{NS})\Big)\Big) \end{split} \ee

\be \begin{split}  \beta_{Y_\nu^{(2)}} = \frac {1} {48} & \Big (2 \Big (35 g_1^4 + 
        70 g_1^2 g_{11 p}^2 + 35 g_{11 p}^4 - 54 g_1^2 g_2^2 - 
        54 g_{11 p}^2 g_2^2 - 138 g_2^4 + 144  \lambda_1^2 + 
        12  \lambda_3^2 \\&  + 460 g_1^2 g_{11 p} g' x_H + 
        460 g_{11 p}^3 g' x_H + 460 g_1^3 g_{1 p1} x_H + 
        460 g_1 g_{11 p}^2 g_{1 p1} x_H + 
        162 g_{11 p} g' g_2^2 x_H  \\& + 162 g_1 g_{1 p1} g_2^2 x_H + 
        288 g_1^2 g'^2 x_H^2 + 574 g_{11 p}^2 g'^2 x_H^2 + 
        572 g_1 g_{11 p} g' g_{1 p1} x_H^2 + 
        574 g_1^2 g_{1 p1}^2 x_H^2 \\&+ 
        288 g_{11 p}^2 g_{1 p1}^2 x_H^2 - 54 g'^2 g_2^2 x_H^2 - 
        54 g_{1 p1}^2 g_2^2 x_H^2 + 460 g_{11 p} g'^3 x_H^3 + 
        460 g_1 g'^2 g_{1 p1} x_H^3  \\&+ 
        460 g_{11 p} g' g_{1 p1}^2 x_H^3 + 
        460 g_1 g_{1 p1}^3 x_H^3 + 35 g'^4 x_H^4 + 
        70 g'^2 g_{1 p1}^2 x_H^4 + 35 g_{1 p1}^4 x_H^4 + 
        504 g_1^2 g_{11 p} g' x_\Phi \\&+ 504 g_{11 p}^3 g' x_\Phi + 
        504 g_1^3 g_{1 p1} x_\Phi + 
        504 g_1 g_{11 p}^2 g_{1 p1} x_\Phi + 
        54 g_{11 p} g' g_2^2 x_\Phi + 
        54 g_1 g_{1 p1} g_2^2 x_\Phi \\& + 
        652 g_1^2 g'^2 x_H x_\Phi + 
        1664 g_{11 p}^2 g'^2 x_H x_\Phi + 
        2024 g_1 g_{11 p} g' g_{1 p1} x_H x_\Phi + 
        1664 g_1^2 g_{1 p1}^2 x_H x_\Phi \\& + 
        652 g_{11 p}^2 g_{1 p1}^2 x_H x_\Phi + 
        54 g'^2 g_2^2 x_H x_\Phi + 
        54 g_{1 p1}^2 g_2^2 x_H x_\Phi  + 
        1664 g_{11 p} g'^3 x_H^2 x_\Phi \\& + 
        1664 g_1 g'^2 g_{1 p1} x_H^2 x_\Phi + 
        1664 g_{11 p} g' g_{1 p1}^2 x_H^2 x_\Phi + 
        1664 g_1 g_{1 p1}^3 x_H^2 x_\Phi + 504 g'^4 x_H^3 x_\Phi \end{split} \nonumber \ee \be \begin{split} \,\,\,\,\,\,\,\,\,\,\,\,\,\,\,\, & + 
        1008 g'^2 g_{1 p1}^2 x_H^3 x_\Phi + 
        504 g_{1 p1}^4 x_H^3 x_\Phi + 374 g_1^2 g'^2 x_\Phi^2 + 
        1574 g_{11 p}^2 g'^2 x_\Phi^2 + 
        2400 g_1 g_{11 p} g' g_{1 p1} x_\Phi^2 \\& + 
        1574 g_1^2 g_{1 p1}^2 x_\Phi^2 + 
        374 g_{11 p}^2 g_{1 p1}^2 x_\Phi^2 + 
        162 g'^2 g_2^2 x_\Phi^2 + 162 g_{1 p1}^2 g_2^2 x_\Phi^2 + 
        3080 g_{11 p} g'^3 x_H x_\Phi^2 \\& + 
        3080 g_1 g'^2 g_{1 p1} x_H x_\Phi^2 + 
        3080 g_{11 p} g' g_{1 p1}^2 x_H x_\Phi^2 + 
        3080 g_1 g_{1 p1}^3 x_H x_\Phi^2 + 
        1574 g'^4 x_H^2 x_\Phi^2  \\& + 
        3148 g'^2 g_{1 p1}^2 x_H^2 x_\Phi^2 + 
        1574 g_{1 p1}^4 x_H^2 x_\Phi^2 + 
        1892 g_{11 p} g'^3 x_\Phi^3 + 
        1892 g_1 g'^2 g_{1 p1} x_\Phi^3 + 
        1892 g_{11 p} g' g_{1 p1}^2 x_\Phi^3 \\& + 
        1892 g_1 g_{1 p1}^3 x_\Phi^3 + 1892 g'^4 x_H x_\Phi^3 + 
        3784 g'^2 g_{1 p1}^2 x_H x_\Phi^3 + 
        1892 g_{1 p1}^4 x_H x_\Phi^3 + 1296 g'^4 x_\Phi^4 \\& + 
        2592 g'^2 g_{1 p1}^2 x_\Phi^4 + 1296 g_{1 p1}^4 x_\Phi^4 + 
        85 g_1^2 y_t^2 + 85 g_{11 p}^2 y_t^2 + 135 g_2^2 y_t^2 + 
        480 g_3^2 y_t^2 + 170 g_{11 p} g' x_H y_t^2 \\& + 
        170 g_1 g_{1 p1} x_H y_t^2 + 85 g'^2 x_H^2 y_t^2 + 
        85 g_{1 p1}^2 x_H^2 y_t^2 + 100 g_{11 p} g' x_\Phi y_t^2 + 
        100 g_1 g_{1 p1} x_\Phi y_t^2 \\& + 100 g'^2 x_H x_\Phi y_t^2 + 
        100 g_{1 p1}^2 x_H x_\Phi y_t^2 + 40 g'^2 x_\Phi^2 y_t^2 + 
        40 g_{1 p1}^2 x_\Phi^2 y_t^2 + 
        15 g_1^2 \textrm {Tr}[Y_\nu Y_\nu^\dag] + 
        15 g_{11 p}^2 \textrm {Tr}[Y_\nu Y_\nu^\dag] \\& + 
        45 g_2^2 \textrm {Tr}[Y_\nu Y_\nu^\dag] + 
        30 g_{11 p} g' x_H \textrm {Tr}[Y_\nu Y_\nu^\dag] + 
        30 g_1 g_{1 p1} x_H \textrm {Tr}[Y_\nu Y_\nu^\dag] + 
        15 g'^2 x_H^2 \textrm {Tr}[Y_\nu Y_\nu^\dag] \\& + 
        15 g_{1 p1}^2 x_H^2 \textrm {Tr}[Y_\nu Y_\nu^\dag] + 
        60 g_{11 p} g' x_\Phi \textrm {Tr}[Y_\nu Y_\nu^\dag] + 
        60 g_1 g_{1 p1} x_\Phi \textrm {Tr}[Y_\nu Y_\nu^\dag] + 
        60 g'^2 x_H x_\Phi \textrm {Tr}[Y_\nu Y_\nu^\dag] \\& + 
        60 g_{1 p1}^2 x_H x_\Phi \textrm {Tr}[
          Y_\nu Y_\nu^\dag] + 
        120 g'^2 x_\Phi^2 \textrm {Tr}[Y_\nu Y_\nu^\dag] + 
        120 g_{1 p1}^2 x_\Phi^2 \textrm {Tr}[Y_\nu Y_\nu^\dag] - 
        162 y_t^4 - 
        54 \textrm {Tr}[Y_\nu Y_\nu^\dag Y_\nu Y_\nu^\dag] \\& - 
        18 \textrm {Tr}[
           y_{NS} y_{NS}^\dag Y_\nu^T Y_\nu^*]\Big) Y_\nu + 
     3 \Big (3 (31 g_1^2 + 31 g_{11 p}^2 + 45 g_2^2 - 
            64  \lambda_1 - 10 g_{11 p} g' x_H - 
            10 g_1 g_{1 p1} x_H  \\& + 31 g'^2 x_H^2 + 
            31 g_{1 p1}^2 x_H^2 + 52 g_{11 p} g' x_\Phi + 
            52 g_1 g_{1 p1} x_\Phi + 52 g'^2 x_H x_\Phi + 
            52 g_{1 p1}^2 x_H x_\Phi + 64 g'^2 x_\Phi^2 \\& - 36 y_t^2 - 
            12 \textrm {Tr}[
              Y_\nu Y_\nu^\dag]) (Y_\nu Y_\nu^\dag Y_\nu ) - 4 (8  \lambda_3 - 11 (g'^2 + g_{1 p1}^2) x_\Phi^2 + 
            3 \textrm {Tr}[y_{NS} y_{NS}^\dag]) (Y_\nu y_{NS}^* y_{NS}^T ) \\& + 
         24 (Y_\nu Y_\nu^\dag Y_\nu Y_\nu^\dag) - 
         2 (Y_\nu y_{NS}^* y_{NS}^T Y_\nu^\dag Y_\nu) - 
         2 (Y_\nu y_{NS}^* y_{NS}^T y_{NS}^* y_{NS}^T)\Big)\Big) \end{split} \ee

\be  \begin{split} \beta_{y_t^{(2)}} = & \frac {1} {432}\Big (2 \Big (1187 g_1^4 + 
        2374 g_1^2 g_{11 p}^2 + 1187 g_{11 p}^4 - 
        162 g_1^2 g_2^2 - 162 g_{11 p}^2 g_2^2 - 1242 g_2^4 + 
        456 g_1^2 g_3^2 \\& + 456 g_{11 p}^2 g_3^2 + 
        1944 g_2^2 g_3^2 - 23328 g_3^4 + 1296  \lambda_1^2 + 
        108  \lambda_3^2 + 7164 g_1^2 g_{11 p} g' x_H + 
        7164 g_{11 p}^3 g' x_H \\& + 7164 g_1^3 g_{1 p1} x_H + 
        7164 g_1 g_{11 p}^2 g_{1 p1} x_H + 
        2106 g_{11 p} g' g_2^2 x_H + 
        2106 g_1 g_{1 p1} g_2^2 x_H - 
        2544 g_{11 p} g' g_3^2 x_H \\& - 
        2544 g_1 g_{1 p1} g_3^2 x_H + 4160 g_1^2 g'^2 x_H^2 + 
        9470 g_{11 p}^2 g'^2 x_H^2 + 
        10620 g_1 g_{11 p} g' g_{1 p1} x_H^2 + 
        9470 g_1^2 g_{1 p1}^2 x_H^2 \\& + 
        4160 g_{11 p}^2 g_{1 p1}^2 x_H^2 - 162 g'^2 g_2^2 x_H^2 - 
        162 g_{1 p1}^2 g_2^2 x_H^2 + 456 g'^2 g_3^2 x_H^2 + 
        456 g_{1 p1}^2 g_3^2 x_H^2 + 7164 g_{11 p} g'^3 x_H^3 \\& + 
        7164 g_1 g'^2 g_{1 p1} x_H^3 + 
        7164 g_{11 p} g' g_{1 p1}^2 x_H^3 + 
        7164 g_1 g_{1 p1}^3 x_H^3 + 1187 g'^4 x_H^4 + 
        2374 g'^2 g_{1 p1}^2 x_H^4 \\& + 1187 g_{1 p1}^4 x_H^4 + 
        4016 g_1^2 g_{11 p} g' x_\Phi + 
        4016 g_{11 p}^3 g' x_\Phi + 4016 g_1^3 g_{1 p1} x_\Phi + 
        4016 g_1 g_{11 p}^2 g_{1 p1} x_\Phi \\& + 
        486 g_{11 p} g' g_2^2 x_\Phi + 
        486 g_1 g_{1 p1} g_2^2 x_\Phi - 
        480 g_{11 p} g' g_3^2 x_\Phi - 
        480 g_1 g_{1 p1} g_3^2 x_\Phi + 
        5524 g_1^2 g'^2 x_H x_\Phi \end{split} \nonumber \ee \be \begin{split} \,\,\,\,\,\,\,\,\,\,\,\,\,\,\,\, & + 
        14344 g_{11 p}^2 g'^2 x_H x_\Phi + 
        17640 g_1 g_{11 p} g' g_{1 p1} x_H x_\Phi + 
        14344 g_1^2 g_{1 p1}^2 x_H x_\Phi + 
        5524 g_{11 p}^2 g_{1 p1}^2 x_H x_\Phi \\& + 
        486 g'^2 g_2^2 x_H x_\Phi + 
        486 g_{1 p1}^2 g_2^2 x_H x_\Phi - 
        480 g'^2 g_3^2 x_H x_\Phi - 
        480 g_{1 p1}^2 g_3^2 x_H x_\Phi + 
        14344 g_{11 p} g'^3 x_H^2 x_\Phi \\& + 
        14344 g_1 g'^2 g_{1 p1} x_H^2 x_\Phi + 
        14344 g_{11 p} g' g_{1 p1}^2 x_H^2 x_\Phi + 
        14344 g_1 g_{1 p1}^3 x_H^2 x_\Phi + 
        4016 g'^4 x_H^3 x_\Phi \\& + 8032 g'^2 g_{1 p1}^2 x_H^3 x_\Phi + 
        4016 g_{1 p1}^4 x_H^3 x_\Phi + 1638 g_1^2 g'^2 x_\Phi^2 + 
        6030 g_{11 p}^2 g'^2 x_\Phi^2 + 
        8784 g_1 g_{11 p} g' g_{1 p1} x_\Phi^2 \\& + 
        6030 g_1^2 g_{1 p1}^2 x_\Phi^2 + 
        1638 g_{11 p}^2 g_{1 p1}^2 x_\Phi^2 + 
        162 g'^2 g_2^2 x_\Phi^2 + 162 g_{1 p1}^2 g_2^2 x_\Phi^2 - 
        192 g'^2 g_3^2 x_\Phi^2 - 192 g_{1 p1}^2 g_3^2 x_\Phi^2  \\& + 
        15160 g_{11 p} g'^3 x_H x_\Phi^2 + 
        15160 g_1 g'^2 g_{1 p1} x_H x_\Phi^2 + 
        15160 g_{11 p} g' g_{1 p1}^2 x_H x_\Phi^2 + 
        15160 g_1 g_{1 p1}^3 x_H x_\Phi^2 \\& + 
        6030 g'^4 x_H^2 x_\Phi^2 + 
        12060 g'^2 g_{1 p1}^2 x_H^2 x_\Phi^2 + 
        6030 g_{1 p1}^4 x_H^2 x_\Phi^2 + 
        4660 g_{11 p} g'^3 x_\Phi^3 + 
        4660 g_1 g'^2 g_{1 p1} x_\Phi^3 \\& + 
        4660 g_{11 p} g' g_{1 p1}^2 x_\Phi^3 + 
        4660 g_1 g_{1 p1}^3 x_\Phi^3 + 4660 g'^4 x_H x_\Phi^3 + 
        9320 g'^2 g_{1 p1}^2 x_H x_\Phi^3 + 
        4660 g_{1 p1}^4 x_H x_\Phi^3 \\& + 1360 g'^4 x_\Phi^4 + 
        2720 g'^2 g_{1 p1}^2 x_\Phi^4 + 1360 g_{1 p1}^4 x_\Phi^4 + 
        765 g_1^2 y_t^2 + 765 g_{11 p}^2 y_t^2 + 
        1215 g_2^2 y_t^2 + 4320 g_3^2 y_t^2 \\& + 
        1530 g_{11 p} g' x_H y_t^2 + 1530 g_1 g_{1 p1} x_H y_t^2 + 
        765 g'^2 x_H^2 y_t^2 + 765 g_{1 p1}^2 x_H^2 y_t^2 + 
        900 g_{11 p} g' x_\Phi y_t^2 \\& + 
        900 g_1 g_{1 p1} x_\Phi y_t^2 + 900 g'^2 x_H x_\Phi y_t^2 + 
        900 g_{1 p1}^2 x_H x_\Phi y_t^2 + 360 g'^2 x_\Phi^2 y_t^2 + 
        360 g_{1 p1}^2 x_\Phi^2 y_t^2 \\& + 
        135 g_1^2 \textrm {Tr}[Y_\nu Y_\nu^\dag] + 
        135 g_{11 p}^2 \textrm {Tr}[Y_\nu Y_\nu^\dag] + 
        405 g_2^2 \textrm {Tr}[Y_\nu Y_\nu^\dag] + 
        270 g_{11 p} g' x_H \textrm {Tr}[Y_\nu Y_\nu^\dag] \\& + 
        270 g_1 g_{1 p1} x_H \textrm {Tr}[Y_\nu Y_\nu^\dag] + 
        135 g'^2 x_H^2 \textrm {Tr}[Y_\nu Y_\nu^\dag] + 
        135 g_{1 p1}^2 x_H^2 \textrm {Tr}[Y_\nu Y_\nu^\dag] + 
        540 g_{11 p} g' x_\Phi \textrm {Tr}[Y_\nu Y_\nu^\dag] \\&+ 
        540 g_1 g_{1 p1} x_\Phi \textrm {Tr}[
          Y_\nu Y_\nu^\dag] + 
        540 g'^2 x_H x_\Phi \textrm {Tr}[Y_\nu Y_\nu^\dag] + 
        540 g_{1 p1}^2 x_H x_\Phi \textrm {Tr}[
          Y_\nu Y_\nu^\dag] + 
        1080 g'^2 x_\Phi^2 \textrm {Tr}[Y_\nu Y_\nu^\dag] \\&+ 
        1080 g_{1 p1}^2 x_\Phi^2 \textrm {Tr}[
          Y_\nu Y_\nu^\dag] - 1458 y_t^4 - 
        486 \textrm {Tr}[Y_\nu Y_\nu^\dag Y_\nu Y_\nu^\dag] - 
        162 \textrm {Tr}[y_{NS} y_{NS}^\dag Y_\nu^T Y_\nu^*]\Big) y_t \\&- 
     9 \Big (-(223 g_1^2 + 223 g_{11 p}^2 + 405 g_2^2 + 
             768 g_3^2 - 576  \lambda_1 - 202 g_{11 p} g' x_H  \\&- 
             202 g_1 g_{1 p1} x_H + 223 g'^2 x_H^2 + 
             223 g_{1 p1}^2 x_H^2 + 100 g_{11 p} g' x_\Phi + 
             100 g_1 g_{1 p1} x_\Phi + 100 g'^2 x_H x_\Phi \\& + 
             100 g_{1 p1}^2 x_H x_\Phi + 64 g'^2 x_\Phi^2 + 
             64 g_{1 p1}^2 x_\Phi^2 - 324 y_t^2 - 
             108 \textrm {Tr}[Y_\nu Y_\nu^\dag]) y_t^3 + 
         12 (-6 y_t^5)\Big)\Big)  \end{split} \ee

\end{appendices}
\bibliographystyle{utphys}
\bibliography{tevportalnew}

\end{document}